\documentclass{aa}  
\usepackage{natbib}
\usepackage{graphicx}
\usepackage{gensymb}
\usepackage[toc,page]{appendix}
\usepackage{pbox}
\usepackage{makecell}
\usepackage[flushleft]{threeparttable}
\usepackage{hyperref}
\usepackage{multirow}
\usepackage{multicol}
\usepackage{subcaption}
\hypersetup{colorlinks=true,breaklinks=true,citecolor=blue}
\usepackage{xcolor}
\usepackage{lscape}
\usepackage{hhline}
\usepackage[export]{adjustbox} 
\usepackage{comment}
\usepackage{placeins}
\usepackage{txfonts}
\usepackage{xspace}
\bibpunct{(}{)}{;}{a}{}{,}

		\newcommand{\z}{\textit{z}\xspace}

            \newcommand{\nirspec}{NIRSpec\xspace}
            \newcommand{\ganifs}{GA-NIFS\xspace}
            
            \newcommand{\snr}{S/N\xspace}

		\newcommand{\oi}{[O{ I}]\xspace}
		\newcommand{\oii}{[O{ II}]\xspace}
		\newcommand{\nii}{[N{ II}]\xspace}
		\newcommand{\sii}{[S{ II}]\xspace}
		\newcommand{\oiii}{[O{ III}]\xspace}
            \newcommand{\Oiii}{{\rm [O{ III}]}\xspace}

		\newcommand{\hal}{H$\alpha$\xspace}
		\newcommand{\hbe}{H$\beta$\xspace}
		\newcommand{\hga}{H$\gamma$\xspace}
		\newcommand{\hde}{H$\delta$\xspace}

		\newcommand{\Msun}{\,{{M}_{\odot}}}
		
		\newcommand{\kms} {\,{\rm km\,s}^{-1}}

		\newcommand{\Mout}{M_{\rm out}}
		\newcommand{\Mstar}{M_{\star}}

		\newcommand{\lumcgs}{erg~s$^{-1}$\xspace}
		\newcommand{\Lumcgs}{\rm erg~s^{-1\xspace}}

		\newcommand{\Lbol}{{L_{\rm bol}}}

		\newcommand{\Vout}{v_{\rm out}}
		\newcommand{\Vmax}{v_{\rm max}}
            
		\newcommand{\Rmax}{R_{\rm max}\xspace}
		\newcommand{\Rout}{R_{\rm out}\xspace}
		\newcommand{\Edot}{\dot{E}_{\rm out}\xspace}
		\newcommand{\Eout}{E_{\rm out}\xspace}

		\newcommand{\Mdot}{\dot{M}_{\rm out}\xspace}

\usepackage{graphicx}
\usepackage{txfonts}

\begin{document}

   \title{GA-NIFS: Mapping $z\simeq3.5$ AGN-driven ionized outflows in the COSMOS field}

   \authorrunning{E.~Bertola et~al.}

   \author{E. Bertola\inst{\ref{OAA}}, 
            G. Cresci\inst{\ref{OAA}},
            G. Venturi\inst{\ref{iNorm}},
            M. Perna\inst{\ref{iCAB}}, 
            C. Circosta\inst{\ref{iram},\ref{ESA}},
            G. Tozzi\inst{\ref{MPE}},
            I. Lamperti\inst{\ref{iUNIFI}, \ref{OAA}},
            C. Vignali\inst{\ref{iUNIBO},\ref{iOAS}},
              S. Arribas\inst{\ref{iCAB}},
              A.~J.~Bunker\inst{\ref{iOxf}},
              S.~Charlot\inst{\ref{iSor}},
              S.~Carniani\inst{\ref{iNorm}},
              R.~Maiolino\inst{\ref{iKav},\ref{iCav},\ref{ucl}},
              B.~Rodr\'iguez Del Pino\inst{\ref{iCAB}},
              H.~\"Ubler\inst{\ref{MPE}},
              C.~J.~Willott\inst{\ref{iNRC}},
          T.~B\"oker\inst{\ref{iESAusa}},
          M.~A.~Marshall\inst{\ref{iLOS}},
          E.~Parlanti\inst{\ref{iNorm}},
          J.~Scholtz\inst{\ref{iKav},\ref{iCav}}
            }

   \institute{
        INAF--OAA, Osservatorio Astrofisico di Arcetri, largo E. Fermi 5, 50127, Firenze, Italy \label{OAA}
   \and
        Scuola Normale Superiore, Piazza dei Cavalieri 7, I-56126 Pisa, Italy\label{iNorm}
    \and
        Centro de Astrobiolog\'ia (CAB), CSIC--INTA, Cra. de Ajalvir Km.~4, 28850 -- Torrej\'on de Ardoz, Madrid, Spain\label{iCAB}
   \and
   Institut de Radioastronomie Millim\'{e}trique (IRAM), 300 Rue de la Piscine, 38400 Saint-Martin-d'H\`{e}res, France \label{iram}
   \and
        European Space Agency, ESAC, Villanueva de la Ca\~{n}ada, E-28692 Madrid, Spain \label{ESA}
    \and
        Max-Planck-Institut f\"ur extraterrestrische Physik (MPE), Gie{\ss}enbachstra{\ss}e 1, 85748 Garching, Germany\label{MPE}
    \and
        Università di Firenze, Dipartimento di Fisica e Astronomia, via G. Sansone 1, 50019 Sesto F.no, Firenze, Italy\label{iUNIFI}
    \and 
        Dipartimento di Fisica e Astronomia ``Augusto Righi'', Universit\`a degli Studi di Bologna, via P. Gobetti 93/2, 40129 Bologna, Italy\label{iUNIBO}
    \and 
        INAF--OAS, Osservatorio di Astrofisica e Scienza dello Spazio di Bologna, via P. Gobetti 93/3, 40129 Bologna, Italy\label{iOAS}
    \and
            Department of Physics, University of Oxford, Denys Wilkinson Building, Keble Road, Oxford OX1 3RH, UK\label{iOxf}
    \and
            Sorbonne Universit\'e, CNRS, UMR 7095, Institut d’Astrophysique de Paris, 98 bis bd Arago, 75014 Paris, France\label{iSor} 
    \and
            Kavli Institute for Cosmology, University of Cambridge, Madingley Road, Cambridge, CB3 0HA, UK\label{iKav}
    \and
            Cavendish Laboratory - Astrophysics Group, University of Cambridge, 19 JJ Thomson Avenue, Cambridge, CB3 0HE, UK\label{iCav}
    \and
        Department of Physics and Astronomy, University College London, Gower Street, London WC1E 6BT, UK\label{ucl}
    \and
            NRC Herzberg, 5071 West Saanich Rd, Victoria, BC V9E 2E7, Canada\label{iNRC}
    \and    
            European Space Agency, c/o STScI, 3700 San Martin Drive, Baltimore, MD 21218, USA\label{iESAusa}
    \and
            Los Alamos National Laboratory, Los Alamos, NM 87545, USA\label{iLOS}
    \\  
    \email{elena.bertola@inaf.it}
             }

 
  \abstract
  {Active galactic nuclei (AGNi) are a key ingredient in galaxy evolution and possibly shape galaxy growth through the generation of powerful outflows. Little is known regarding AGN-driven ionized outflows in moderate-luminosity AGNi ($\log(\Lbol/\Lumcgs)<47$) beyond cosmic noon ($z\gtrsim3$). 
In this work we present the first systematic analysis of the ionized outflow properties of 
a sample of seven X-ray-selected AGNi ($\log(L_{\rm X}/\Lumcgs)>44$) from the COSMOS-Legacy field at $z\simeq3.5$ and with $\log(\Lbol/\Lumcgs)=45.2-46.7$ by using JWST NIRSpec/IFU near-IR spectroscopic observations as part of the ``Galaxy Assembly with NIRSpec IFS'' (GA-NIFS) program. We spectrally isolated and spatially resolved the ionized outflows by performing a multi-component kinematic decomposition of the rest-frame optical emission lines. JWST/\nirspec IFU data also revealed a wealth of close-by companions, of both non-AGN and AGN nature, and ionized gas streams likely tracing tidal structures and large-scale ionized gas nebulae extending up to the circumgalactic medium. Ionized outflows were detected in all COS-AGNi targets, with outflow masses in the range $1.5-11\times10^6\Msun$, outflow velocities in the range $\simeq570-3200~\kms$, and mass outflow rates in the range $\simeq1.4-40~\Msun~\rm yr^{-1}$.
  We compared the outflow properties of AGNi presented in this work with previous results from the 
  literature up to $z\simeq3$, which were opportunely (re-)computed for a coherent comparison. We normalized outflow energetics ($\Mdot$, $\Edot$) to the outflow density in order to standardize the various assumptions that were made in the literature. Our choice is equal to assuming that each outflow has the same gas density. We find GA-NIFS AGNi to show outflows consistent with literature results, within the large scatter shown by the collected measurements, thus suggesting no strong evolution with redshift in terms of total mass outflow rate, energy budget, and outflow velocity for fixed bolometric luminosity. Moreover, we find no clear redshift evolution of the ratio of mass outflow rate and  kinetic power over AGNi bolometric luminosity beyond $z>1$. In general, our results indicate no significant evolution of the physics driving outflows beyond $z\simeq3$. } 
   


   \maketitle
%

\section{Introduction}
A key phase of galaxy evolution is the fast transition ($<2$~Gyr) from early galaxy and black-hole assembly at cosmic dawn (i.e., $z>6$) to the peak of both the cosmic star formation (SF)  rate and the supermassive black hole (SMBH) accretion density at cosmic noon \citep[$z\simeq1-3$; e.g.,][and references therein]{madaudickinson2014}. 
Despite the present evidence for a key role of active galactic nuclei (AGNi; \citep[e.g.,][]{Magorrian1998,strateva2001,mcconell2011,harrison2017,caglar2023}), a comprehensive understanding of AGN-galaxy coevolution  and feedback processes is still far off \citep[e.g.,][]{kormendy2013,harrison2018,harrison_ramosalmeida2024}. 

Accretion onto an SMBH is a very powerful process that can release great amounts of energy into the interstellar medium (ISM) of galaxies, giving rise to AGN-driven winds at all scales and in all gas phases \citep[e.g.,][]{cicone2018_NatAs,costa2020,harrison_ramosalmeida2024,ward2024}. 
Powerful AGN-driven outflows are indeed one of the main channels to establish the AGN-galaxy coevolution  \citep[e.g.,][]{kingpounds2015,costa2018b}. Especially in the past fifteen years, great effort has been expended by the community to characterize AGN-driven winds and their impact on AGNi host galaxies \citep[e.g., see][for recent reviews]{crescimaiolino2018NatAs,veilleux2020,laha2021,harrison_ramosalmeida2024}. 
Such winds originate from the inner regions of galaxies and are observable in X-rays \citep[e.g.,][]{cappi2006,igo_2020,chartas2021,matzeu2023}, and they expand up to galactic scales as neutral \citep[e.g.,][]{morganti2016_mrk231}, ionized \citep[e.g.,][]{harrison2012,genzel2014,carniani2015,carniani2017,foersterschreiber2019,kakkad2020,cresci2023,tozzi2024}, and molecular outflows \citep[e.g.,][]{feruglio2017,brusa2018}, which are traced in the optical and submillimeter rest-frame bands \citep[e.g.,][]{cicone2018_NatAs}. Studies in the local Universe \citep[e.g.,][]{fluetsch2019} up to cosmic dawn \citep[e.g.,][]{bischetti2022} have demonstrated the ubiquitous presence of such AGN-driven winds, which often carry a kinetic power that could indeed affect the SF of their hosts \citep[$\Edot/\Lbol>0.5-5\%$,][]{dimatteo2005,hopkins2010}. 

The ionized outflow component has been traced mainly through the \oiii and \hal optical emission lines \citep[e.g.,][]{cano-diaz2012,perna2015_z1.5,harrison2016,leung2019_mosdef}, probing gas motions up to kiloparsec scales from the AGN. On the one hand, tracing outflows with the \hal emission line offers the advantage of converting the line luminosity into ionized gas mass without assumptions regarding the gas metallicity \citep{cresci2015a,carniani2015}, but it risks possible contamination of emission from the broad-line region (BLR) in Type 1 AGNi. On the other hand, the \oiii doublet is a forbidden line transition that can only originate in low-density regions. Thus in principle, it is free of contamination by emission from a higher-density region, such as that of the BLR, yet the gas metallicity has to be known (or assumed) in order to convert its luminosity into ionized gas mass. Moreover, \hal emission is often less affected by outflows and therefore can be used to better identify rotating motions in the ISM (e.g., \citealp{perna2022}) as well as gas ionized by star-forming regions in the AGN host. The latter is particularly important to unveiling possible anticorrelations between outflowing gas and SF gas and thus to testing feedback scenarios \citep[e.g.,][but see also \citealp{scholtz2020}]{cresci_2015b_magnum,crescimaiolino2018NatAs}. 
While spatially integrated spectra can efficiently probe ionized outflows beyond the local Universe \citep[e.g.,][]{perna2015_z2.5,leung2019_mosdef,temple2024}, only integral field spectroscopy (IFS) allows unperturbed motions (e.g., rotation) to be disentangled from galaxy-wide outflows \citep[e.g.,][]{venturi2018,gallagher2019_manga,zanchettin2023,travascio2024_pdsmuse,ulivi2024,speranza2024}. However, ground-based near-infrared (NIR) integral field units (IFUs) can observe the rest-frame ionized emission only up to $z\simeq2.6$ in the \hal line and up to $z\simeq3.5$ in the \oiii line, while for higher redshift, these lines are redshifted out of the K band \citep[e.g.,][]{cano-diaz2012,harrison2012,carniani2015,carniani2016,davies2020,kakkad2020,vayner2021a,vayner2021b,tozzi2024}. 

Powerful AGNi are expected to drive the most prominent outflows \citep[e.g.,][]{cicone2014,fiore2017,fluetsch2019,musiimenta2023}. They are also the least challenging systems to observe beyond the local Universe and have allowed in-depth analyses up to $z\gtrsim3$ \citep[e.g.,][]{bischetti2017,perrotta2019}. However, it is critical to investigate AGN feedback also in galaxies hosting AGN of less extreme power ($\log(\Lbol/\Lumcgs)<47$) since they are the majority of the AGNi population. This was well addressed at cosmic noon by the KASH\textit{z} \citep[KMOS AGN Survey at High redshift, PI: D. Alexander;][Scholtz et al., in prep.]{harrison2016} and the SUPER\footnote{\url{http://www.super-survey.org/}} \citep[SINFONI Survey for Unveiling the Physics and Effect of Radiative feedback, PI: V. Mainieri;][]{circosta2018} surveys. Investigating AGN-driven outflows and feedback effects in sources that were selected solely based on their X-ray luminosity, both surveys showed that powerful AGN-driven ionized outflows are common at cosmic noon \citep[][see also \citealp{vietri2020} for UV-traced winds]{harrison2016,kakkad2020,tozzi2024}, but the surveys were limited to $z\lesssim2.6$ by the goal of also mapping the unobscured SF in the \hal line.

Until recently, outflows in the $z\simeq3-7$ redshift range could only be probed through spatially integrated rest-frame UV data \citep[e.g.,][]{yang2021,vietri2022} and in the submillimeter bands \citep{bischetti2019b,puglisi2021}. 
The James Webb Space Telescope (JWST; \citealp{gardner2023_jwst}) now provides the first optical rest-frame view of AGNi in this redshift range and the additional benefit of IFS at sub-kiloparsec resolution using the \nirspec IFU mode \citep{boeker2022_nirspec,jakobsen2022_nirspec}. The capabilities of JWST range from observing the ionized gas component in host galaxies of high-redshift quasars up to $z>7$ \citep[e.g.,][]{Harikane2023,cresci2023,loiacono2024,liu2024_nirspec_z7} to resolving the rich environment in the immediate neighborhood of AGNi, in which case it often shows companions and merger features \citep[e.g.,][]{perna2025dualsamp,perna2023dualwiggles,marshall2024arXiv}. 

In this work, we present the analysis of a sample of AGNi selected from the X-ray COSMOS-Legacy deep field \citep{marchesi2016} and targeted by the JWST Near Infrared Spectrograph (\nirspec) within the 
Guaranteed Time Observations (GTO) program ``Galaxy Assembly with NIRSpec IFS'' (GA-NIFS\footnote{\url{https://ga-nifs.github.io}}; PI: R. Maiolino and S. Arribas).
We present our sample in Sect. \ref{sec:sample}, the data reduction in Sect. \ref{sec:data_red}, and the data analysis in Sect. \ref{sec:data_analysis}. We discuss our results in Sect. \ref{sec:results}, where we present the properties of each source and the derivation of outflow properties. In Sect. \ref{sec:discussion}, we compare the outflows of COS-AGNi with results present in the literature, and we summarize our results in Sect. \ref{sec:summary}. Throughout this work, we adopt a \cite{Chabrier2003} initial mass function ($0.1-100~M_\odot$) and a flat ${\Lambda}$CDM cosmology \citep{planckcoll2020}, with $H_{\rm 0}=67.7\ {\rm km\ s^{-1}\ Mpc^{-1}}$ and $\Omega_{\rm m,0}=0.31$ throughout the paper. For all the maps shown in this work, north is up and east is to the left. 

\begin{table*}[!t]
    \centering
    \caption{Summary of AGN and host galaxy properties of the COS-AGNi GA-NIFS sample.}
    \label{tab:sed}
       \resizebox{\textwidth}{!}{
        \begin{tabular}{lccccccccccc}
	\hline\hline
	Target & RA & DEC &	$z_{\rm spec}$ & Type & $\log N_{\rm H}$&  $\log L_{\rm 2-10 keV}$ &  $\log\Lbol$                & $\log\Mstar$             & SFR                     & $A_{\rm V}^{\rm tot}$ & $A_{\rm V}^{\rm nar}$  \\ 
&   (deg)      &   (deg)   &	  &  & ($\rm cm^{-2}$) &  ($\Lumcgs$)  &  ($\Lumcgs$)                & ($\Msun$)                & ($\Msun\rm~yr^{-1}$)       & (mag)                  & (mag)                 \\ 
(1)       & (2)             & (3)            &     (4)                     &    (5)                     & (6) 					    & 		(7)			    & (8)       & (9)       & (10)       & (11)     & (12)    \\
\hline
COS590              & 149.755412 &+2.73853 & 3.52385$\pm$0.00002    & 1 & $<23.2$   & 44.41 & 45.57 $\pm$ 0.14                              & 10.7  $\pm$ 0.07    & $<11$  										    & $0.6\pm0.1$	& $1.3\pm	0.1 $ \\ 
COS1118             &149.879192 &+2.22584 & 3.64300$\pm$0.00003     & 1 & $<20$     & 44.30 & 45.2  $\pm$ 0.15                              & 10.48 $\pm$ 0.17    & $<43$  											& $1.1\pm0.1$	& $0.9\pm	0.1 $ \\ 
COS349              & 150.004377 &+2.03892 &  3.5093$\pm$0.0003     & 1 & $<20$    & 44.26  & 46.17 $\pm$ 0.02                              & $<10.56$            & $<116$ 						 					& $0.9\pm0.1$$^a$	&  --		      \\ 
COS1656-A           & 150.271546 &+1.61383 & 3.5101$\pm$0.0004    & 2 & --        & 44.43\textsuperscript{\textdaggerdbl} & $<45.84$                                      & 10.95 $\pm$ 0.04    & <80  											& $2.2\pm0.9$	&  --           \\ 
COS2949             & 150.402917 &+1.87889 & 2.0478$\pm$0.0003    & 2 &  23.65$^{+0.28}_{-0.23}$$^{\star}$  &  44.48$^{+0.24}_{-0.26}$$^{\star}$ & 45.36 $\pm$ 0.11                              & 10.44 $\pm$ 0.1     & 57 $\pm$ 2 		                                & $3.2\pm0.1$$^b$	 & -- \\ 
COS1638$^\ast$      & --        &   --     & --        & -- & $<23.5$\textsuperscript{\textdaggerdbl}   & 44.50\textsuperscript{\textdaggerdbl} & 46.78 $\pm$ 0.02                              & 11.08 $\pm$ 0.19      & 1947 $\pm$ 58                                &  --           &   --            \\ 
\hspace{0.5em}COS1638-A           & 150.735557 &+2.19953 & 3.5057$\pm$0.0001    & 1 & --        & -- & 46.7 $\pm$ 0.3\textsuperscript{\textdagger}   & --  & -- 													& $0.9\pm0.1$$^a$	& $1.2\pm	0.1 $ \\ 
\hspace{0.5em}COS1638-B           & 150.735847 &+2.19962 & 3.5109$\pm$0.0001    & 2  &--        & -- & 46.2 $\pm$ 0.3\textsuperscript{\textdagger}   & --  & -- 													& $4.2\pm0.8$	&  --           \\ 

	\hline\end{tabular}
       }
 \tablefoot{(1) Target ID; (2)-(3) Coordinates; (4) Spectroscopic redshift determined from the narrow component of the \oiii line; (5) AGN Type; (6)-(7) Column density and X-ray luminosity from \citet[][]{marchesi2016}; (8)-(10) SED fitting results (AGN bolometric luminosity, stellar mass, star formation rate; Circosta et al., in prep.); (11)-(12) Dust extinction for total emission and narrow components, respectively. 
 $^\ast$: The parameters of COS1638 refer to X-ray data and SED fitting performed on photometry that are spatially integrated over AGNi A and B of this dual system. \textsuperscript{\textdagger}: Bolometric luminosity as measured in \citet{perna2025dualsamp}. \textsuperscript{\textdaggerdbl}: Dual AGN that is not spatially resolved in COSMOS-Legacy, X-ray parameters are referred to the integrated system. $^{\star}$: X-ray properties estimated from dedicated fitting of the X-ray spectrum using the spectroscopic redshift measured in this work. $^a$: Dust extinction measured as the median of the other Type 1 COS-AGNi (COS1118, COS590).  $^b$: Dust extinction measured as the median of the other Type 2 COS-AGNi (COS1638-B, COS1656-A). 
 }
\end{table*}
\section{COS-AGNi in GA-NIFS: an uncharted $\Lbol$ range at $z>3$}
\label{sec:sample}
\begin{figure*}[!h]
    \centering
    \includegraphics[width=\textwidth]{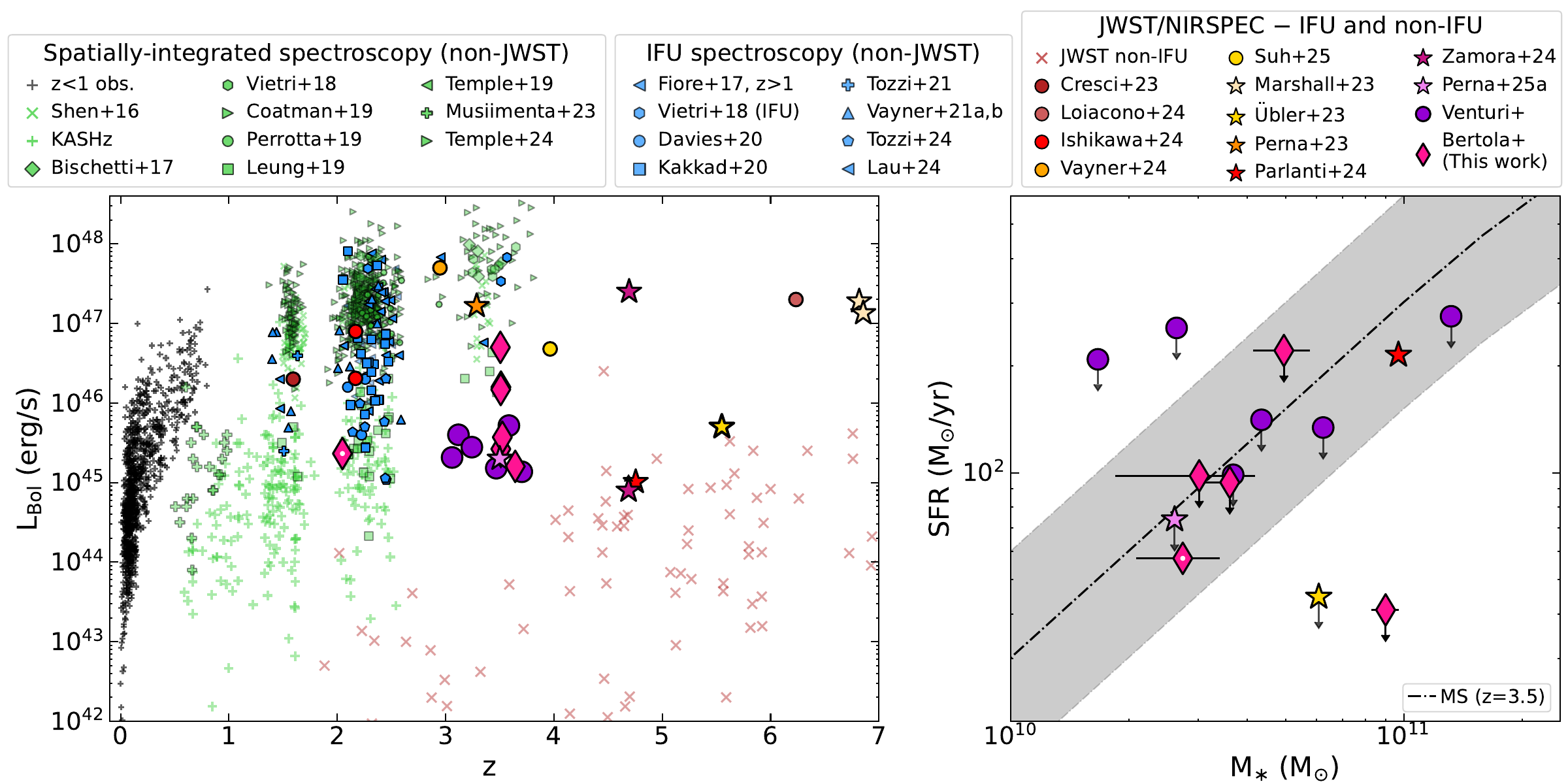}
    \caption{General properties of \ganifs AGNi and summary of IFS observations from the literature. \textit{Left:} Bolometric luminosity versus redshift distribution of the COS-AGNi presented in this work (magenta diamonds). Building on the compilations of \citet{fiore2017} and \citet{circosta2018} (see their Fig. 1), we also show literature AGN for which optical rest-frame spectroscopy is available and has allowed ionized outflows to be probed, to highlight the new bolometric luminosity range probed by GA-NIFS AGNi at $z>3$. Literature targets are color-coded as follows: a (non-exhaustive) compilation of AGNi at $z<1$ is shown as black crosses (see Sect. \ref{sec:sample}); 
    AGN studied through spatially integrated spectra are shown as green symbols (see legend and Sect. \ref{sec:sample} for details); AGNi studied through IFS data are shown as blue symbols (see legend and Sect. \ref{sec:sample} for details). 
    Stars and circles mark JWST/NIRSpec IFU observations from GA-NIFS and other surveys, respectively (see legend and Sect. \ref{sec:sample} for details). GOODS-S AGNi from GA-NIFS that will be presented in Venturi et al. in prep are shown as purple circles. 
     Red x signs show a compilation of AGNi and AGNi candidates newly discovered by JWST through spatially integrated NIRSpec spectra \citep{Scholtz2023arXiv,maiolino2024,Harikane2023,kocevski2023,Carnall2023,Greene2024,Furtak2024,Chisholm2024}. 
    \textit{Right:} Star formation rate versus stellar mass plot of GA-NIFS AGNi to show their relation with the main sequence of star forming galaxies. Both quantities are measured from SED fitting. Color coding is the same as the left panel. A white dot marks COS2949, the COS-AGNi at $z\simeq2.05$. The black dash-dotted line is the main sequence of \citet{schreiber2015} at $z=3.5$, with the shaded area showing the 0.3~dex scatter.}
    \label{fig:sample}
\end{figure*}

The GA-NIFS program targeted 55 objects (AGNi and galaxies) at $z\sim2-11$ with NIRSpec IFU during JWST cycles 1 and 3, totaling more than 300~hr of exposure time. 
In this work we focus on the AGNi selected from COSMOS-Legacy, which consist of six X-ray-selected systems at $z\simeq3.5$ \citep[see][]{perna2025dualsamp}. Such a narrow redshift range was specifically selected to simultaneously observe all the primary optical emission lines, from \oii$\lambda3726,3728$ to \sii$\lambda\lambda6717,6730$, within one single NIRSpec grating (G235H/F170LP). Two of the observed systems are dual AGNi (COS1656 and COS1638), thus increasing the sample size to eight AGNi. We exclude the secondary AGN of the COS1656 system (COS1656-B) because of its complex line profiles, which could be due to tidal interactions, outflows or simply the superposition  with another system along the line of sight \citep{perna2025dualsamp}. Moreover, one AGN (COS2949) was wrongly identified as a $z\simeq3.5$ object in the COSMOS-Legacy catalog, and JWST/NIRSpec data unambiguously uncovered a $z\simeq2$ AGN spectrum (see Sect. \ref{sec:single_sources}).  Despite its lower redshift ($z\simeq2$), we do not exclude COS2949 from our sample, and we highlight it in the plots to differentiate it from the other COS-AGNi at $z\simeq3.5$.  
The sample of COSMOS AGNi used in this work (COS-AGNi sample henceforth) comprises only X-ray-selected AGN (and their secondary, if dual), totaling seven AGN with median redshift $z\simeq3.5$: four Type 1s (COS1118, COS590, COS1638-A, COS349) and three Type 2s (COS1656-A, COS1638-B, COS2949). AGN classification in Type 1/Type 2  based on the presence/absence of AGN broad lines in the rest-frame UV/optical data is part of the COSMOS-Legacy catalog of \citet{marchesi2016} and JWST data have confirmed it. 
AGN and host galaxy properties were retrieved through dedicated SED fitting with CIGALE \citep{boquien2019_cigale,yang2020_xcigale,yang2022_cigale}, using photometry from the rest-frame UV-optical to FIR, including that available from JWST imaging in the COSMOS field. 
Photometry collection and SED fitting of the full GA-NIFS AGNi sample will be presented in a forthcoming paper (Circosta et al., in prep.), including the derivation of host galaxy properties and AGN bolometric luminosities, which are found to range between L$_{\rm bol}$ $= 10^{45}-10^{47}$ \lumcgs. Table \ref{tab:sed} summarizes the properties of COS-AGNi presented in this work and Figure \ref{fig:sample} (right) shows the SFR--$\Mstar$ distribution of GA-NIFS AGNi, including those selected from GOODS-S (\citealp{Ubler2023a,parlanti2024,perna2025_gs133}; Venturi et al., in prep). 

Figure \ref{fig:sample} (left) shows the $\Lbol$--\z distribution of GA-NIFS AGNi. Building on the compilation in \citet{circosta2018} and \citet{Scholtz2023arXiv}, we also show in the $\Lbol$--\z panel AGNi from the literature that were studied for their AGN-driven ionized outflows. We show both targets investigated with spatially integrated spectra \citep[green markers;][]{shen2016,bischetti2017,vietri2018,coatman2019,perrotta2019,leung2019_mosdef,temple2019,temple2024,musiimenta2023} and IFU data \citep[blue markers;][]{vietri2018,davies2020,kakkad2020,vayner2021a,vayner2021b,tozzi2021,tozzi2024,lau2024_ifuperrotta}\footnote{We refer to \citet{saccheo2023} for the bolometric luminosities of AGNi selected from the WISE/SDSS selected Hyper-luminous quasars sample (WISSH; \citealp{bischetti2017,vietri2018}).}. The compilation by \citet{fiore2017}\footnote{The compilation of \citet{fiore2017} includes results from the following studies: 
\citet{nesvadba2006,nesvadba2008,alexander2010,cano-diaz2012,harrison2012,maiolino2012,harrison2014,rupke_veilleux2013,liu2013a,liu2013b,genzel2014,brusa2015a,cresci2015a,carniani2015,perna2015_z1.5,perna2015_z2.5,cicone2015,zakamska_2016,brusa2016,wylezalek2016,bischetti2017,duras2017_wissh}.} includes both spatially integrated and IFU spectra of AGNi from the local Universe up to $z\lesssim3.4$. We further expand the $z<1$ compilation (black markers) with the following works, not included in \citet{fiore2017}: \citet{liu2014,husemann2013,husemann2014,husemann2017,harrison2014,karouzos2016,rupke2017,perna2017,baron2019,bessiere2024}. For the sources presented in \citeauthor{bessiere2024} that were already included in the compilation of \citet{fiore2017}, we consider the more recent analysis of \citet{bessiere2024}.
We also show AGNi recently targeted by JWST/NIRSpec IFU \citep[][]{cresci2023,suh2025,vayner2024_q3d,loiacono2024}, including results from the GA-NIFS program \citep[][]{marshall2023,Ubler2023a,perna2023dualwiggles,parlanti2024,zamora2024,perna2025_gs133}. Moreover, we include the AGNi newly discovered by JWST through spatially integrated NIRSpec spectra \citep[][]{Scholtz2023arXiv,maiolino2024,Harikane2023,kocevski2023,Carnall2023,Greene2024,Furtak2024,Chisholm2024} up to $z\simeq7$ for showing purposes. 
Figure \ref{fig:sample} highlights the capabilities of JWST/NIRSpec IFU and the novelty of the GA-NIFS sample: we are probing for the first time the ionized gas properties of galaxies harboring AGNi of $45.2<\log(\Lbol/\rm erg\, s^{-1})<46.7$ at $z>2.6$, a $\Lbol$ range that is  well studied only at lower redshifts. Moreover, GA-NIFS AGNi are the first sample of AGNi at $z>3$ for which spatially resolved rest-frame optical spectroscopy is available. We note that this $z-\Lbol$ range is partly probed by Keck/MOSFIRE (spatially integrated) spectra that only cover the \hbe+\oiii complex  ($\log(\Lbol/\rm erg s^{-1})>45.7$, \citealp{trakhtenbrot2016}), the low \snr and lower spectral resolution of which hampered the investigation of outflows. 

\section{Observations and data reduction}
\label{sec:data_red}
\begin{figure*}[!h]
    \centering
    \includegraphics[width=\textwidth]{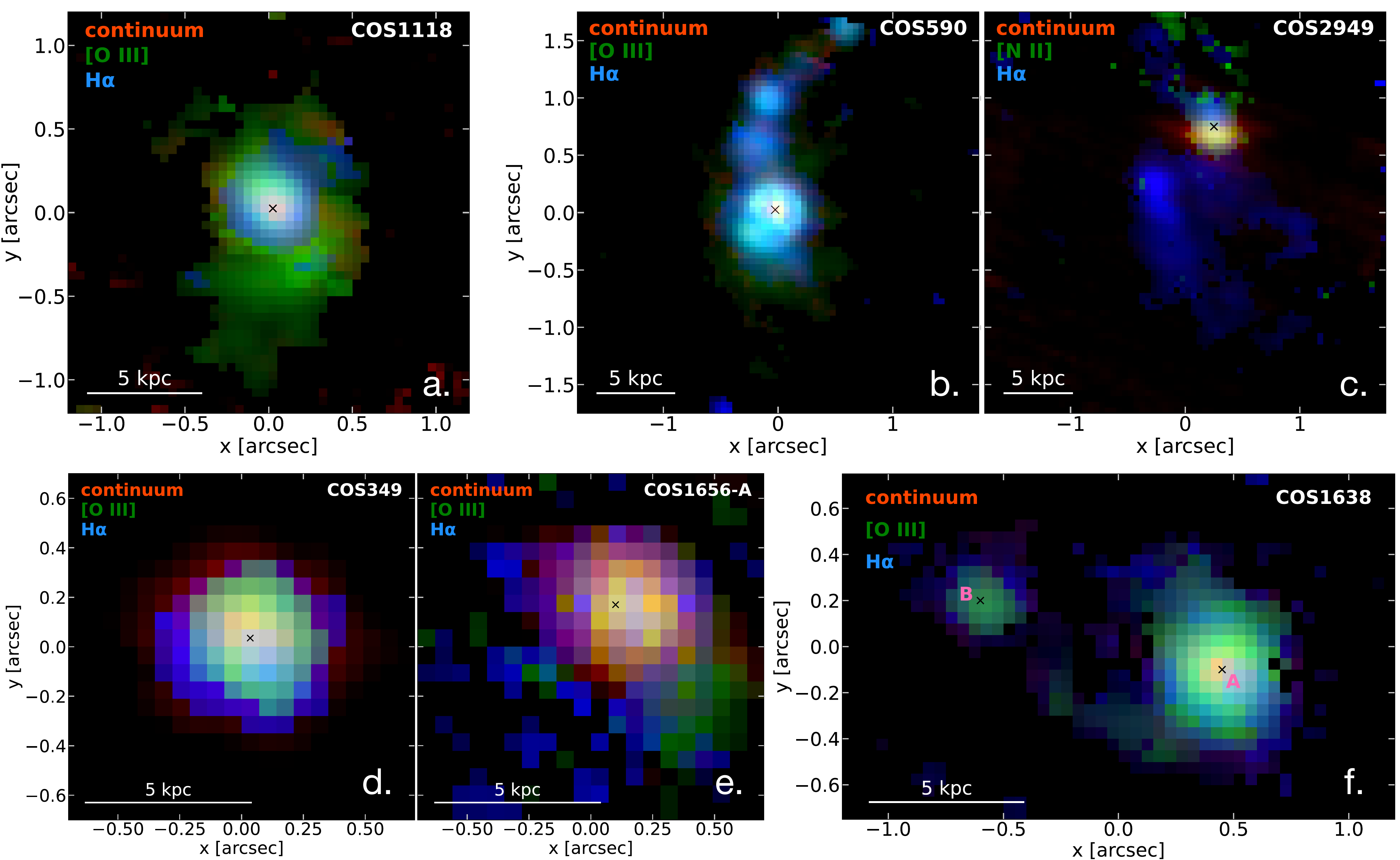}
    \caption{Three-color images of COS-AGNi from GA-NIFS: Panel a. COS1118; Panel b. COS590; Panel c. COS1949; Panel d. COS349; Panel e. COS1656-A; and Panel f. COS1638-A (right) and -B (left). Red is the continuum, green is total \oiii emission, and blue is total \hal emission. For COS2949, green is total \nii emission, to highlight the position of the AGN. Black crosses mark the position of the AGN. The continuum of COS1118, COS1656-A and COS2949 is measured over the full range probed by NIRSpec ($\simeq3500-7000\AA$ for the first two, $6500-10000\AA$ for the latter), after masking the emission lines. The continuum of COS1638-A and COS349 is measured in the $\simeq5500-6300\AA$ range because lower wavelengths are affected by strong Fe~II emission. The continuum of COS590 is measured in the $\simeq$3500--5200$\AA$ range to exclude the noisier channels of the cube. For this target, we also subtract the field noise in the continuum map as the median of the signal for each pixel column. } 
    \label{fig:rgb_images}
\end{figure*}

COS-AGNi were observed for 3560s of on-source time each between April and May 2023 (PID: 1217, PI: Nora L\"utzgendorf) with NIRSpec IFU in the G235H/F170LP grating/filter configuration (resolving power $R\simeq2700$). These observations were tailored at spatially and spectrally isolating fast AGN-driven outflows from the systemic emission of the host galaxy. They cover rest-frame optical nebular emission lines (e.g., \oii$\lambda3726,3728$, H$\beta$, [OIII]$\lambda4959,5007$, H$\alpha$, [NII]$\lambda6548,6583$ and [SII]$\lambda\lambda6717,6730$) providing the unprecedented spatial resolution of 350pc/spaxel at $z\simeq4$ in the NIR (for a 0.05$''$ spaxel size). We present the three-color images of COS-AGNi in Figure \ref{fig:rgb_images}. 

Raw data files were downloaded from the MAST archive and subsequently processed with the JWST Science Calibration pipeline version 1.8.2 under CRDS context jwst\_1068.pmap. We made several modifications to the default three-stages reduction to increase data quality, which are described in detail by \cite{perna2023dualwiggles} and briefly reported here. We patched the pipeline to avoid oversubtraction of the elongated cosmic ray artifacts during Stage1. The individual count-rate frames were further processed at the end of the Stage1 pipeline, to correct for different zero-level in the individual (dither) frames, and subtract the vertical 1/f noise. In addition, we made the following corrections to the *cal.fits files after Stage 2: we masked pixels at the edge of the slices (two pixels wide) to conservatively exclude those with unreliable sflat corrections; we removed regions affected by leakage from failed open shutters; finally, we used a modified version of LACOSMIC \citep{vanDokkum2001} to reject strong outliers before the construction of the final data cube \citep{DEugenio2024NatAs}. 
The final cubes were combined using the drizzle method with pixel scales of 0.05\arcsec, for which we used an official patch to correct a known bug, as implemented in the pipeline v1.9.0 and higher. 

The spatial under-sampling of the point spread function (PSF) of \nirspec produces the so-called ``wiggles'' in the single-spaxel spectra of bright targets. Such an effect is mitigated when extracting 1-D spectra over larger apertures, as a reference, typically larger than 0.2$''$-0.5$''$. \citet{perna2023dualwiggles} developed an algorithm\footnote{The algorithm by \citet{perna2023dualwiggles} is available at \url{https://github.com/micheleperna/JWST-NIRSpec\_wiggles}.} to correct for this effect since there is no  available correction in the data reduction pipeline. We therefore applied this algorithm to correct for the wiggles in our cubes. We briefly summarize it here and refer the interested reader to \citet{perna2023dualwiggles} for additional details. We model the wiggles as a sinusoidal function in the spectrum extracted from the brightest spaxel of the cube. Wiggle modeling is performed after masking the narrow emission lines, the broad line profiles (e.g., we mask the full \hal complex) and the \nirspec spectral gap between the two detectors. Masking the lines is a necessary step to avoid mistaking the broad line wings for an under-sampling-induced modulation of the signal. The wiggles model is built in an iterative approach over small portions of the wavelength range, until the corrected single-spaxel spectrum of the brightest spaxel fairly resembles the integrated spectrum from a larger aperture, in which the effects of wiggles disappear. For our targets, we extract the integrated 1-D spectrum from an aperture centered at the \hal emission peak and of 5-pixel radius (0.25$''$), which we find to be large enough for the wiggles effect to disappear in our data. 

The cubes of COS1638 and COS1656 show the most prominent wiggles, likely because they are the brightest targets in the sample. 
COS1118, COS590 and COS349 present much less prominent wiggles, yet they would still hamper a correct spaxel-by-spaxel fitting and as such we correct their cubes for this effect. We do not observe wiggles in the \nirspec cube of COS2949 and thus such a correction is deemed unnecessary.

\section{Data analysis}
\label{sec:data_analysis}
We present in this section the analysis of the NIRSpec IFU data, which was carried out both in a spaxel-by-spaxel approach (Sect. \ref{sec:spaxelfitting}) and on 1-D spectra (Sect. \ref{sec:int_spec}) integrated over entire spatial extension of the outflows to best determine their properties. Spectroscopic redshifts are estimated from the peak of the \oiii narrow emission component in the integrated spectra. 

\subsection{Spaxel-by-spaxel spectral fitting}
\label{sec:spaxelfitting}
We follow the procedure outlined in \citet{marasco2020} and \citet{tozzi2021} to derive the kinematic properties of ionized gas in the NLR, star-forming regions, and any kpc-scale structure surrounding the host galaxies (e.g., tidal tails and companions)
of  COS-AGNi, as traced by the main rest-frame optical emission lines like H$\beta$, [OIII]$\lambda4759,5007$, H$\alpha$, [NII]$\lambda6548,6583$ and [SII]$\lambda\lambda6717,6730$. This fitting procedure aims at producing cubes of the ionized-gas component by modeling and subtracting any (stellar and/or AGN) continuum and unresolved AGN BLR emission, if present. Hereafter, we will refer to the cubes produced after continuum and AGN BLR subtraction as the ionized-gas cubes. 

As described in \citet{tozzi2021}, this fitting approach is comprised of three main steps for Type 1 AGNi: \textit{i)} first, we extract a high signal-to-noise ratio (\snr) spectrum from the nuclear region of the AGN to fit the broad lines and produce a BLR template; \textit{ii)} we fit the full data cube spaxel-by-spaxel, including a polynomial or power-law function to reproduce the continuum and letting the BLR template vary only in amplitude, and subtract the BLR+continuum best-fit; 
\textit{iii)} we fit the  BLR+continuum-subtracted cube with a spaxel-by-spaxel approach to derive the gas kinematical properties at galaxy scales. For Type 2 AGNi, the fitting procedure comprises only step \textit{ii)} and \textit{iii)} given the absence of emission from the BLR. Both for Type 1 and Type 2 AGNi, we use the nuclear spectrum to measure the source redshift (as listed in Table \ref{tab:sed}) from the narrow peak of the \oiii$\lambda5007$ line in a multi-Gaussian fit (with one to three components based on the least number of components returned by the Kolmogorov-Smirnov test in step \textit{iii)}). We use the \oiii$\lambda5007$ line for all targets but COS2949, whose redshift was anchored to the \hal narrow peak since the \oiii falls blueward of the \nirspec's wavelength range, and for COS1638-A, for which we use the narrow \hal peak since the \oiii line is dominated by the outflows also in the nuclear spectrum. 

\subsubsection{Modeling of BLR lines and continuum template}
In step \textit{i)}, we model the BLR emission lines using a nuclear-emission spectrum integrated from a circular region of radius 1 spaxel (0.05$''$) centered on the spaxel with the highest \hal flux. We use a broken power-law model convolved with a Gaussian core to best reproduce the broad emission line profiles and their wings (\hal, \hbe, \hga, \hde, HeI$\lambda5876$, HeII$\lambda4687$, when present). 
We anchor the line shape to the highest \snr broad line in our data, which for all our Type 1 AGNi is the \hal line. We find the broken power-law best fit from \hal, and then fit the rest of the broad lines by scaling its peak to match the intensity of each of them \citep[e.g.,][]{cresci2023}. We model the broad Fe~II emission using templates produced with the semi-analytic model released by \citet{kovacevic2010}, following the approach of \citet{marasco2020} and \citet{tozzi2021}. To best model the shape of the broad lines, we included also the NLR contribution as two or three Gaussian components, linked across all the emission lines included in the model, to fully reproduce the line profiles and best disentangle the contribution of BLR and NLR. 
Throughout our fitting process, we set the flux ratios of \oiii$\lambda5007/\lambda4959$ (\oiii henceforth), \nii$\lambda6583/\lambda6548$ (\nii henceforth) and \oi$\lambda6300/\lambda6364$ to the theoretical expectation of 3\footnote{\href{https://www.nist.gov/pml/nist-atomic-spectra-bibliographic-databases}{https://www.nist.gov/pml/nist-atomic-spectra-bibliographic-databases}}.  The AGN continuum was modeled either as a polynomial of 2nd or 3rd order or as a power law, depending on the continuum shape of the specific target. The resulting BLR template only includes the best fit of the broad Balmer lines and the Fe~II templates. Figures \ref{fig:blr_of_1}-\ref{fig:blr_of_2} show the best-fit of the nuclear spectra of the Type 1 AGNi used to produce the BLR template, including continuum and BLR emission.

\begin{figure*}[!h]
    \centering
    \includegraphics[width=\textwidth]{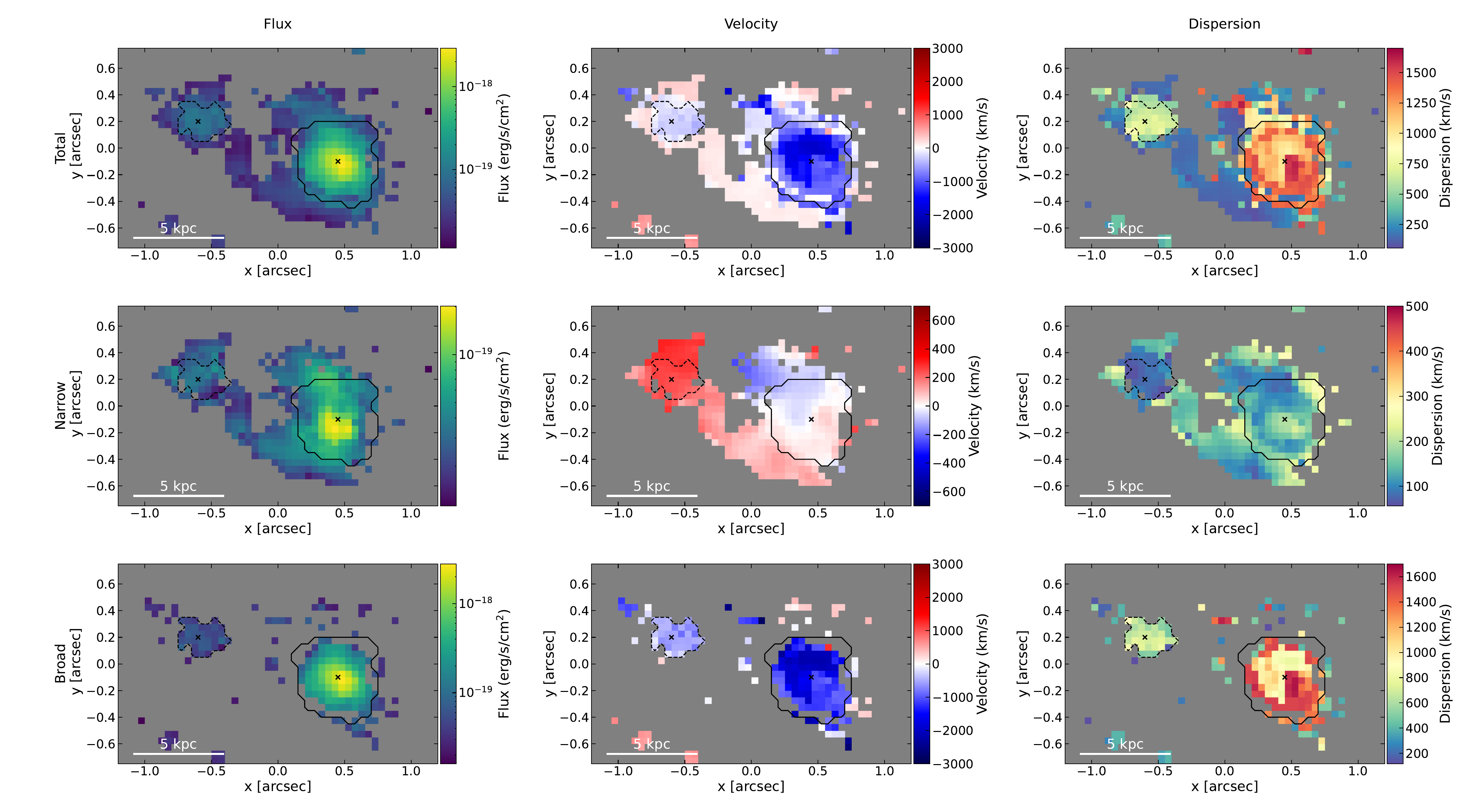}
    \\
    \includegraphics[width=0.45\textwidth]{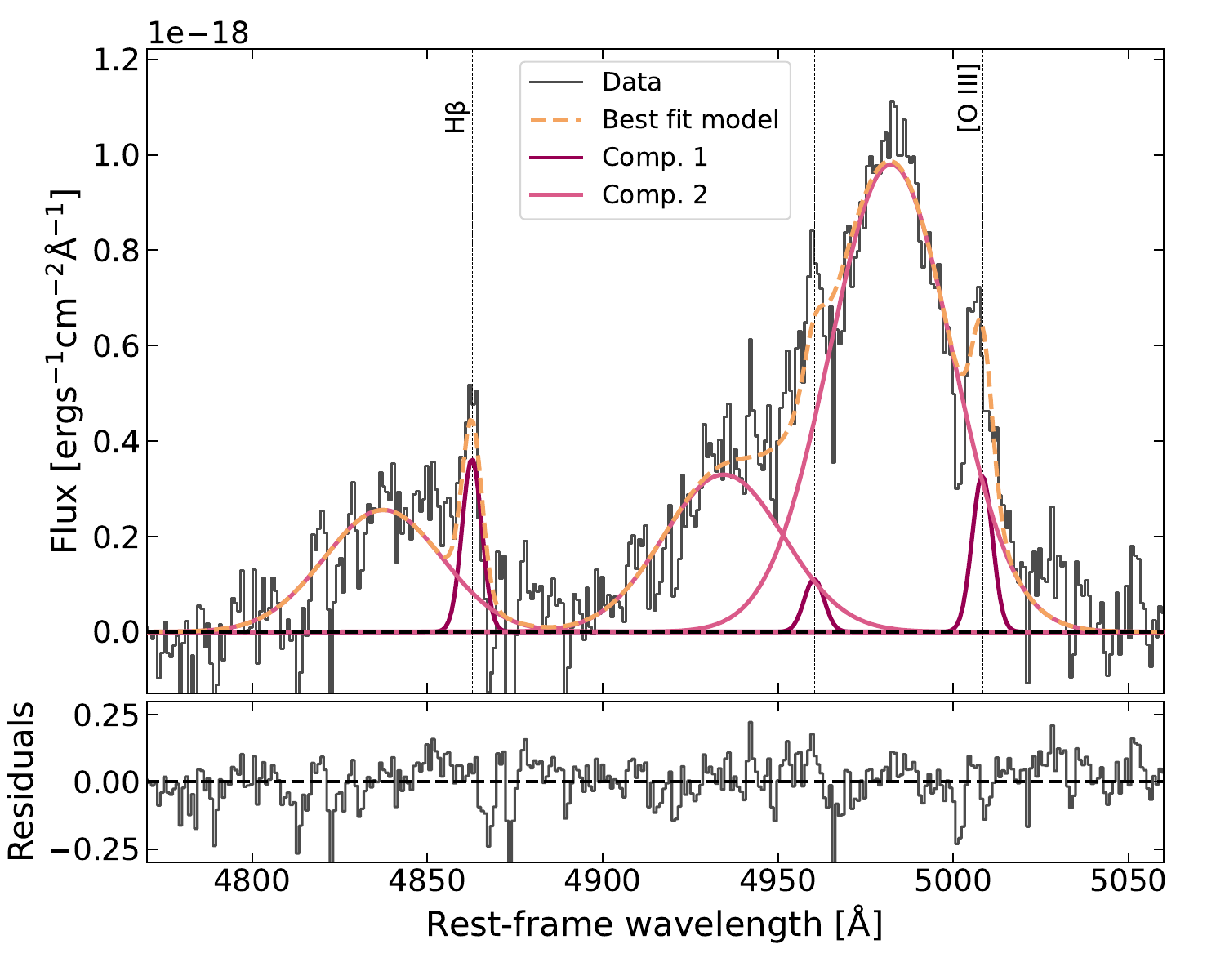}
    \hfil
    \includegraphics[width=0.45\textwidth]{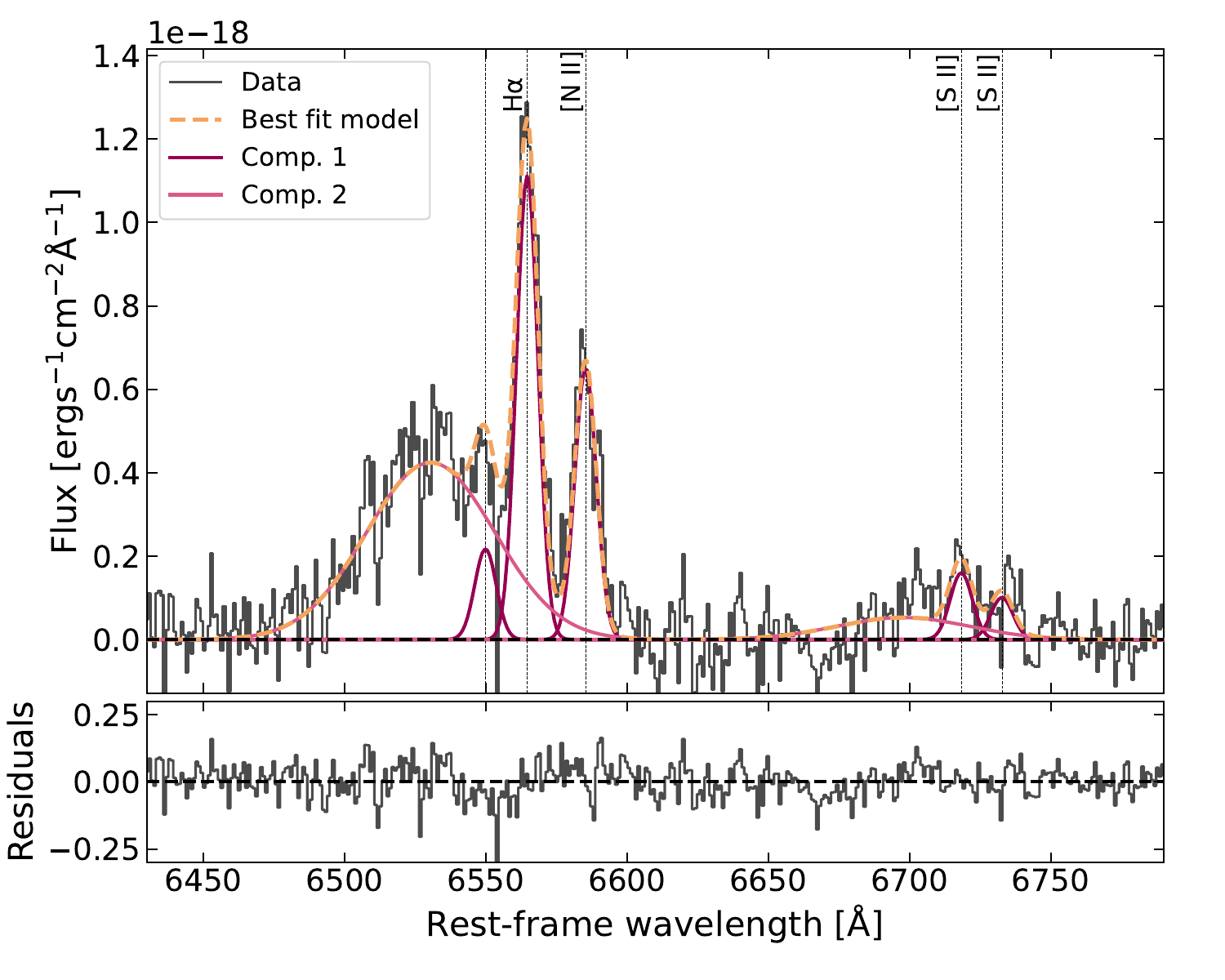}
    \caption{\textit{Top}: \hal maps of COS1638 AGNi A and B. From top to bottom: Total, narrow, and broad components. From left to right: Flux, velocity, and velocity dispersion. Solid (dashed) black contours mark the outflow region used to produce the outflow-integrated spectra of COS1638-A (COS1638-B). \textit{Bottom}: Close-up on the \hbe+\oiii complex (left) and the \hal+\nii+\sii complex (right) of the continuum- and BLR-subtracted spectrum of COS1638 AGN A integrated over the $\snr>3$ mask of the broad \oiii emission at the position of AGN A (solid black line). Data and residuals are in black, total model is in dashed orange, single-line components are shown as dark to light purple. Component 1 corresponds to the narrow component; additional components add up to the broad component (component 2). }
    \label{fig:maps2_cos1638_ha}
\end{figure*}

\subsubsection{Modeling of the total emission}
In step \textit{ii)}, we map the AGNi and galaxy emission by fitting the full data cube spaxel-by-spaxel with the  Penalized PiXel-Fitting (pPXF) method  \citep{cappellari2004_ppxf,cappellari2023_ppxf}. The aim of this step is to model the total continuum emission and, when present, the BLR contribution using the BLR template produced in step \textit{i)} to then subtract them and produce the ionized-gas cube. The BLR emission is produced in the innermost regions of the AGN, and thus it is unresolved in our data. As such, we allowed for the BLR template to scale in intensity throughout the field of view without modification to the shape of its emission lines. The continuum is fit with a polynomial of second or third order, plus the power law continuum of the AGN accretion disk if indicated as best BLR model in step \textit{i)}.
As in step \textit{i}, we included additional Gaussian components, with velocity shift and broadening linked between all the considered emission lines, to best reproduce the complex line profiles arising from the ionized gas and thus model the underlying continuum at best. We ran step \textit{ii} three times, including from one to three Gaussian components, and saved the best fit of each spaxel for each run. Following the Kolmogorov-Smirnov test as implemented by \citet{marasco2020}, we then selected the best model of each spaxel by comparing the residuals of the \oiii and \hal lines, i.e. those of interest for our study, modeled with one to three Gaussians in step \textit{ii}. 
 
With this procedure, we obtained the best-fitting model of each spaxel in the full field of view. We then used the models to subtract, spaxel-by-spaxel, the continuum and, when present, the BLR contribution in order to produce the ionized-gas data cubes. 

\subsubsection{Modeling of the ionized-gas cubes}
To increase the \snr in the ionized-gas cubes, we smooth the signal of each cube plane with a Gaussian kernel of size smaller than the instrumental PSF: $\sigma_{\rm smooth}=0.05''$ compared to a $FWHM_{\rm PSF}=0.1''-0.17''$ \citep{DEugenio2024NatAs}. 
Despite the careful data reduction, the cube of COS1656 still shows some spurious signal, possibly due to contamination from open MSA shutters. We thus do not smooth the cube of COS1656 to avoid the contaminating signal affecting more spatial pixels.
In step \textit{iii}, we fit the ionized-gas data cubes spaxel-by-spaxel. As done in step \textit{ii}, we use from one to three Gaussian components linking velocity shift and broadening across the lines with the aim of best reproducing the full line profiles. At this stage, we focus on best modeling our data and do not associate a physical meaning to each component yet. This approach allowed us to correctly reproduce the ionized gas emission of all the analyzed cubes, except for COS1638. In fact, the asymmetries of the \oiii and \hal line profiles are not similar enough to be fit together, in the sense that the best fit produced from the \oiii line does not fully reproduce the bluest side of the broad component of \hal, and viceversa, for both COS1638-A and COS1638-B. Such different asymmetries may well be caused by an imperfect wiggles correction. For this reason, we fit this cube including two sets of one to three Gaussian components with velocity shift and broadening tied among different lines: one set is tied between \hbe and \oiii, while a second set, with the same number of Gaussian components as the first one, is tied between \hal, \nii and \sii. 

As in step \textit{ii}, we then compare the residuals of each model in each spaxel and, applying the Kolmogorov-Smirnov test, we select the minimum number of gas components required to best reproduce the ionized gas emission in each spaxel separately. Three components are typically needed in our data to reproduce the highest \snr spaxels, corresponding to the emission peak at the target position. 
Having obtained the best model of the ionized gas emission, we separate narrow components, typical of systemic motions, from broader ones that are associated with 
peculiar, possibly disturbed, gas kinematics (``broad'' component) in each spaxel. We identify as a narrow component those Gaussians with velocity shift $|\Delta v|<300\kms$ and broadening $\sigma<300\kms$, and flag as ``broad'' every other component not matching this requirement. Such thresholds roughly correspond to the expected rotation velocity and velocity dispersion for star-forming galaxies of $\log(\Mstar/\Msun)\simeq10.5-11$ \citep[e.g., see][for a review]{foersterschreiber2020}. 

We then produce the moment maps of order 0, 1 and 2 of the total model, the narrow component and the broad component(s) for each line included in the fitting process. We use the moment 0 maps of the broad component of \hal and \oiii to characterize properties and size of the outflows in each target. Maps only show spaxels with S/N > 3, where the S/N is calculated based on the integrated flux over the $\pm3\sigma$ velocity range of each line component. The systemic velocity used for the velocity maps is given by the redshift estimated from the nuclear spectra and reported in Table \ref{tab:sed}. 

\subsection{Outflow-integrated spectra }
\label{sec:int_spec}
To study the kinematic properties of outflows, we are interested in the characterization of the broad components of \hal and \oiii. Yet, a proper characterization of other emission lines, like \hbe and \sii, is necessary to measure all the needed physical quantities, for instance the density and temperature of the outflowing gas and the dust extinction of the outflow. Unfortunately, the strength of these other lines is not high enough to allow for a spatially resolved characterization of their emission in the outflow component. For this reason, we rely on spatially integrated spectra to measure some of the outflow properties. \\
We produce outflow-integrated spectra from the on-target 3$\sigma$ mask of the \oiii or \hal broad emission flux maps, depending on which of the two lines shows the most spatially  extended broad component emission. For COS590, we exclude the spaxels at the position of the other components northern of the AGN ($\delta>0.4''$ in Fig. \ref{fig:maps2_cos590_oiii}) because this emission is possibly associated to tidal tails and not outflows. 
Since wiggles are significant only at single-spaxel level and disappear when integrating over large-enough apertures, we extract integrated spectra from the datacubes that were not corrected for wiggles, to avoid any interpolation. 

\subsubsection{Fitting of the outflow-integrated spectra }
Integrated spectra are fitted following the same steps used for cube fitting and described in Sect. \ref{sec:spaxelfitting}. For Type 1 AGNi, we use the BLR template built in the cube-fitting process because it is less contaminated by large-scale, ionized-gas emission. We produce the ionized-gas integrated spectra by fitting and subtracting the continuum and, for Type 1 AGNi, the BLR component. We then fit the ionized-gas integrated spectra using from one to three Gaussian components, with velocity and width linked between the various emission lines. In contrast to the spaxel-by-spaxel approach, integrated spectra of COS1638-A and COS1638-B are well fitted by tying the Gaussian components between all the lines, further supporting the possibility of such asymmetries being due to an imperfect wiggles correction. We note that four Gaussian components are needed to best reproduce both the \hal and the \oiii blue wings with linked kinematics between all the emission lines included in the fit (see Fig. \ref{fig:spectra_type_2}. 
We then select the best-fit model as the one with the lowest number of Gaussian components through the Kolmogorov-Smirnov test on the residual distribution. All spectra are best fitted using two or three Gaussian components. Figure \ref{fig:maps2_cos1638_ha} shows a close-up of the spectrum of COS1638-A around the \oiii and \hal lines as an example, the spectra of the other targets are presented in Appendix \ref{app:maps_specs}. We then identify the ``narrow'' component of each spectrum as the Gaussian component with velocity shift $|\Delta v|<300\kms$ (with respect to the source rest frame) and broadening $\sigma<300\kms$, and flag the remaining one or two Gaussian components as ``broad''. 

\subsubsection{Dust extinction}
\label{sec:extinc}
We measure the extinction correction from the spectra integrated in the outflow region, that is, in the broad-component 3$\sigma$ masks. The broad components of the \hbe line, if any, are too faint and thus we do not estimate the extinction correction for the broad components. We compute the Balmer decrements for the total model and for narrow components separately, for those spectra where the narrow components are well detected and decoupled from the other emission lines (that is, in COS1638-A, COS590 and COS1118).   
We use the {\tt extinction} v0.4.6 python package \citep{barbary2016_extinction.v0.3.0} and the extinction law of \citet{cardelli1989} with an intrinsic Balmer
decrement H$\alpha$/H$\beta$ of 2.86. We estimate the uncertainty on E(B-V) by propagating the uncertainty on the Balmer decrement. 

Table \ref{tab:sed} summarizes the measured dust extinction. As can be expected, we find Type 2 AGNi to be more extincted than Type 1 AGNi. Only COS1118 and COS590, both Type 1 AGNi, allowed for the determination of dust extinction from both their total models and narrow components. However, while for COS1118 these two are consistent within errors, the total extinction measured in COS590 is lower than that derived from narrow components only. This could be due to the quality of the data or to an outflow that is overall less dust extincted than the rest of the galaxy. 
Regarding the Type 2 AGNi, we find that COS1638-B is heavily extincted ($A_{\rm V}^{\rm tot}=4.2\pm0.8\rm~mag$) while COS1656-A presents a more moderate dust extinction ($A_{\rm V}^{\rm tot}=2.2\pm0.9\rm~mag$). 
For COS349, the strong degeneracy between the BLR and NLR emission lines hampers a meaningful determination of its dust extinction. Such a degeneracy, even if less strong, is also present in COS1638-A, for which we could measure the dust extinction of the narrow components, that are well separated from the outflows, but not for the total model and for the broad components. For these two targets, we assume as dust extinction the median of the total extinction measured in the other two Type 1 AGNi of our sample, COS1118 and COS590. We note that dust extinction of the narrow component in COS1638-A is slightly higher than the median of the sample (see Table \ref{tab:sed}) but would lead to overall consistent results. 
Lastly, we consider as dust extinction of COS2949 the median of that measured in the other Type 2 AGNi of our sample (COS1638-B and COS1656-A) since its \hbe line falls outside the \nirspec wavelenght range. 

For all targets, we deredden the total measured fluxes using the dust extinction listed in Table \ref{tab:sed}. To correct for extinction the fluxes of the broad components, we use the dust extinction as derived for the total fluxes.

\subsubsection{Gas density}
\label{sec:gas_dens}
The ratio of the [SII]$\lambda\lambda6717,6731$ lines is often used to estimate the electron density \citep[e.g.,][]{osterbrockferland2006.book,cresci2023}. Before the advent of JWST, the \sii doublet could be observed only up to $z\simeq2.5$ with ground-based facilities. 
Unfortunately, \sii emission of COS-AGNi is generally weak and is mostly undetected in single spaxels. In the spectra integrated over the outflow region, the \sii doublet is undetected only in COS349. However, in the other targets the \snr of the \sii doublet hampers any detection of blue wings like the prominent ones of the \oiii and \hal lines (see Figs. \ref{fig:maps2_cos1118_oiii}--\ref{fig:maps1_cos2949_ha}). 
For this reason, we fit the \sii emission lines in the spectra integrated over the outflow region with a single Gaussian component to best measure the \sii line ratio and convert it to gas density following \citet{osterbrockferland2006.book}. We do it in two ways: \textit{i)} we fit the \sii lines alone with one Gaussian component each without linking their velocity parameters to the other lines, and \textit{ii)} we used one Gaussian component with velocity shift and broadening set to those of the narrow component of the bright lines. 
We find that the two methods yield consistent results for all targets but COS1638-A and COS2949. In fact, these two AGNi are the only ones that show broader and blueshifted emission in the \sii lines (see Figs. \ref{fig:maps2_cos1638_ha} and \ref{fig:maps1_cos2949_ha}  for COS1638-A and COS2949, respectively); however, they do not share velocity shift and broadening of the outflows in the bright lines as an effect of the lower \snr. We thus fit the \sii lines of these two AGNi with two Gaussian components, for both fitting methods described above. Yet, as an effect of the degeneracy intrinsic to using a total of four Gaussian components to fit the low \snr \sii lines, the yielded best models include a broad component only in one of the two \sii lines, leading to an unphysical determination of the \sii ratio in the broad components and precluding the determination of the electron density of the outflows. 

In all cases, the \sii ratio uncertainties are large and in some cases span the full density range the \sii ratio is sensitive to. For this reason, here we only quote the density regime of our targets following \citet{osterbrockferland2006.book}. The \sii ratio range of COS2949 and COS1656-A indicates gas in a high density regime (\sii~$\lambda\lambda 6717,6731\lesssim0.6$, i.e., $n_{\rm e}\gtrsim3300\rm~cm^{-3}$); the \sii ratios of COS1118 and COS590 indicate gas in a low density regime ( \sii~$\lambda\lambda 6717,6731\gtrsim1.2$, i.e., $n_{\rm e}\lesssim200 \rm~cm^{-3}$); within the uncertainties, the  \sii ratio of COS1638-A and -B covers the full density range the \sii ratio is sensitive to, even though at face value COS1638-B might host more dense gas ($n_{\rm e}\simeq1000\rm~cm^{-3}$) than COS1638-A ($n_{\rm e}\simeq260\rm~cm^{-3}$). 
\begin{figure}[!h]
    \centering
    \includegraphics[width=0.5\textwidth]{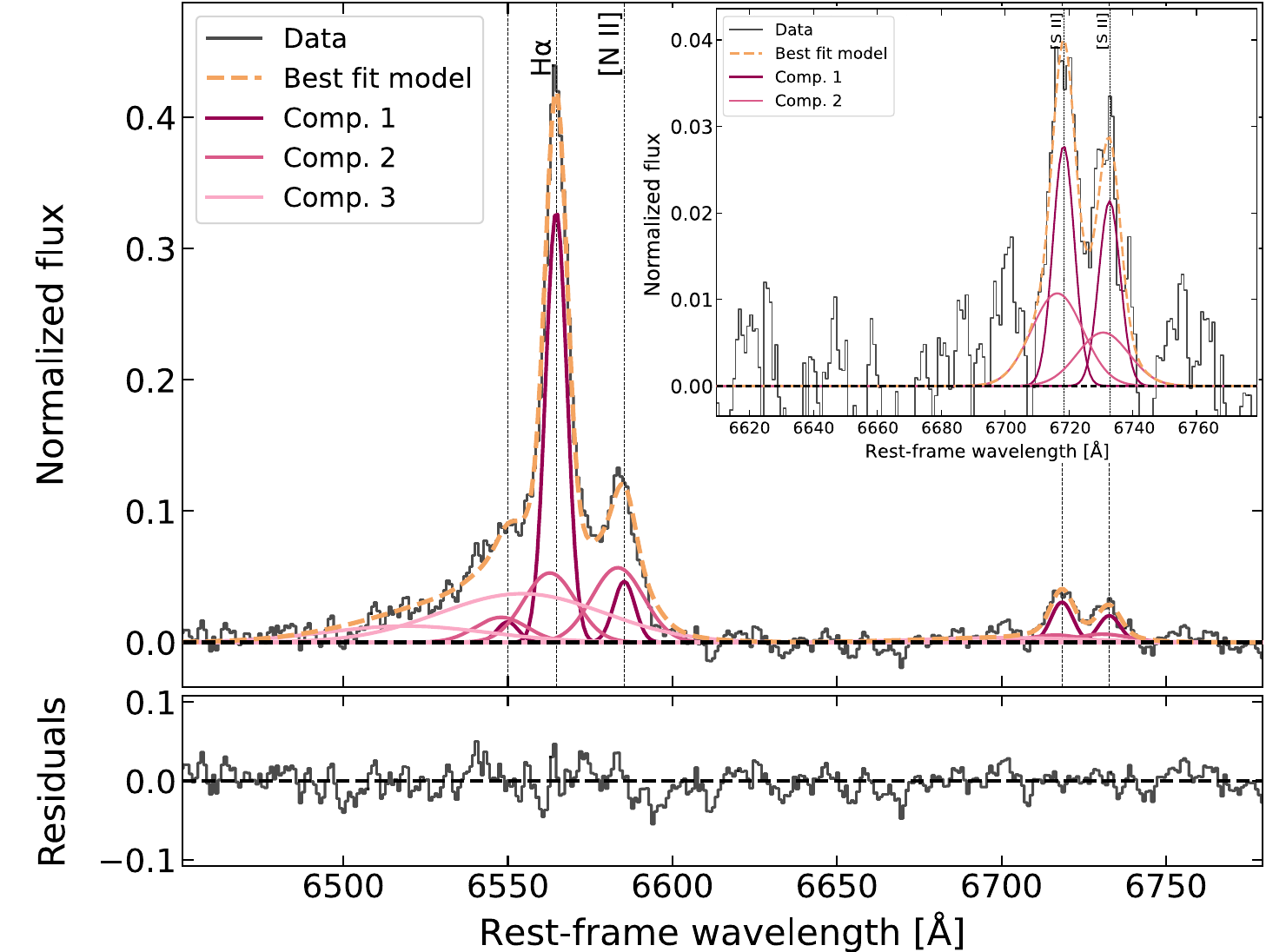}
    \caption{ Fit of stack spectrum. Data and residuals are in black, the total model is in dashed orange, and single line components are shown as dark to light purple. Inset: Close-up on the \sii emission lines. The color coding is as in the main figure. }
    \label{fig:stack}
\end{figure}

With the aim of increasing the \snr on the \sii lines, we also analyze the spectrum obtained by stacking all COS-AGNi targets, except for COS349 because undetected in the \sii lines. We weigh each BLR- and continuum-subtracted spectrum integrated over the outflow region on its errors, we normalize each weighted spectrum to the peak of the \sii$\lambda$6717\AA\ emission line before stacking and then we normalize the stack spectrum to its peak. 
We fit \hal, \nii and \sii lines in the normalized stack spectrum using three Gaussian components with linked velocity dispersion and shift (see Fig. \ref{fig:stack}). The \sii line ratio in the stack spectrum is typical of gas in the low density regime (\sii~$\lambda\lambda 6717,6731\gtrsim1.2$, $n_{\rm e}\lesssim200\rm~cm^{-3}$), both for the narrow and the broad Gaussian components. A similar result is obtained also when allowing the \sii line ratio to vary only in the 0.4-1.4 range, that is, excluding the asymptotic regimes where the \sii ratio becomes insensitive to gas density variations. 

In summary, we find that the \sii ratio of COS2949, COS1656-A and COS1638-B indicates a high density regime for the bulk of the gas in the galaxy ($n_{\rm e}\gtrsim3300\rm~cm^{-3}$ in the first two, $n_{\rm e}\simeq1000\rm~cm^{-3}$ in the third), while the \sii ratio of COS1118, COS590 and COS1638-A is indicative of a global low electron density ($n_{\rm e}\lesssim260\rm~cm^{-3}$). These results might hint to Type 2 AGNi harboring denser gas compared to Type 1 AGNi, yet our sample size is small and the \sii ratio of COS1638-A and -B becomes consistent to the full density range the \sii ratio is sensitive to considering the uncertainty range. 
Results from the stack spectrum indicate a global electron density of $n_{\rm e}\lesssim200\rm~cm^{-3}$ for both narrow and broad components of the \sii emission lines. However, we did not observe in the \sii lines of the stack spectrum the same prominent blueshifted components shown by \hal. We thus consider our results from the stack as indicative of the global mean electron density of COS-AGNi, which is likely dominated by sources in the low density regime and affected by a fitting degeneracy between narrow and broad components, but not as representative of the density within the prominent outflows observed in our targets.

\section{Results}
\label{sec:results}
We present the line ratios and diagnostic diagrams in Sect. \ref{sec:diagn}, discuss the properties of single sources in Sect. \ref{sec:single_sources} and compute the physical properties of the outflows in Sect. \ref{sec:outflow_prop}. Full spectra are shown in Fig. \ref{fig:blr_of_1}-\ref{fig:spectra_type_2}, while maps and close-up spectra in the \hbe+\oiii and \hal+\nii regions are shown in Fig. \ref{fig:maps2_cos1118_oiii}-\ref{fig:maps1_cos2949_ha}. 

\begin{figure*}[!h]
    \centering
    \includegraphics[width=\textwidth]{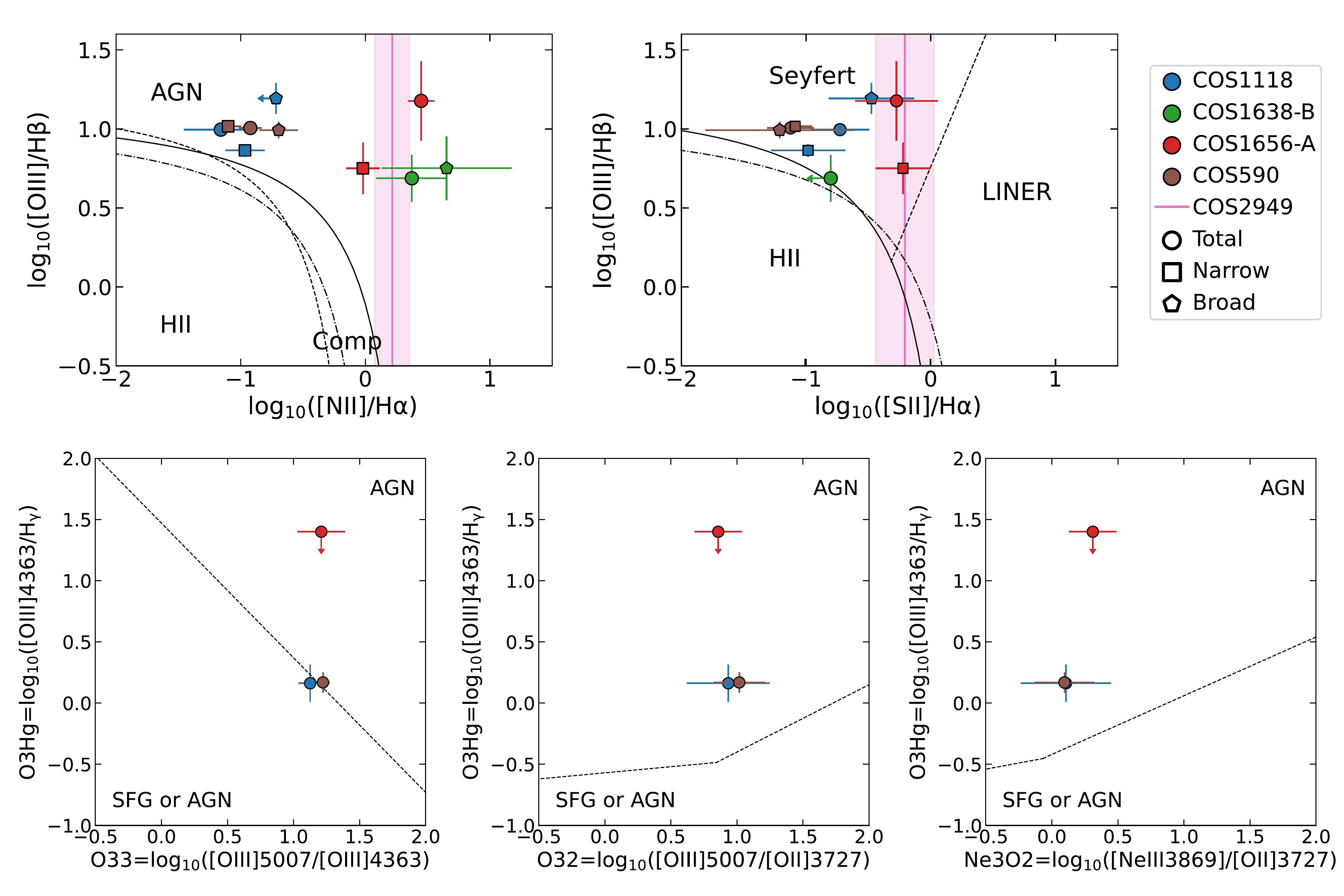}
    \caption{Diagnostic diagrams. \textit{Top, left to right}: Standard optical log(\oiii/\hbe) versus log(\nii/\hal) and log(\oiii/\hbe) versus log(\sii/\hal) diagnostic diagrams. Different ionization regimes are separated by solid and dashed black lines, taken from \citet{Kewley2001_bpt1,Kewley2006_bpt2} and \citet{Kauffmann2003_bpt}. Dashed-dotted black lines show the observed position of $z\simeq2-3$ star-forming galaxies in \citet{strom2017}. \textit{Bottom , left to right}: Diagnostic diagrams log([O~III]$\lambda4363\AA$/\hga) versus log([O~III]$\lambda5007\AA$/[O~III]$\lambda4363\AA$), log([O~III]$\lambda4363\AA$/\hga) versus log([O~III]$\lambda5007\AA$/[O~II]$\lambda3727\AA$), and log([O~III]$\lambda4363\AA$/\hga) versus log([Ne~III]$\lambda3869\AA$/[O~II]$\lambda3727\AA$) and demarcation lines from \citet{mazzolari2024}. Total, narrow, and outflow component line fluxes are shown as circles, squares, and pentagons. COS2949 is marked as a pink line since NIRSpec/IFU data do not cover the \hbe+\oiii range. Error bars are dominated by the uncertainties resulting from extinction correction. }
    \label{fig:diagns}
\end{figure*}
\subsection{Diagnostic diagrams}
\label{sec:diagn}
First, we produce diagnostic diagrams to inspect the ionization nature of the total, narrow (i.e., systemic) and broad (i.e., outflow) components. We employ both standard optical diagnostic diagrams  \citep[BPT and VO87 diagrams; e.g.,][]{baldwin1981,Veilleux1987,kewley2013_bpt}, and the novel diagnostics suggested by \citet{mazzolari2024}. Figure \ref{fig:diagns} shows the following diagnostic parameter spaces: \textit{i)} log([O~III]$\lambda5007$/\hbe) versus log(\nii/\hal); \textit{ii)} log([O~III]$\lambda5007$/\hbe) versus log(\sii/\hal); \textit{iii)} log([O~III]$\lambda4363$/\hga) versus log([O~III]$\lambda5007$/[O~II]$\lambda3727$); \textit{iv)} log([O~III]$\lambda4363$/\hga) versus log([Ne~III]$\lambda3869$/[O~II]$\lambda3727$)\footnote{[O~II]$\lambda3727$ refers to the sum of [O~II]$\lambda\lambda3726,3728$ as in \citet{mazzolari2024}.}; \textit{v)} log([O~III]$\lambda4363$/\hga) versus log([O~III]$\lambda5007$/[O~III]$\lambda4363$). We plot line ratios as measured from the spectra integrated over the outflow region of all the AGNi for which the needed emission lines are detected in the total, narrow and broad components  (see Figs. \ref{fig:maps2_cos1118_oiii}-\ref{fig:maps1_cos2949_ha} and \ref{fig:spectra_type_2}). We exclude COS1638-A and COS349 because of the strong degeneracy between the BLR and NLR line components due to strong broad Balmer lines \citep[see also][]{perna2025dualsamp}.

All AGNi that allow for the computation of the line ratios needed for the diagnostic diagrams fall in the AGN region (see Fig. \ref{fig:diagns}). We note that also the outflow components for which the line ratios can be computed fall in the AGN region, as expected for outflows driven by AGNi. COS1118, COS590 and COS1656-A show the needed lines for the diagnostic diagrams presented in \citet{mazzolari2024}, 
providing further proof of the validity of the \oiii$\lambda$4363\AA~ diagrams in discerning AGNi and star forming galaxies. 
Standard line ratios are known to vary with redshift and JWST results have recently highlighted how the standard diagnostic diagrams fail to identify high-\z AGNi residing in low-mass and low-metallicity host galaxies, often missing X-ray emission \citep[e.g.,][]{sanders2023,shapley2025}. However, such an effect does not apply to our sources. 
The AGNi analyzed in the present work were selected in the X-rays for showing $L_{2-10}>10^{44}\Lumcgs$, that is, as ``standard'' AGNi,  and thus their position in both the BPT and VO87 optical diagnostic diagrams and in those of \citet{mazzolari2024} is in line with expectations. Lastly, we note that COS-AGNi reside in galaxies that are massive and close to the main sequence of star forming galaxies (see Fig. \ref{fig:sample}, right) and are not placed at such a high redshift to be expected to be metal poor (unlike other GA-NIFS AGNi, for instance GS\_3073 presented in \citealt{Ubler2023a}). 

\subsection{Notes on individual sources}
\label{sec:single_sources}
\paragraph{COS1118.}
This target is a Type 1 AGN at $z=3.643$ \citep[][]{trump2009,brusa2009}, as confirmed by this work. The target is unobscured in the X-rays and shows a rest-frame 2--10 keV luminosity $L_{2-10}\simeq3\times10^{44}\Lumcgs$ \citep{marchesi2016}. 
\nirspec/IFU spectra of COS1118 are rich in emission lines, including nebular lines such as \oii$\lambda\lambda3726,3728$, \hga, \hde, [Ne~III]$\lambda3869\AA$, [O~III]$\lambda4363\AA$ and He~II$\lambda4687\AA$. However, many of these lines are detected only in a few spaxels of the cube and therefore do not allow for a spatially resolved analysis. Nevertheless, they are well seen in the integrated spectra (see Fig. \ref{fig:blr_of_1}), thus enabling computation of the line ratios needed for the diagnostic diagrams in Fig. \ref{fig:diagns}. This target shows spatially extended narrow emission both in the \hal and \oiii lines (see Fig. \ref{fig:maps2_cos1118_oiii} for the \oiii line). The North-West region shows ionized gas that is detected in both lines, sharing similar blueshifted kinematics in their narrow components, while redshifted emission extending South-East of the target is present only in the  \oiii line, without a counterpart in \hal. The total extent of the \oiii nebula is slightly less than 2$''$ in major axis, corresponding to $\lesssim15$~kpc, while the \hal emission is much less extended ($\simeq0.7''$ in major axis, corresponding to $\simeq5$~kpc). 
Emission from perturbed gas, traced by the outflow components, is much less extended and detected only in the proximity of the AGN (within 0.5$''$ both in \oiii and \hal). The integrated spectrum in the outflow region is dominated by the gas at rest with the galaxy, yet broad components are needed to reproduce the broad line wings visible both in the \oiii and \hal lines and in all the brighter lines (see Fig. \ref{fig:maps2_cos1118_oiii}). The kinematic properties of the ionized nebula are thus not clearly linked to the outflow detected in this AGN and more data would be needed to probe its origin. Unfortunately, no other spatially resolved data are available in the archives (e.g., ESO, ALMA, VLA) for this target, thus hampering a more detailed study of this \oiii-traced nebula.  

\paragraph{COS349.}
This target is a Type 1 AGN \citep[redshift measured with VLT/VIMOS: $z=3.5078$,][]{lilly2007_zcos} at $z=3.5093\pm0.0003$ (as measured in this work) that is unobscured in the X-rays and shows a rest-frame 2--10 keV luminosity $L_{2-10}\simeq1.8\times10^{44}\Lumcgs$ \citep{marchesi2016}. The rest-frame optical properties of COS349 were first analyzed in \citet{trakhtenbrot2016}, focusing on the BLR properties and SMBH mass derivation, through low-\snr Keck/MOSFIRE K-band spectra (alias of source ID: CID-349) covering only the \hbe+\oiii complex. No outflow component was detected by \citeauthor{trakhtenbrot2016} in the \oiii lines, probably due to the lower spectral resolution and lower \snr and to an overestimated contribution from Fe~II owing to the limited rest-frame wavelength range covered by the MOSFIRE spectrum ($\simeq4500-5200\AA$). 
JWST/NIRSpec IFU data show the presence of prominent Fe~II emission as well as complex and perturbed kinematics suggested by the need of multiple Gaussian components (see Fig. \ref{fig:blr_of_2}). Emission line maps show a compact structure for the ionized gas in this AGN, which is the most compact observed in the GA-NIFS AGNi sample (roughly 3.5~kpc in diameter), both in the narrow and in the broad gas components, as can be seen from the maps in Figs. \ref{fig:rgb_images} and \ref{fig:maps2_cos349_oiii}. Degeneracy between BLR and NLR components is strong, likely producing an oversubtraction of the BLR component of the \hbe line (Fig. \ref{fig:maps2_cos349_oiii}) and strong degeneracy between \hal and \nii in the fitting of the spectrum integrated in the outflow region. For this reason, we do not include this target in the diagnostic diagrams and compute its mass outflow rate using also the \oiii emission line, in addition to the \hal line (see Sect. \ref{sec:outflow_prop}).

\paragraph{COS590.}
This target is a Type 1 AGN at $z=3.52385$ \citep[previous redshift $z=3.535\pm0.013$,][]{trump2009} that could be partially obscured in the X-rays ($N_{\rm H}<1.6\times10^{23}\rm cm^{-2}$) and shows a rest-frame 2--10 keV luminosity $L_{2-10}\simeq2.6\times10^{44}~\Lumcgs$ \citep{marchesi2016}. 
COS590 presents a clumpy morphology, with at least two companions at slightly higher redshift than the central AGN, as traced by the redshifted emission to the north of the AGN in Fig. \ref{fig:maps2_cos590_oiii}. All the components are well traced by both \oiii and \hal narrow emission lines ($|v|<300$~km/s, $\sigma<300$~km/s), with brighter emission in the \oiii line. Continuum emission is also present in these additional components and the elongated morphology of COS590 suggests an ongoing merger. Moreover, the additional components do not show broad wings in the Balmer lines and are undetected in both \sii and \nii, thus we have no means to discriminate between SFGs or AGNi. A more in-depth analysis of the other components and diffuse emission is outside the scope of the present paper. However, as for COS1118, no other spatially resolved data are available for this system, hampering a multiwavelength characterization of the companions revealed by \nirspec/IFU. The central AGN shows broad emission at its core and in its proximity, with \oiii-traced gas bridging the AGN and the closest companion. Broad \oiii emission is also detected in a few spaxels close to the position of one of the northern clumps. The integrated spectrum in the outflow region (i.e., the black contour enclosing the AGN in the broad emission map of Fig. \ref{fig:maps2_cos590_oiii}) exhibits prominent narrow emission and a blueshifted broad component that is present in all the brightest emission lines. 

\paragraph{COS1638-A and -B.}
This target is a Type 1 AGN \citep[redshift from the SDSS quasar catalog: $z=3.5026$,][]{paris2014_sdss} at $z=3.5057$ (as measured in this work) that could be partially obscured in the X-rays ($N_{\rm H}<3.2\times10^{23}\rm cm^{-2}$) and shows a rest-frame 2--10 keV luminosity $L_{2-10}\simeq2.6\times10^{44}~\Lumcgs$ \citep{marchesi2016}. 
The \hbe+\oiii complex of COS1638 was analyzed in \citet{trakhtenbrot2016}, focusing on the BLR properties, through a low-\snr Keck/MOSFIRE K-band 1-D spectrum (alias of source ID: LID-1638). Prominent \oiii asymmetric lines are visible also in the MOSFIRE spectrum, which the authors interpreted as possibly due to prominent outflows or as indicative of a dual AGN candidate. 
JWST/\nirspec IFU indeed confirms the presence of extremely prominent blueshifted outflows, as argued by the \citeauthor{trakhtenbrot2016}. 
The exquisite sensitivity of JWST/\nirspec data also allowed \citet{perna2025dualsamp} to identify COS1638 as a dual AGN: COS1638-A is a Type 1 AGN at $z=3.5057$, while its companion, COS1638-B, is a type 2 AGN  at $z=3.5109$. Given the similarity of the MOSFIRE spectrum with that of COS1638-A \citep{trakhtenbrot2016}, it is likely that COS1638-B was not in the MOSFIRE slit. Both targets are brighter in \hal than in the \oiii line  and both their nuclear spectra and those integrated over the outflow region are dominated by very broad emission lines tracing prominent outflows and highly perturbed gas kinematics (see Figs. \ref{fig:maps2_cos1638_ha}, \ref{fig:blr_of_2} and \ref{fig:maps2_cos1638_oiii}). 
Both AGNi are also characterized by extended emission in \hal, whose narrow component bridges the two AGNi (Figs. \ref{fig:rgb_images} and \ref{fig:maps2_cos1638_ha}). The \oiii emission from COS1638-A shows an extension with a C-shape structure, with an arm extending toward COS1638-B, yet without connecting the two (see Figs. \ref{fig:rgb_images} and \ref{fig:maps2_cos1638_oiii}). The emission around the positions of both AGNi in this dual system is characterized by extremely broad \oiii emission lines that are much more prominent than the gas at rest in the galaxy 
(Figs. \ref{fig:maps2_cos1638_ha} and \ref{fig:maps2_cos1638_oiii}, for spectra of COS1638-A and -B, respectively). In fact, the total emission of both AGNi is completely dominated by blueshifted gas at $v<-1000~\kms$, 
while the narrow component is almost undetectable at the spaxel level, probably too weak and thus drowned in the prominent outflow component. These peculiar spectral shapes are also observed in the sub-millimeter galaxy companion of BR1202-0725 at  $z\simeq4.7$ \citep{zamora2024}. 

\begin{table*}[!t]
	\centering
	\caption{Outflow properties of the GA-NIFS COS-AGNi sample as traced by the \hal emission line. }
	\label{tab:of_prop_all}
	 \resizebox{\textwidth}{!}{
	\begin{tabular}{lcccccc|cc}
		\hline\hline
  Target   &    $L_{\rm H\alpha,out}$&   $R_{\rm out,fw}$ &    $v_{\rm out}$ &    $\Mout$         &    $\Mdot^{\rm }$                &    $\Edot^{\rm }$ &  $R_{\rm max}$  &    $v_{\rm max}$ \\
            &    ($10^{41} \rm ~erg~s^{-1}$)  & (kpc)         &     ($\kms$)        & ($10^6~\Msun$) & ($\rm \Msun~yr^{-1})$  &  ($10^{41} \rm ~erg~s^{-1} )$ & (kpc)         &     ($\kms$)     \\
            (1)   & (2)   & (3)         &     (4)       &    (5)  & (6) & (7) & (8)  & (9) \\ 
		\hline
COS590	    &  12.1	$\pm$ 0.2	& 1.0	$\pm$  0.4	 &   569   $\pm$ 9	&  3.9  $\pm$	1.9 &  6.6	$\pm$    4.1	&  6.8	    $\pm$ 4.2	   & 2.6 $\pm$ 0.7 &796 $\pm$  5     \\
COS1118	    &  6.3	$\pm$ 0.2	& 0.7	$\pm$  0.4	 &   707   $\pm$ 4	&  2.0  $\pm$	1.0 &  5.9	$\pm$    4.1	&  9.3	    $\pm$ 6.5	   & 2.6 $\pm$ 0.7 &823 $\pm$  3     \\
COS349	    &  7.9	$\pm$ 0.4	& 1.0	$\pm$  0.4	 &   1278  $\pm$ 13	&  2.5  $\pm$	1.3 &  10.4	$\pm$    6.6	&  53.5	    $\pm$ 34.2	   & 2.6 $\pm$ 0.7 &1384 $\pm$ 8    \\
COS1656-A	&  18.4	$\pm$ 1.6	& 1.2	$\pm$  0.4	 &   950   $\pm$ 9	&  5.9  $\pm$	3.0 &  14.8	$\pm$    8.9	&  42.2	    $\pm$ 25.4	   & 2.6 $\pm$ 0.7 &1326 $\pm$ 6     \\
COS2949	    &  5.7	$\pm$ 0.1	& 1.0	$\pm$  0.4	 &   732   $\pm$ 5	&  1.8  $\pm$	0.9 &  4.3	$\pm$    2.9	&  7.3	    $\pm$ 4.9	   & 2.1 $\pm$ 0.9 &901 $\pm$  3     \\
COS1638-A	&  33.9	$\pm$ 0.3	& 0.9	$\pm$  0.4	 &   3199  $\pm$ 201&  10.9 $\pm$	5.4 &  120.8 $\pm$   79.7	&  3900.8	$\pm$ 2676.5   & 3.0 $\pm$ 0.7 &3551 $\pm$ 131  \\
COS1638-B	&  307.3 $\pm$ 0.9	& 1.1   $\pm$  0.4	 &   2066  $\pm$ 90 &  98.3 $\pm$   49.2 & 568.0 $\pm$ 345.0	&  7652.3   $\pm$ 4754.5   & 1.8 $\pm$ 0.7 &2623 $\pm$ 56   \\
		\hline\end{tabular}
	 }
\tablefoot{Outflow properties in columns (5) to (7) are computed with flux-weighted outflow radii, outflow velocity $\Vout={\rm max}(v_{10},v_{90})$ and $n_{\rm e}=1000 \rm~ cm^{-3}$. The last two columns list the parameters needed to scale the outflow properties to the framework used for the comparison with the literature. 
Columns: (1) target name; (2) \hal luminosity in units of $10^{41}~\Lumcgs$; (3) flux-weighted outflow radius in kpc units;  (4) outflow velocity $\Vout={\rm max}(v_{10},v_{90})$ in $\kms$;  (5) outflow mass computed from Eq. \ref{eq:of_mass_cresci17} in units of solar masses; (6) mass outflow rate computed in biconical geometry (Eq. \ref{eq:cone}) in units of solar masses per year; (7) outflow kinetic power in units of $10^{41}~\Lumcgs$; (8) maximum outflow radius in kpc units; (9) maximum outflow velocity $\Vmax=|v_{\rm nar}-v_{\rm bro}|+2\sigma_{\rm bro}$ in $\kms$.} 
\end{table*}
\paragraph{COS1656-A.}
This target is a Type 2 AGN with a rest-frame 2--10 keV luminosity $L_{2-10}\simeq2.7\times10^{44}\Lumcgs$ at $z=3.5101$ \citep[as measured in this work and roughly consistent with the redshift $z=3.512$ from the COSMOS collaboration,][]{marchesi2016,hasinger2018_deimos_cosmos} and for which obscuration in the X-rays could not be measured in COSMOS-Legacy \citep{marchesi2016}. 
\nirspec/IFU data of COS1656 recently allowed \citet{perna2025dualsamp} to resolve the target into a dual system (projected separation of 1.4$''$ corresponding to $\simeq10$~kpc), with both sources identified as Type 2 AGN. COS1656-B probably has a close galaxy companion that contaminates its emission, and as such we left it out of our analysis due to the impossibility of disentangling emission of the companion from possible broad components in the spectrum of AGN-B. 
COS1656-A shows extended ionized gas traced in both \hal and \oiii narrow and broad emission lines (see Fig. \ref{fig:maps2_cos1656a_oiii} for the \oiii maps). Interestingly, the narrow components of both lines seem to trace the core of the AGN host and a close by clump South-East of COS1656-A characterized by blueshifted emission with lower velocity dispersion compared to the AGN host. Such a clump is also clearly visible in the three-color images maps of Fig. \ref{fig:rgb_images}. However, the broad components of both \hal and \oiii only partially involve such a clump as they both mostly extend North-East of COS1656-A (Fig. \ref{fig:maps2_cos1656a_oiii}).  

\paragraph{COS2949.}
This target is a Type 2 AGN that was included in the GA-NIFS program because it is listed as a $z\simeq3.5$ AGN in the COSMOS-Legacy catalog, based on a spectroscopic redshift \citep[$z=3.571$ estimated from a VIMOS spectrum][]{marchesi2016}. Yet, \nirspec observations revealed that COS2949 is definitely a lower redshift source: \nirspec/IFU data only shows the \hal complex in its bluest side, and its wavelength range ends red-ward of the \oiii emission line.
We estimate the new redshift as $z=2.0478$, which is much more similar to the photometric redshift of the COSMOS-Legacy catalog \citep[$z\simeq2.55$;][]{marchesi2016}. We thus recomputed  the X-ray properties of COS2949 at the redshift measured with JWST (see Table \ref{tab:sed}). 
The AGN associated with the X-ray emission detected in COSMOS-Legacy corresponds to the only source of this NIRSpec field emitting in \nii, traced in both  its narrow and broad components, and in continuum (see Fig. \ref{fig:rgb_images}). Such an AGN is surrounded by at least one \hal-emitting companion in the South-East direction (projected separation 0.7$''$ corresponding to $\simeq6$~kpc) and by other clumps traced by narrow, low-dispersion \hal emission (see Fig. \ref{fig:maps1_cos2949_ha}). The AGN itself is dominated by blueshifted ($v<-350$~km/s), large-dispersion ($\sigma>300$~km/s) emission in its Southern side, that is, the one toward the brightest companion, and narrow, lower-dispersion emission in the Northern side, that is, connected to a receding \hal-emitting gas stream.

\subsection{Outflow properties}
\label{sec:outflow_prop}
We compute the physical properties of the outflows in GA-NIFS COS-AGNi using the \hal emission line to avoid the additional assumptions regarding gas metallicity. For instance, using the \oiii emission line as tracer of ionized gas mass in AGNi outflows was found to produce a factor 2--3$\times$ lower mass outflow rates \citep[e.g.,][]{cano-diaz2012,carniani2015,cresci2023,venturi2023_teacup}. Given the degeneracy of BLR and NLR components in COS349, for this target we derive the mass outflow rate also from the \oiii emission line and discuss the possible differences. 

\subsubsection{Outflow gas density}
As discussed in Sect. \ref{sec:gas_dens}, our data do not allow us to measure the gas density in a spaxel-by-spaxel way and the gas densities estimated from the spectra integrated over the outflow region should be considered as representative of the total system. The majority of COS-AGNi show a total gas density of $n_{\rm e}\lesssim200\rm~ cm^{-3}$, with two targets having much higher density ($n_{\rm e}\gtrsim3300\rm~ cm^{-3}$ ). As a result, the density that can be measured from the \sii line ratio in the stack spectrum is dominated by the contribution of the targets showing a lower gas density ($n_{\rm e}\lesssim200\rm~ cm^{-3}$). In any case, the quality of the data did not allow for the density of the outflow to be measured. 

At the redshift of our AGNi, few other sources have electron density measurements \citep{isobe2023,cresci2023,marconcini2024a_jd1,li2024_glass_density}, two of which are star-forming galaxies  part of the GA-NIFS sample \citep{lamperti2024,rodriguezdelpino2024}. Galaxies at $z\simeq3-4$ show a median gas density of 200--300 $\rm cm^{-3}$, and thus the tentative measurements from our work are roughly consistent with results in the literature. However, these values are referred to gas at rest in star-forming galaxies and it is reasonable to expect outflows, especially if AGN driven, to be denser, as found at lower redshift \citep[e.g.,][]{perna2017,foersterschreiber2019}. Other works focusing on AGN-driven outflows have found a variety of results \citep[e.g.,][]{mingozzi2019,foersterschreiber2019,daviesRic2020,cresci2023,speranza2024}, with values ranging from few hundreds to few thousands of atoms per $\rm cm^3$. Here we assume an outflow gas density of $n_{\rm e}=1000\rm~ cm^{-3}$ with a 50\% uncertainty to best consider the available measurements at cosmic noon, which mainly come from the KMOS$^{\rm 3D}$ survey \citep{foersterschreiber2019}. 

\subsubsection{Outflow radius}
We estimate the outflow radius in two ways. First, we take advantage of the spatially resolved maps available for our sample and compute the flux-weighted outflow radius ($R_{\rm out,fw}$) as follows: 
\begin{equation}
	\label{eq:flux_weighted_r}
    R_{\rm out,fw} = \frac{\sum_{i,j}~d(X_{\rm peak},X_{i,j})~f_{i,j}^{\rm broad}}{\sum_{i,j}~f_{i,j}^{\rm broad}},
\end{equation}
where $d(X_{\rm peak},X_{i,j})$ is the distance between the peak of the total line flux of a target ($X_{\rm peak}$) and the spaxels with \snr>3 broad component emission ($X_{i,j}$), and $f_{i,j}^{\rm broad}$ is the flux of the broad component in the corresponding spaxel. 
Secondly, we also measure the maximum outflow radius $R_{\rm out, max}$ as the maximum distance between the flux peak and the furthest spaxel showing broad emission at \snr>3 to compare our results with those of the literature, in particular to \citet{fiore2017} and later works employing the same framework. 
In few of our targets, the \hal line shows a lower overall \snr in the maps compared to the \oiii line. We thus estimated the spatial extension of the outflow as the maximum between that observed in \hal and \oiii, or \hal and \nii for COS2949 given the unavailability of the \oiii line. We correct both the flux-weighted radius and the maximum radius of the outflows by subtracting the PSF FWHM/2 in quadrature, using the modeling of NIRSpec PSF FWHM as in \citet{DEugenio2024NatAs}. We note that the extension of the outflows measured as flux-weighted radii ($R_{\rm out,fw}\simeq0.1''-0.16''$) is roughly twice the PSF FWHM and that the maximum radial extent ($R_{\rm max}\simeq0.25''-0.4''$) is at least 2.5$\times$ the PSF. 

\subsubsection{Mass outflow rate and kinetic power}
\label{sec:mout_formulas}
We computed the \hal outflow mass assuming an electron temperature of $T_{\rm e}\simeq10^{4}$~K, typical of optically emitting warm ionized gas, and applying the formula from \citet{cresci2017}:
\begin{equation}
    \Mout({\rm H}\alpha) = 3.2\times10^5 \frac{L_{[40]}({\rm H}\alpha)}{n_{\rm e [2]}} \Msun
    \label{eq:of_mass_cresci17},
\end{equation}
where $L_{[40]}({\rm H}\alpha)$ is the dereddened \hal luminosity of the outflow (see Sect. \ref{sec:extinc}) in units of $10^{40}$~\lumcgs and $n_{\rm e [2]}$ is the outflow gas density in units of $10^2~ \rm cm^{-3}$. 
As mentioned, for COS349, we computed the \oiii outflow mass following \citet{cano-diaz2012} and again assumed an electron temperature of $T_{\rm e}\simeq10^{4}$~K:
\begin{equation}
   \Mout(\Oiii) = 5.33\times10^7~\frac{C~L_{[44]}(\Oiii)}{n_{\rm e [3]}~10^{\rm[O/H]/[O/H]_{\odot}}}
   \label{eq:mout_oiii}.
\end{equation}
Here,  $L_{[44]}(\Oiii)$ is the dereddened \oiii luminosity of the outflow (see Sect. \ref{sec:extinc}) in units of $10^{44}$~\lumcgs, $n_{\rm e [3]}$ is the outflow gas density in units of $10^3~ \rm cm^{-3}$, $C$ is the condensation factor that we assumed is equal to one (i.e., we assumed that all ionizing gas clouds have the same density), and $10^{\rm[O/H]}$ gives the oxygen abundance in solar units that we assumed as solar. 

None of the COS-AGNi allow the outflow geometry to be resolved. We then computed the mass outflow rate assuming a biconical outflow geometry considering recent works both at low redshift \citep[e.g.,][]{musiimenta2023,zanchettin2023} and high-redshift \citep[e.g.,][]{kakkad2020,tozzi2024}: 
\begin{equation}
	\Mdot^{\rm cone} = 3 \Mout \frac{\Vout}{\Rout}
	\label{eq:cone},
\end{equation}
with $\Vout={\rm max}(v_{10},v_{90})$ \citep[e.g.,][]{tozzi2024}, where $v_{10}$ and $v_{90}$ are the 10th and 90th percentiles of the velocity line profile of the broad \hal, $\Rout$ is the flux-weighted radius $R_{\rm out,fw}$ from Eq. \ref{eq:flux_weighted_r}, and the factor 3 corresponds to the assumption of a constant average volume density in the cone \citep[e.g.,][]{lutz2020}. 
We estimate the uncertainty of $\Vout$ as the standard deviation of the mean of its spatially resolved maps from cube-fitting results, restricted to the spaxels above 3$\sigma$ in the broad flux map and having excluded the spaxels associated to random noise. We note that this procedure returns a larger uncertainty for the outflow velocity in COS1638-A, due to the very large velocity values in each spaxel, ranging from $\gtrsim2000~\kms$ up to $\simeq3500~\kms$. We consider one spaxel of uncertainty ($0.05''$) for the flux-weighted radius, that translates to $\simeq0.4$~kpc at $z\simeq3.5$. We then compute the kinetic power $\Edot$ of the outflows as $\Edot=\frac{1}{2}\Mdot\Vout^2$.

Table \ref{tab:of_prop_all} summarizes the outflow properties derived from \hal and computed with $\Vout={\rm max}(v_{10},v_{90})$, $\Rout=R_{\rm out,fw}$ and $n_{e}=1000\rm~cm^{-3}$. For COS349, we obtained $M_{\rm out}(\rm [OIII])=(3.0\pm1.5)\times 10^{5}\Msun$, that is roughly one order of magnitude lower than the value measured from the \hal broad emission component. This could be indicative of an overestimation of the \hal broad component in the best fit of the ionized-gas spectrum (see Fig. \ref{fig:maps2_cos349_oiii}) or of wrong assumptions in our computation of the outflow mass. For instance, Eq. \ref{eq:mout_oiii} assumes a temperature of $T\simeq10^{4}$~K, typical of NLR, for which the emissivity of the \oiii has a weak dependence on the temperature. Moreover, one of the strongest assumptions behind Eq. \ref{eq:mout_oiii} is that most of oxygen in the ionized outflow is in the form O$^{2+}$. Lastly, it could also be due to a wrong estimation of the dust reddening in COS349. In fact, due to the strong degeneracy between BLR and NLR lines, we could not estimate the Balmer decrement of this target, and thus we used the median of the other Type 1 AGNi of our sample. We note that the outflow velocity estimated from \oiii and from \hal for this target are consistent with each other. Thus, any difference in the outflow properties originates from the outflow mass determination.  

For a consistent comparison with literature results in Sect. \ref{sec:discussion}, in particular with the $\Lbol$ versus $\Vmax$, $\Mdot$ versus $\Lbol$ and $\Edot$ versus $\Lbol$ scaling relations of \citet{fiore2017} and \citet{musiimenta2023}, we re-compute the outflow energetics with the definition of outflow velocity and extent as adopted by those authors. In particular, we use $\Vmax=|v_{\rm nar}-v_{\rm bro}|+2\sigma_{\rm bro}$  \citep[as defined in][]{rupke_veilleux2013} as the outflow velocity and the maximum extension of the outflow as radius ($\Rmax$). We estimate $v_{\rm bro}$ and $\sigma_{\rm bro}$ as the moment 1 and moment 2 of the \hal broad line profile and $v_{\rm nar}$ as the moment 1 of the \hal narrow line profile in the outflow spectra, and compute their uncertainties as the spaxel-by-spaxel standard deviation of such parameters in the spatially resolved cube fitting, as done for $\Vout$. We measure the maximum radius $\Rmax$ of the outflow as the maximum distance of the broad component emission from the AGN in the 3$\sigma$ flux maps. We assign to $\Rmax$ an uncertainty of 2 spaxels, i.e., 0.1$''$, to take into account the asymmetrical extension  of the broad emission. 
We then compute the mass outflow rate assuming a biconical outflow geometry (Eq. \ref{eq:cone}). 
Table \ref{tab:of_prop_all} also summarizes the parameters used to recompute the outflow properties following the assumptions of \citet{fiore2017}, which are the maximum outflow velocity $v_{\rm max} $ and the outflow maximum extension $R_{\rm max}$.

\begin{figure*}[!t]
	\centering
	\includegraphics[width=\textwidth]{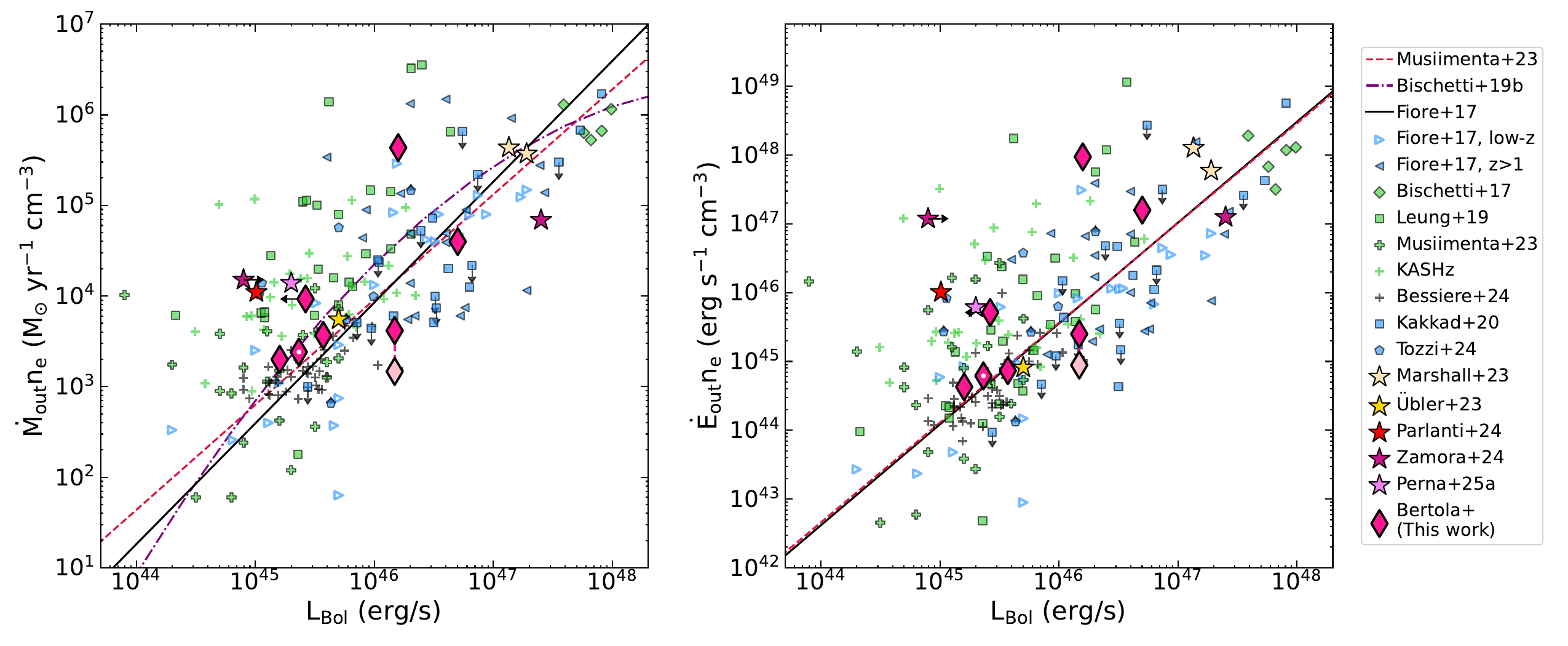}
	\caption{Comparison of outflow properties of \ganifs COS-AGNi and literature samples in units of gas density. \textit{Left:} Mass outflow rate against AGN bolometric luminosity. 
		\textit{Right:} Kinetic power as a function of AGN bolometric luminosity. 
        The color coding of the single measurements is as in Fig. \ref{fig:sample}. Black solid lines mark the best-fit relation of \citet{fiore2017}, the purple dash-dotted line marks that from \citet{bischetti2019a} and the red dashed lines mark the best fit relation of \citet{musiimenta2023}. We mark COS2949 with an additional white dot, being the only target at $z<3.5$ in our sample. Pink diamonds show the outflow properties of COS349 as computed from the \oiii line and the magenta dashed line connects them with the estimates from the \hal line.  }
	\label{fig:outflow_prop1}
\end{figure*}
\section{Discussion}
\label{sec:discussion}
The  GA-NIFS COS-AGNi are among the first targets having spatially resolved, rest-frame optical observations of AGNi outflows at $z>3$ in the luminosity range $\log(\Lbol/\Lumcgs)=45.2-46.7$. In this section we compare them with results and scaling relations present in the literature. 
Given that some assumptions are needed to compute outflow properties for both  our AGNi and  those in the literature, we try to mitigate the systematics and limitations of such a comparison by recomputing literature measurements in a consistent way. 

\subsection{Close environment of COS-AGNi}
JWST/\nirspec IFU data showed that almost all the analyzed sources show clumps and/or reside in rich environments, a subset of which also present extended emission. 
Three of our COS-AGNi (COS2949, COS1118, COS590) are embedded in ionized gas extending on kpc scales that, given the typical sizes of galaxies at $z\simeq3.5$ \citep[$r_{\rm e}\simeq0.3$~kpc; e.g.,][]{allen2017}, reaches up to the circumgalactic medium. 
Four sources (COS2949, COS590, COS1638, COS1656) out of the six analyzed systems show companions, both as close-by AGNi (dual AGN: COS1638, COS1656), as galaxies (COS590; continuum is detected also at the location of the companions, see Fig. \ref{fig:rgb_images}), and line-emitter clumps (COS2949). 
These could be tracing merging systems, given the close proximity of the companions and the ionized gas nebulae surrounding them. Moreover, the spatial distribution of the \hal emission in COS2949 could be tracing the tidal tails produced by the close encounter of the AGN and (at least) one galaxy. 
However, we cannot investigate the relation between outflow strength and presence of companions or merger signatures due to the small size of the COS-AGNi sample. 
The \oiii nebula of COS1118 could resemble those observed in lower-\z (and high-$\Lbol$) AGNi by \citet{liu2013a,liu2013b}. However, given the observed kinematics (see Fig. \ref{fig:maps2_cos1118_oiii}) and that the nebula is seen only in its \oiii emission, the present data offer no means to link it to the observed outflow. 

\subsection{Homogenization of outflow measurements}
\label{sec:of_hom}
Given that the mass outflow rate and the kinetic power depend on the 1st- and 2nd-power of the outflow velocity, respectively, one of the main discriminant choice for a consistent recomputation of these outflow properties relies in the adopted definition of the outflow velocity, along with the availability of the outflow luminosity. The non-parametric fitting approach is very useful in determining phenomenological outflow properties \citep[e.g.,][]{harrison2016,temple2019}, yet it is not trivial to build a coherent literature sample because of the various line percentiles (2nd, 5th, 10th) at which the outflow velocity is measured at in the different works  \citep[e.g.,][]{perrotta2019,coatman2019}. 
We adopt here the definition of outflow velocity as $\Vmax$ (see Sect. \ref{sec:mout_formulas}), following \citet{rupke_veilleux2013}, which allows for a larger sample of literature results to be included compared to percentile velocities. 
For GA-NIFS COS-AGNi, the median $\Vmax$ is approximately 1.2$\times$ larger than the median $\Vout={\rm max}(v_{10},v_{90})$ in our sample, leading to mass outflow rates and kinetic powers that are approximately 1.5$\times$ and 1.9$\times$ larger, respectively, than those computed with $\Vout={\rm max}(v_{10},v_{90})$.  

Another debated parameter is the outflow radial extent. In our comparison, we assume as outflow extent the maximum extent of the outflow relative to the source position, as defined in Sect. \ref{sec:mout_formulas}, that is, $R_{\rm max}$.  
We compared the impact of different definitions of such parameters in our COS-AGNi. We find that $R_{\rm max}$ are roughly 2--3 times larger than flux-weighted radii. 

Another key parameter in the outflow mass and energy budget computation is the gas electron density. 
This is one of the most difficult parameters to measure and is often assumed based on previous results, as also done in the present work. The determination of the best representative value of an outflow gas density is still currently widely debated in the literature. In fact, the outflow gas densities that can be found in the literature span at least one order of magnitude \citep[from hundreds to thousands of $\rm cm^{-3}$,][]{mingozzi2019,foersterschreiber2019,daviesRic2020,speranza2024}. Moreover, there is indication that the density of the perturbed gas in a single object can vary from a factor 3 up to more than one order of magnitude, as recently seen in XID2028 \citep{cresci2023} and the Teacup galaxy \citep{venturi2023_teacup}, respectively. 
To overcome this limitation, we compare outflow properties as a function of $n_{\rm e}$, that is, we multiply each outflow parameter (mass outflow rate, kinetic power) by its outflow density as reported in the corresponding literature study. This approach is equivalent to assuming a gas density common to each outflow, as done for instance by \citet{fiore2017} and \citet{musiimenta2023}, without selecting a representative value of the outflow density.  

The parameter that most clearly affects the determination of the outflow mass, and thus also the mass outflow rate and kinetic power, is the gas density \citep[e.g.,][]{harrison2018}. In fact, assuming a gas density of 200 $\rm cm^{-3}$, as done for the scaling relations of \citet{fiore2017} and \citet{musiimenta2023}, instead of 1000 $\rm cm^{-3}$, as measured by \citet{foersterschreiber2019} in the KMOS$^{\rm 3D}$ stack, increases the outflow properties by a factor of five.

\subsection{Literature measurements}
\label{sec:of_lit}
Out of all the works listed in Sect. \ref{sec:sample}, we include in our comparison sample single measurements from the compilation in \citet{fiore2017}, from \citet{bessiere2024},  \citet{bischetti2017}, \citet{leung2019_mosdef}, \citet{kakkad2020}, \citet{musiimenta2023}, and \citet{tozzi2024}, as well as from previous GA-NIFS works \citep{marshall2023,Ubler2023a,parlanti2024,zamora2024,perna2025_gs133}. 
For all the listed works, if needed, we recomputed the outflow properties as $\Mout n_{\rm e}$ and $\Eout n_{\rm e}$, using Eq. \ref{eq:cone} with outflow parameters as described above. Results of \citet{bessiere2024} were included scaling their measurements of $v_{05}$ to $v_{\rm max}$ based on the $v_{\rm max}$ versus $v_{05}$ relation measured for COS-AGNi ($\log (v_{\rm max}/(1000~{\rm km/s})=0.12 + 0.81(\log (v_{\rm 05}/1000~{\rm km/s})$). 
The results collected in \citet{fiore2017} come both from slit spectroscopy and from spatially resolved data collected with ground-based telescopes. 
Regarding the results from slit spectroscopy, it is not possible to measure the outflow extent and so different works have made different assumptions. For instance, \citet{bischetti2017} assume $\Rmax=7$~kpc based on the analysis of the 2D spectra and \citet{leung2019_mosdef} build the radial profile of the outflows in the 2D spectra and measure its extent as the distance at which the outflow emission flux is 1/10 the strength of the maximum ($\Rmax=0.3-11$~kpc), while \citet{musiimenta2023} assume as outflow extent for their $0.5 < z < 1$ eROSITA AGNi the effective radius of the host galaxies ($\Rmax=2-10$~kpc), that is the galaxy half-light radius as estimated from AGN-host galaxy image decomposition. Lastly, \citet{bessiere2024} assume as outflow radius the mean deprojected outflow size measured in \citet{fischer2018} from a combination of HST narrow-band imaging and STIS long-slit spectroscopy. 
Regarding the results from ground-based IFS, the AGNi of \citet{harrison2012} and \citet{carniani2015} were observed with SINFONI/VLT, as were SUPER AGNi \citep{kakkad2020,tozzi2024} but in adaptive-optics assisted mode, resulting in better spatial resolution. While \citeauthor{carniani2015} estimate the outflow radius from spectroastrometry analysis, \citet{harrison2012} and \citeauthor{tozzi2024} use $R_{\rm max}$, the latter additionally correcting for the PSF, and \citet{kakkad2020} assume $R=1$~kpc for all targets. 
We note that the differences between $R_{\rm max}$ and the flux-weighted radii for our GA-NIFS targets are approximately within a factor 2--3$\times$, while radii in \citeauthor{carniani2015} and \citeauthor{kakkad2020} are roughly 5--7$\times$ smaller than the $R_{\rm max}$ of the rest of the compilation in \citet{fiore2017}. The choice of \citet{kakkad2020} was driven by the fact that most of the SUPER Type 1 AGNi studied in their work show outflows that are unresolved compared to the PSF size. With the aim of best estimating the maximum outflow radii, we recompute their mass outflow rates using the PSF size for the unresolved targets and $D_{600}$, that is the maximum projected spatial extent when considering the \oiii emission at $w_{80}>600\kms$, for resolved targets. 
Lastly, we include KASHz AGNi \citep{harrison2016} using the updated fitting results that will be presented in Scholtz et al. (in prep). KASHz AGNi are typically unresolved because observed in seeing-limited conditions. Based on the results in \citet{scholtz2020}, we assumed $\Rmax=3$~kpc as maximum outflow extent, which roughly corresponds to a median PSF of the KMOS observations presented in \citet{scholtz2020}.

Part of the outflows we collected from the literature are traced in the \oiii emission line. Following the calibration of \citet{fiore2017}, we rescale these measurements by assuming that the true outflow mass is 3$\times$ that measured using \oiii as tracer.
Moreover, in the works of SUPER \citep{kakkad2020,tozzi2024}, the authors measure the flux of the outflow from the wings of the multi-Gaussian line profiles ($|v|>300~\kms$), while we consider the full line profile of the broad components. 
\citeauthor{kakkad2020} and \citeauthor{tozzi2024} compared the two methods for SUPER Type 1 and Type 2 AGNi and report a difference of a factor 2$\times$, which we use to rescale their results. 

Lastly, we also include other recent results from the GA-NIFS GTO program by \citet{marshall2023}, \citet{Ubler2023a}, \citet{parlanti2024}, \citet{perna2025_gs133} and \citet{zamora2024}, recomputing the outflow parameters if needed to align them to the assumptions used in this work.

\begin{figure}[!h]
	\centering
	\includegraphics[width=0.5\textwidth]{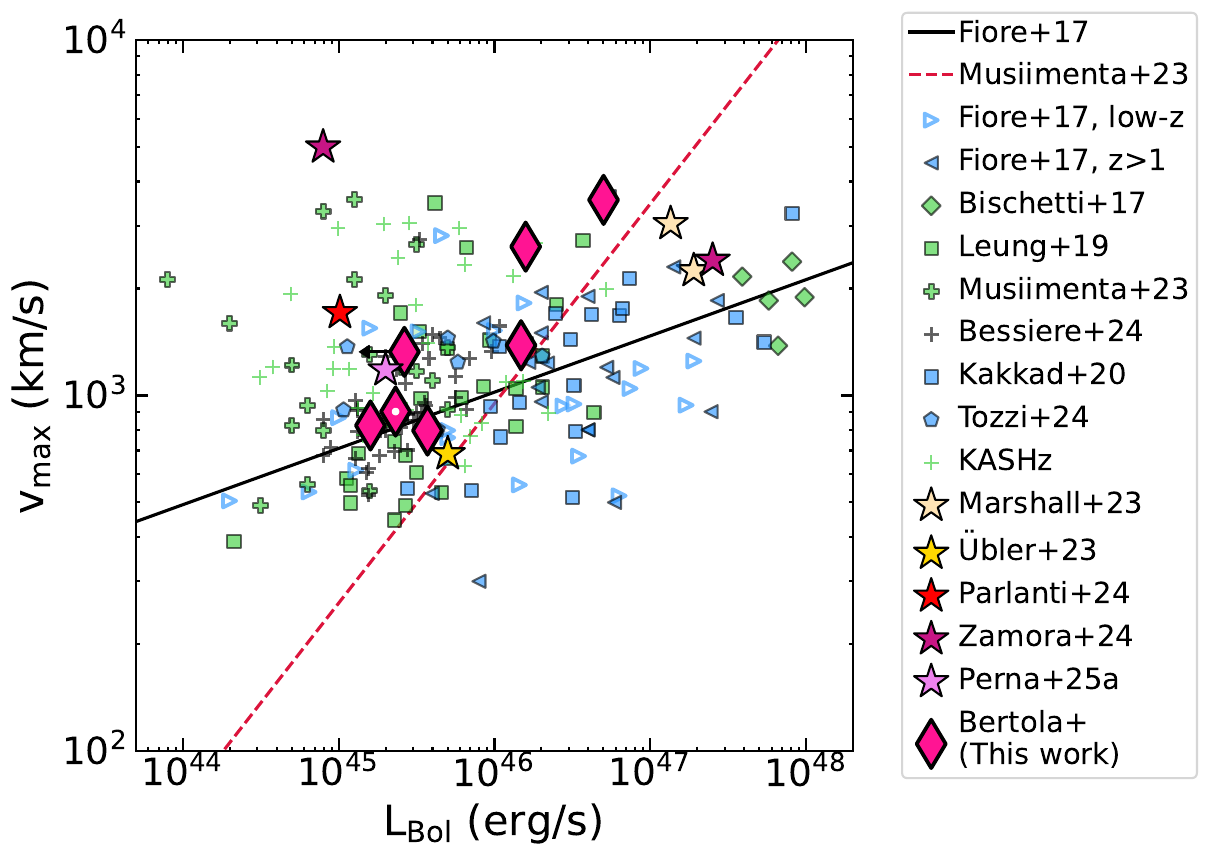}
	\caption{Outflow velocity $v_{\rm max}$ as a function of bolometric luminosity of \ganifs COS-AGNi and literature samples used in this work. The color coding is the same as Fig. \ref{fig:sample}. The solid line marks the best fit relation of \citet{fiore2017} and the dash-dotted line the best fit relation of \citet{musiimenta2023}.  }
	\label{fig:outflow_prop2}
\end{figure}

\subsection{Comparison with the literature}
We show mass outflow rate and kinetic power against bolometric luminosity in Fig. \ref{fig:outflow_prop1}, and outflow maximum velocity $v_{\rm max}$ against bolometric luminosity in Fig. \ref{fig:outflow_prop2}. We report the $\Mdot$ versus $\Lbol$, $\Edot$ versus $\Lbol$ and the inverted $\Lbol$ versus $\Vmax$ scaling relations of \citet{fiore2017}, that were built using the shown literature compilation homogenized in gas density, outflow radius and outflow geometry. We also show the more recent scaling relations of \citet{musiimenta2023}, which only considered AGNi beyond the local Universe ($z>0.5$). 
The latter relations were obtained for outflows of eROSITA AGNi and literature measurements homogenized in all parameters but the outflow velocity. Moreover, as mentioned in Sect. \ref{sec:of_lit}, \citet{musiimenta2023} assume the effective radius of the host galaxies as outflow extent for their eROSITA AGNi. The main difference between \citet{musiimenta2023} and \citet{fiore2017} relations is a steeper $v_{\rm max}$ versus $\Lbol$  relation of the former compared to that of the latter. We include the $\Mdot$ versus $\Lbol$ relation derived by \citet{bischetti2019b}, which predicts a flattening in the growth of the mass outflow rate in very bright sources ($\log(\Lbol/\Lumcgs)>47$). All the shown scaling relations were scaled in gas density as done for the single measurements. 

\begin{figure}[!h]
    \centering
    \includegraphics[width=0.45\textwidth]{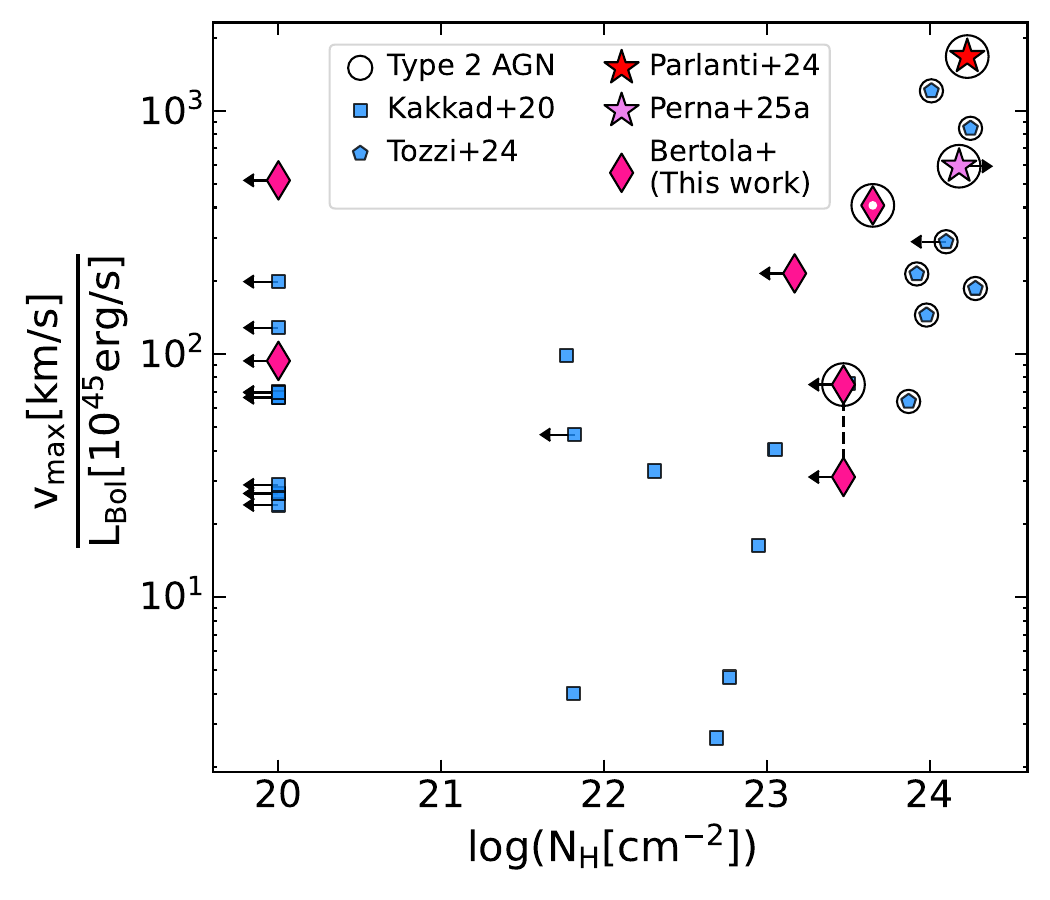}
    \caption{Outflow maximum velocity normalized by the bolometric luminosity (in units of $10^{45}~\Lumcgs$) versus column density adapted from \citet{tozzi2024}. The color coding and references are as in Fig. \ref{fig:sample}. Type 2 AGNi are marked with a black circle. COS1638-A (Type 1) and COS1638-B (Type 2) are connected by a dashed black line since the bolometric luminosity for this dual AGN was obtained from X-ray data integrated over the two components. 
    COS2949, the only COS-AGNi at $z\simeq2$ and for which $N_{\rm H}$ is constrained, confirming its high-obscuration, is marked with an additional white dot. We note that SUPER AGNi \citep{kakkad2020,tozzi2024} cover the redshift range $z=2-2.6$ and GA-NIFS AGNi (except COS2949) span the $z=3.5-4.8$ range \citep[from][\citealp{perna2025_gs133} and this work]{parlanti2024}. }
    \label{fig:voutnh_tozzi}
\end{figure}
The outflows observed in $z\sim3.5$ COS-AGNi and in other GA-NIFS AGNi previously analyzed \citep{marshall2023,Ubler2023a,parlanti2024,zamora2024,perna2025_gs133} span a range of $\Mdot$ and $\Edot$ similar to that of literature measurements and are roughly consistent with the scaling relations in the $\Mdot$ versus $\Lbol$ plane. We note that roughly half of the GA-NIFS AGNi shown in Figs. \ref{fig:outflow_prop1}-\ref{fig:outflow_prop2} show energetics ($\Mdot$, $\Edot$) larger than those predicted by the scaling relations of \citet{fiore2017}, \citet{bischetti2019a} and \citet{musiimenta2023}. Given the strong fitting degeneracy between the BLR and the NLR components in COS349, we show its outflow properties as measured through both \hal and \oiii (see Sect. \ref{sec:mout_formulas}), the latter corrected by a factor of 3 as done for the literature outflows traced in the \oiii line. The two measurements for COS349 are fairly in agreement with each other and consistent with the scatter of the literature measurements (see Fig. \ref{fig:outflow_prop1}). There are two objects from GA-NIFS that are well above the predictions, especially for what concerns the $\Edot$ versus $\Lbol$ plane. These are COS1638-B from this work and the AGN-host submillimeter galaxy companion of the QSO BR1202-0725 \citep{zamora2024}. Both targets show \oiii profiles that are totally dominated by the outflowing gas and could be identified as ``blue-outliers'' \citep[e.g.,][]{zamanov2002,marziani2003,cracco2016,perna2021,lanzuisi2024}. 
GA-NIFS COS-AGNi are also well consistent with the parameter space spanned by literature AGNi in the $\Vmax$ vs $\Lbol$ plane, with a general tendency to lie at faster velocities than those predicted by the relations of \citet{fiore2017} and \citet{musiimenta2023}. Given the uncharted $z-\Lbol$ parameter space covered by COS-AGNi, this result indicates no significant evolution of the physics driving outflows beyond $z\simeq3$.

We note that GA-NIFS COS-AGNi are the only non-extreme AGNi at $z>3$ that were studied so far for their outflow properties. In fact, even though only the results for WISSH AGNi allowed us to compute the outflow energetics following the approach of \citet{fiore2017}, only observations of extremely bright AGNi ($\log(\Lbol/ \Lumcgs)>46.7$ have been able to probe AGN-driven outflows, due to sensitivity limitations and lower \snr even in spatially integrated spectra. We note that \citet{suh2025} recently presented JWST/\nirspec IFU data of a low-mass AGNi at $z\lesssim4$ that is accreting at super-Eddington rates and shows prominent outflows, whose properties were computed using assumptions that cannot be homogenized to ours. 

Outflows detected in COS-AGNi are less extended than what assumed in literature works using unresolved slit spectroscopy, especially at these redshifts  \citep[e.g., $\Rmax\simeq7$~kpc,][]{bischetti2017}. These are equal to twice the maximum radii ($\langle\Rmax\rangle\simeq2.5$~kpc) and seven times the flux-weighted radii ($\langle\Rmax\rangle\simeq1$~kpc) of COS-AGNi, while the radii reported in \citet{shen2016} can be even larger ($R\simeq5-15$~kpc)\footnote{We note that these values are mentioned in the main text of \citet{shen2016} but are not tabulated in the provided catalog. We thus chose not to include outflow measurements from this work in our comparison given the absence of such an info target by target.}. There is also indication that Extremely-Red Quasars (ERQs) show more compact (by roughly a factor $2\times$) and faster outflows (by roughly a factor $3\times$) compared to blue AGNi \citep{perrotta2019,lau2024_ifuperrotta}, and that Type 2 AGNi show more extended outflows compared to Type 1 AGNi \citep{tozzi2024}. Moreover, \citet{tozzi2024} showed that Compton-Thick AGNi (i.e., highly obscured, $\log N_{\rm H}>24$, in the X-rays) harbor faster outflows compared to their Type 1 and less-obscured counterparts matched in AGN bolometric luminosity, suggesting a relevant role of radiation pressure on dust in outflow acceleration. The findings of \citeauthor{perrotta2019}, \citeauthor{lau2024_ifuperrotta} and \citeauthor{tozzi2024} can be interpreted in terms of the evolutionary scenario from obscured accretion to shining AGNi through the ``blow-out'' phase \citep[e.g.,][]{lapi2014} at different AGN power. 

We show in Fig. \ref{fig:voutnh_tozzi} the distribution of SUPER and GA-NIFS AGNi in terms of maximum outflow velocity over bolometric luminosity as a function of nuclear column density. 
Figure \ref{fig:voutnh_tozzi} shows that GA-NIFS AGNi at $z>3$ presented in previous studies \citep{parlanti2024,perna2025_gs133} and in this work (see Tables \ref{tab:sed} and \ref{tab:of_prop_all}) tentatively follow the ``faster outflows in highly obscured AGNi'' trend seen by \citet{tozzi2024} for AGNi at $z=2-2.5$. The Kolmogorov-Smirnov test applied to Type 1 versus Type 2 AGNi in GANIFS (or in the COS-AGNi sample only) in inconclusive, due to the very small sample size.
A larger sample of AGNi would be needed to confirm this behavior at $z>2.5$. We note that the bolometric luminosity estimated from SED fitting for COS1118 (highest $\Vmax/\Lbol$ for $\log N_{\rm H}<20$) might not be well constrained due to a poor photometric coverage of the target, mainly dominated by upper limits in the IR regime, leading to an overestimated $\Vmax/\Lbol$ in Fig. \ref{fig:voutnh_tozzi}. In fact, using the relations derived by \citet{duras_2020}, we find an X-ray-based bolometric luminosity $\log(\Lbol/\Lumcgs)=45.6$, that is $\simeq0.5$~dex larger than that obtained from SED fitting (see Table \ref{tab:sed}), which would move COS1118 closer to the other Type 1 AGNi with no obscuration in the X-rays. 

\begin{figure}[!h]
	\centering
	\includegraphics[width=0.5\textwidth]{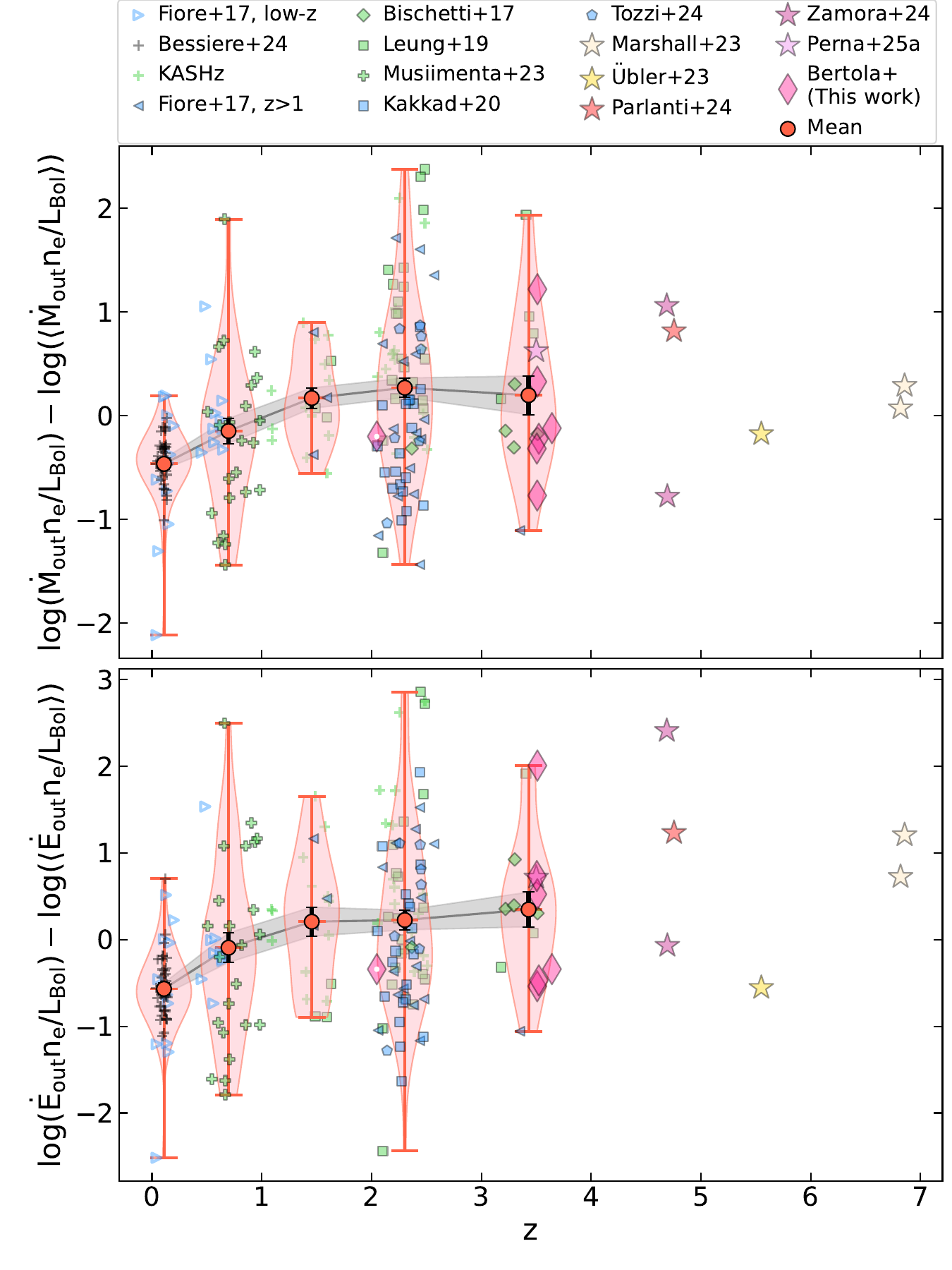}
	\caption{Ratio of mass outflow rate times gas density over bolometric luminosity (top) and ratio of kinetic power times gas density over bolometric luminosity (bottom), normalized to their sample means, versus redshift. Samples and color coding are the same as Fig. \ref{fig:outflow_prop1}. Violin plots show the distribution in each considered redshift bin, red circles mark the mean of each redshift bin and black vertical error bars mark the standard deviation of the mean. Points at $z>4$ are shown to highlight the few measurements available. We note that only other two AGNi at $z>4$ were observed by JWST \citep{loiacono2024,suh2025} and are not part of the plot due to the impossibility of homogenizing their measurements with our assumptions in Sect. \ref{sec:of_hom}. }
	\label{fig:outflow_violins}
\end{figure}
Lastly, we investigate the possibility of a redshift evolution of the strength of outflows up to redshift $z=4$, where measurements become too sparse. Figure \ref{fig:outflow_violins} shows the redshift evolution of the logarithm of the ratio of $\Mout n_{\rm e}$ (left), that is the mass outflow rate for a parameterized gas density, and of $\Eout n_{\rm e}$ (right), that is the outflow kinetic power for a parameterized gas density, over AGN bolometric luminosity $\Lbol$, normalized at their mean values computed across all redshifts. We divide the collected measurements in five redshift bins ($0.08<z<0.45$, $0.45\leq z<1$, $1\leq z<2$, $2\leq z<3$, $3\leq z<4$) and compute the mean $\Mout n_{\rm e}/\Lbol$ in each bin. While the scatter is large, as the violin plots indicate, the mean values do not show any evidence of evolution beyond $z>1$. We note that repeating this exercise for the mass outflow rate and the bolometric luminosity separately returns an increasing trend of the mean values for increasing redshift. Outflows at $z\simeq0$ in Fig. \ref{fig:outflow_violins} appear to be less energetic than those at $z>1$ by more than one, but less than two, standard deviations. This behavior holds also when excluding the outlier at lower $\Mout$($\Eout$) in the $z\simeq0$ bin and the outlier at larger $\Mout$($\Eout$) in the $z\simeq0.5$ bin. However, we caution that there are many other works at $z<0.5$ that could not be included because the reported measurements could not be homogenized with our assumptions. 
We also caution that there are many caveats in the results shown Fig. \ref{fig:outflow_violins}, the main one being that the collected measurements allow for a rather sparse or non-exhaustive sampling of some redshift bins, like the at $z\simeq1.5$ and at $0.08<z<0.45$, respectively. 
Moreover, having parameterized the outflow gas density, the comparison as is in Fig. \ref{fig:outflow_violins} is equal to assuming that each outflow has the same gas density. 

\citet{maiolino2025} recently investigated the \textit{Chandra}-JWST properties of AGNi selected based on the presence of the BLR. They find that the majority of such JWST-selected AGNi at $z>4$ do not show evidence of prominent outflows, and claim it is in contrast with lower redshift or higher luminosity, at odds with our result of no redshift evolution. One possible explanation is that COS-AGNi are a sample of X-ray selected AGNi and the majority of the literature AGNi are X-ray detected, while the bulk of JWST-selected AGN are X-ray undetected, as also remarked by \citeauthor{maiolino2025}. 
This could support the scenario in which the X-ray weakness of JWST-selected AGNi could be due to weaker outflows (possibly because of lower metallicity, which is the key factor in driving both nuclear -line locking- and extended -dust driven- outflows). Hence, more gas would linger around the BH in these JWST-selected AGNi, causing large covering factor of Compton thick material which makes them X-ray weak. The AGNi that are X-ray selected (or X-ray detected) are instead those in which the outflow managed to develop and clear some of the nuclear gas to make X-rays easier to detect, possibly because such X-ray detected AGNi are more metal rich or because they are more evolved, giving radiation pressure more time to interact with the ISM. 

\section{Summary and conclusions}
\label{sec:summary}
We have presented the analysis of JWST/\nirspec IFU data from the GA-NIFS GTO program of seven $z\simeq3.5$ AGNi with $45.2<\log(\Lbol/\Lumcgs)<46.7$ (two of which are part of a dual-AGN system; \citealp{perna2025dualsamp}) that were selected in X-rays ($\log(L_{\rm X}/\Lumcgs)>44$) from the COSMOS-Legacy field \citep{marchesi2016}. The capabilities of JWST/\nirspec IFU allowed an uncharted $\Lbol-z$ parameter space to be probed, as highlighted in Fig. \ref{fig:sample}. In fact, we have sampled for the first time the rest-frame optical properties of a sample of AGNi at $z\simeq3.5$ and $\log(\Lbol/\Lumcgs)=45.2-46.7$, collecting spectra of a high \snr and broad wavelength range and with the added benefit of the spatially resolved information provided by IFU data. 

\begin{itemize}
	\item We investigated the ionized gas properties of our AGNi both in a spaxel-by-spaxel fitting approach and from integrated spectra, with the main aim of studying their outflows. Outflow properties were derived from spectra integrated over the ``outflow region,'' identified as the 3$\sigma$ emission region of the \hal or \oiii broad component maps (or \hal or \nii broad component maps for COS2949) based on whichever was most extended. 
	\item We find that the majority of our targets show close-by companions and clumps, both in the form of galaxies (COS590, COS2949) and dual AGNi systems (COS1656, COS1638; see also \citealp{perna2025dualsamp}). Moreover, a good fraction of the objects are embedded in diffused ionized emission characterized by low-velocity dispersion and moderate-velocity displacement with respect to the AGNi rest frame (COS1638, COS590, COS2949, COS1118) and with kinematics that appear to be unrelated to the outflows, which are mostly confined to the proximity of the AGN. 
	\item We derived the outflow properties of our targets and compared our results with those present in the literature, both from slit spectroscopy and IFS data (see Fig. \ref{fig:outflow_prop1}-\ref{fig:outflow_prop2}). We homogenized the literature measurements to those of \citet{fiore2017} and \citet{musiimenta2023} for a consistent comparison with their scaling relations. Given the very different results and assumptions that are present in the literature regarding the outflow gas electron density, we scaled the outflow properties by the outflow gas density to obtain a comparison that is not dependent on the outflow gas density. 
	\item We find that outflows observed in COS-AGNi are well consistent with the previous results in the $\Mdot$ versus $\Lbol$, $\Edot$ versus $\Lbol$, and $\Vmax$ versus $\Lbol$ parameter spaces and that they mostly fall close to or above previous scaling relations \citep[e.g.,][]{fiore2017,bischetti2019b,musiimenta2023}. 
	\item The GA-NIFS AGNi (COS-AGNi complemented with GA-NIFS AGNi from the literature; \citealp{Ubler2023a,parlanti2024,perna2025_gs133}) tentatively follow the trend discovered by \citet{tozzi2024} in AGNi at $z\simeq2-2.5$, that is, showing faster outflows for the thickest absorbers as traced in the X-rays, while we find no clear redshift evolution of $\Mout n_{\rm e}/\Lbol$ and $\Edot n_{\rm e}/\Lbol$ beyond $z>1$ (Fig. \ref{fig:outflow_violins}). 
	\item  In general, the trends shown by GA-NIFS AGNi at $z>3$ are similar to those observed at lower redshifts, suggesting that the nature of AGN feedback does not significantly evolve beyond $z\simeq3$. 
\end{itemize}

The unprecedented capabilities of JWST have opened a new window in extragalactic astrophysics, enabling for the first time to characterize the ionized gas properties as traced by optical emission lines like \hal and \oiii in high-\z galaxies up to cosmic dawn. In the field of AGN feedback and AGN/galaxy coevolution,  future observations with JWST/NIRSpec will provide statistical samples 
provide statistical samples of AGN at $z>4$ to assess the properties and effects of AGN-driven ionized outflows in the key phase of galaxy evolution, that is the cosmic epoch between cosmic noon and cosmic dawn.

\begin{acknowledgements}
EB and GC acknowledge financial support from INAF under the Large Grant 2022 ``The metal circle: a new sharp view of the baryon cycle up to Cosmic Dawn with the latest generation IFU facilities''. EB acknowledges financial support from INAF through the “Ricerca Fondamentale 2024” program (mini-grant 1.05.24.07.01).
AJB acknowledges funding from the “FirstGalaxies” Advanced Grant from the European Research Council (ERC) under the European Union’s Horizon 2020 research and innovation program (Grant agreement No. 789056).
H\"U acknowledges funding by the European Union (ERC APEX, 101164796). Views and opinions expressed are however those of the authors only and do not necessarily reflect those of the European Union or the European Research Council Executive Agency. Neither the European Union nor the granting authority can be held responsible for them. 
RM acknowledges support by the Science and Technology Facilities Council (STFC), from the ERC Advanced Grant 695671 "QUENCH", and funding from a research professorship from the Royal Society. 
JS acknowledges support by the Science and Technology Facilities Council (STFC), by the ERC through Advanced Grant 695671 ``QUENCH'', and by the UKRI Frontier Research grant RISEandFALL.
MP, SA, and BRdP acknowledge grant PID2021-127718NB-I00 funded by the Spanish Ministry of Science and Innovation/State Agency of Research (MICIN/AEI/10.13039/501100011033). MP also acknowledges the grant RYC2023-044853-I, funded by  MICIU/AEI/10.13039/501100011033 and European Social Fund Plus (FSE+).
IL acknowledges support from  PRIN-MUR project “PROMETEUS”  financed by the European Union -  Next Generation EU, Mission 4 Component 1 CUP B53D23004750006.
SC and GV acknowledge support from the European Union (ERC, WINGS,101040227). 
G.T. acknowledges financial support from the European Research Council (ERC) Advanced Grant under the European Union’s Horizon Europe research and innovation programme (grant agreement AdG GALPHYS, No. 101055023).
\end{acknowledgements}

%
\bibliographystyle{aa} 
\bibliography{mybiblio} 
%

\begin{appendix}
\onecolumn
\FloatBarrier
\section{Maps and spectra}
\label{app:maps_specs}
\begin{figure*}[!h]
    \centering
    \includegraphics[width=1\textwidth]{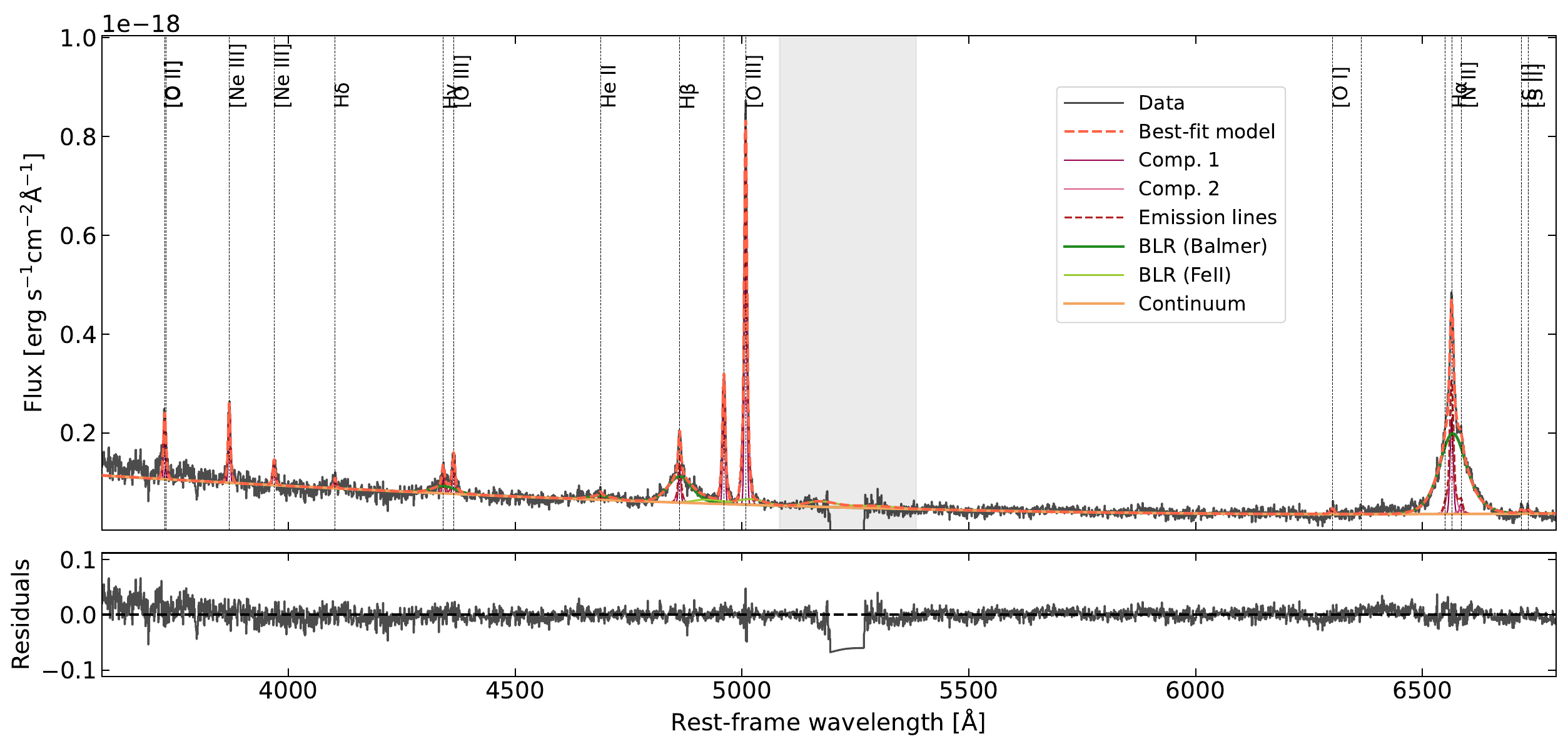}
    \includegraphics[width=1\textwidth]{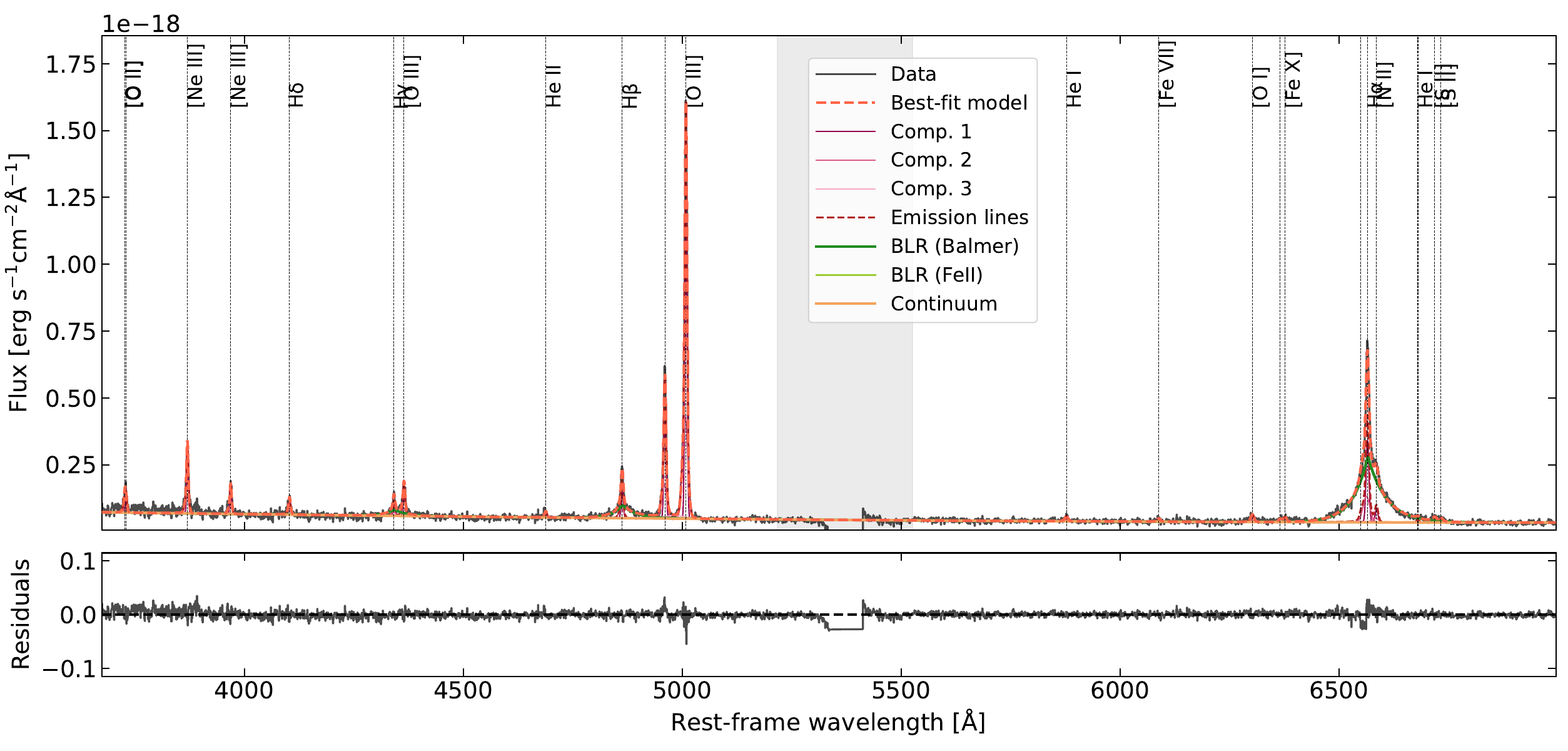}\\
    \caption{Nuclear spectra of COS1118 (Top) and COS590 (Bottom) used to produce the BLR template. Data and residuals are in black, total model is in dashed red, continuum is in orange, broad Balmer lines and FeII emission from the BLR are in dark and light green, respectively. FeII emission in these two targets is weak, and in COS590 almost marginal. Additional Gaussian emission line components are shown as dark to light purple and the total emission line model is shown as a dashed purple line. The gray shaded region marks the wavelength range that was excluded in the fitting process. }
    \label{fig:blr_of_1}
\end{figure*}

\begin{figure*}[!h]
	\centering
	\includegraphics[width=0.98\textwidth]{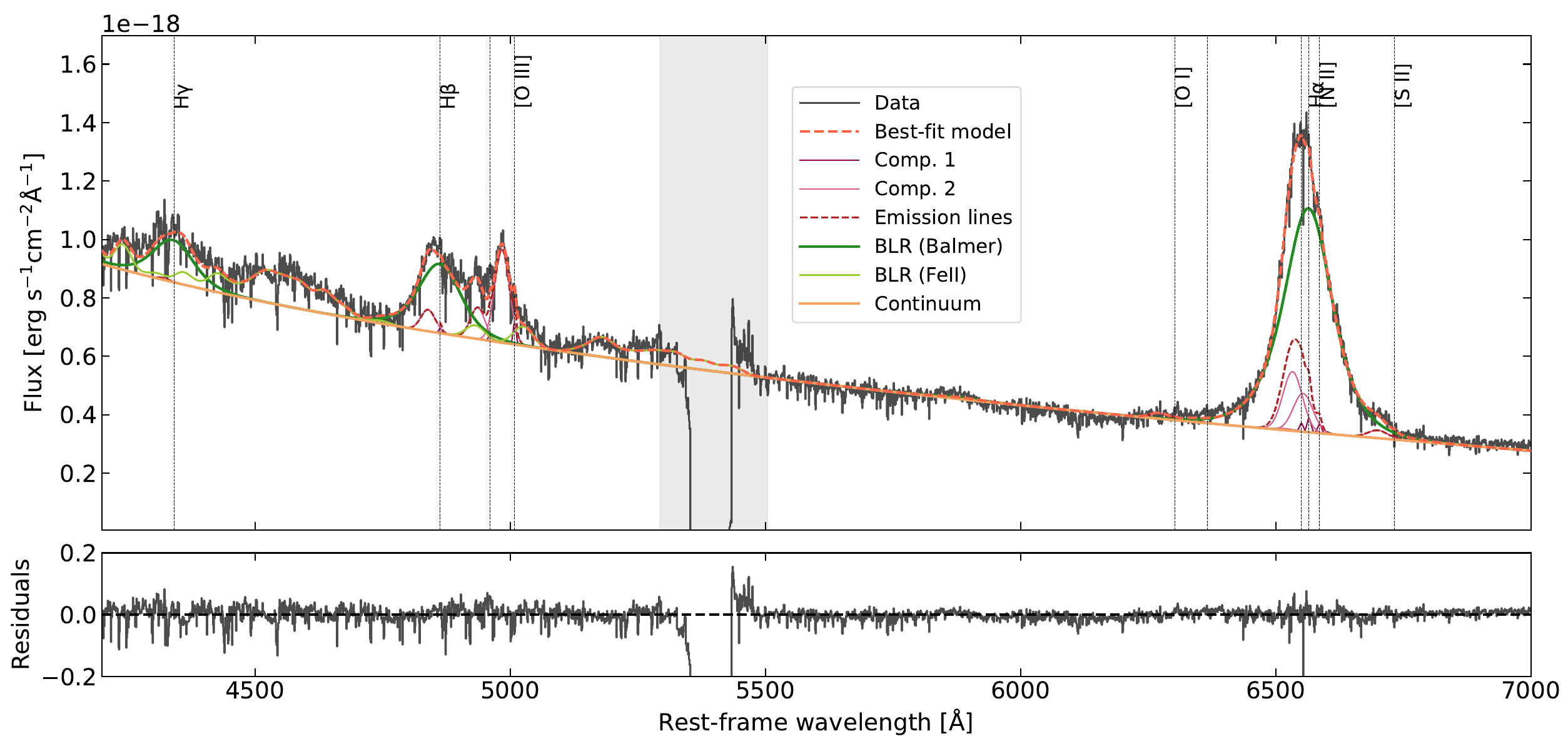}\\
	\includegraphics[width=0.98\textwidth]{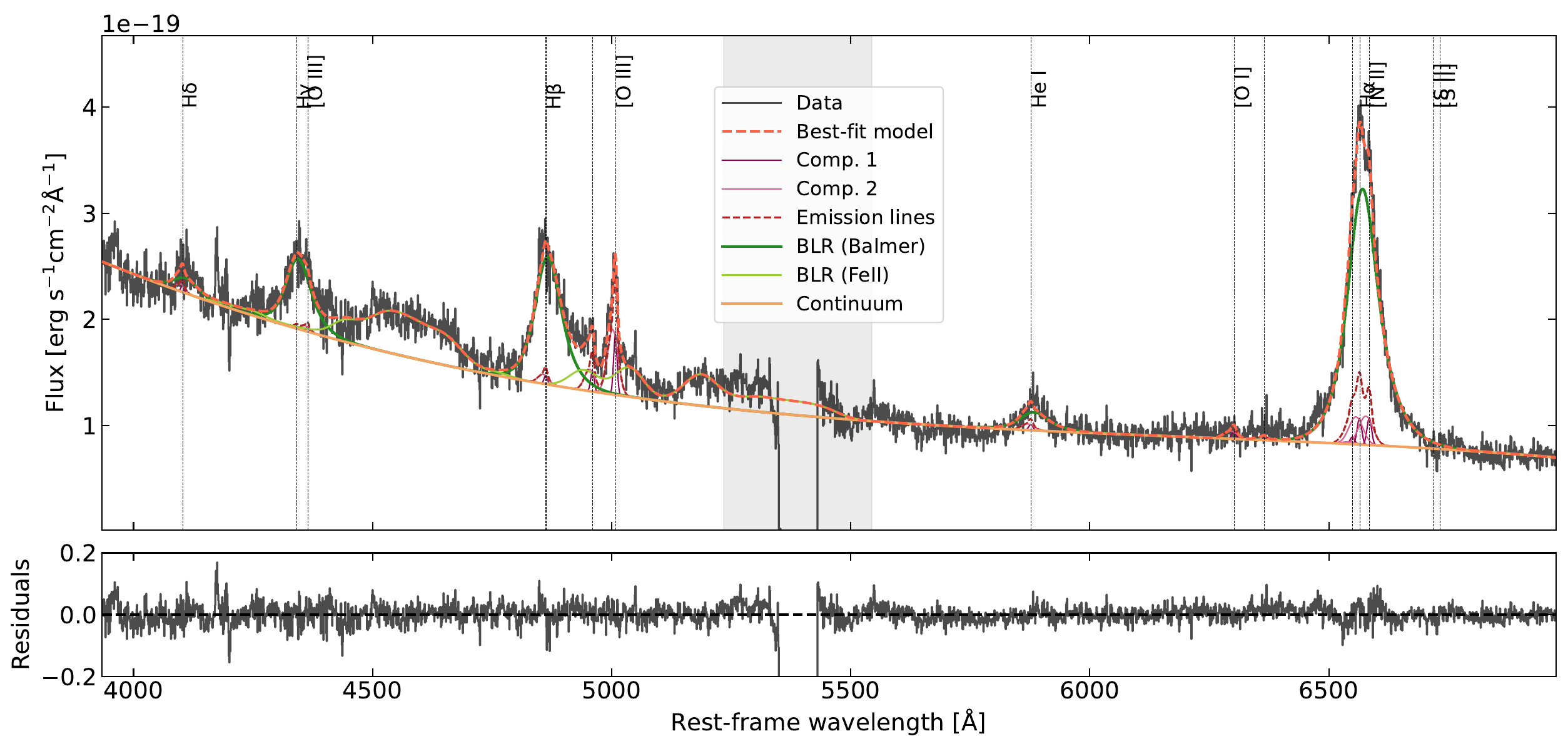}\\
	\caption{Nuclear spectra of COS1638-A (Top) and COS349 (Bottom) used to produce the BLR template. Color coding is as in Fig. \ref{fig:blr_of_1}. }
	\label{fig:blr_of_2}
\end{figure*}

\begin{figure*}[!h]
    \includegraphics[width=0.98\textwidth]{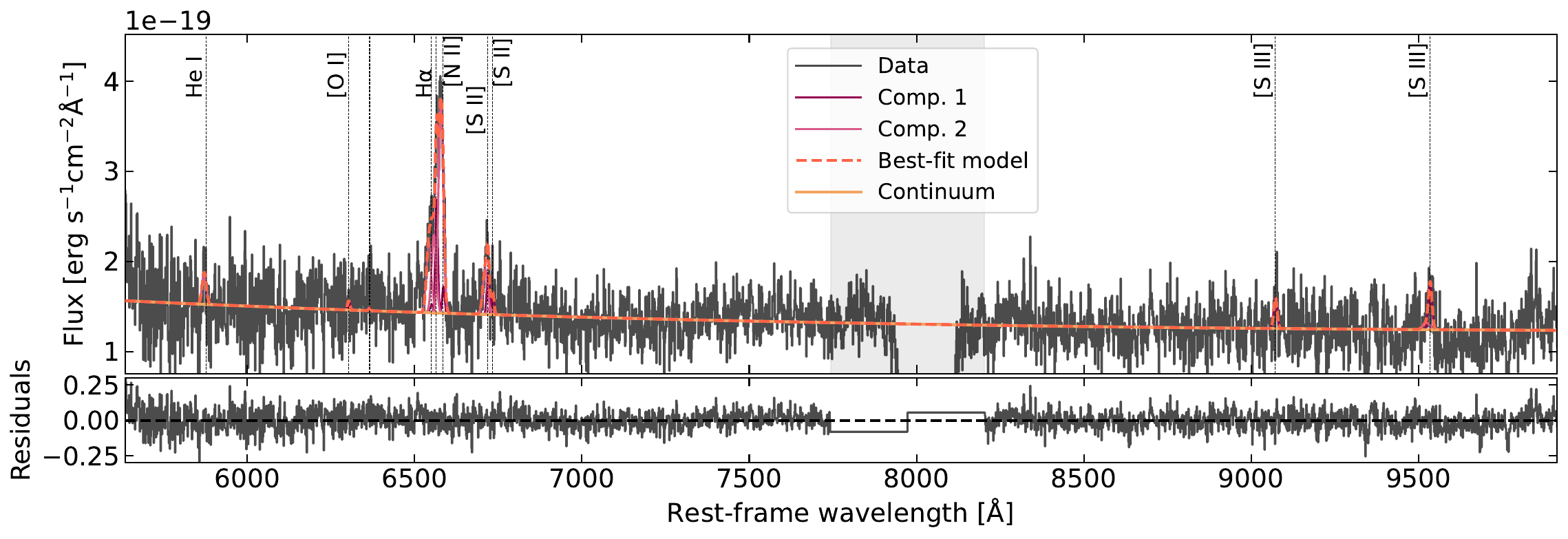}\\
    \centering
    \includegraphics[width=\textwidth]{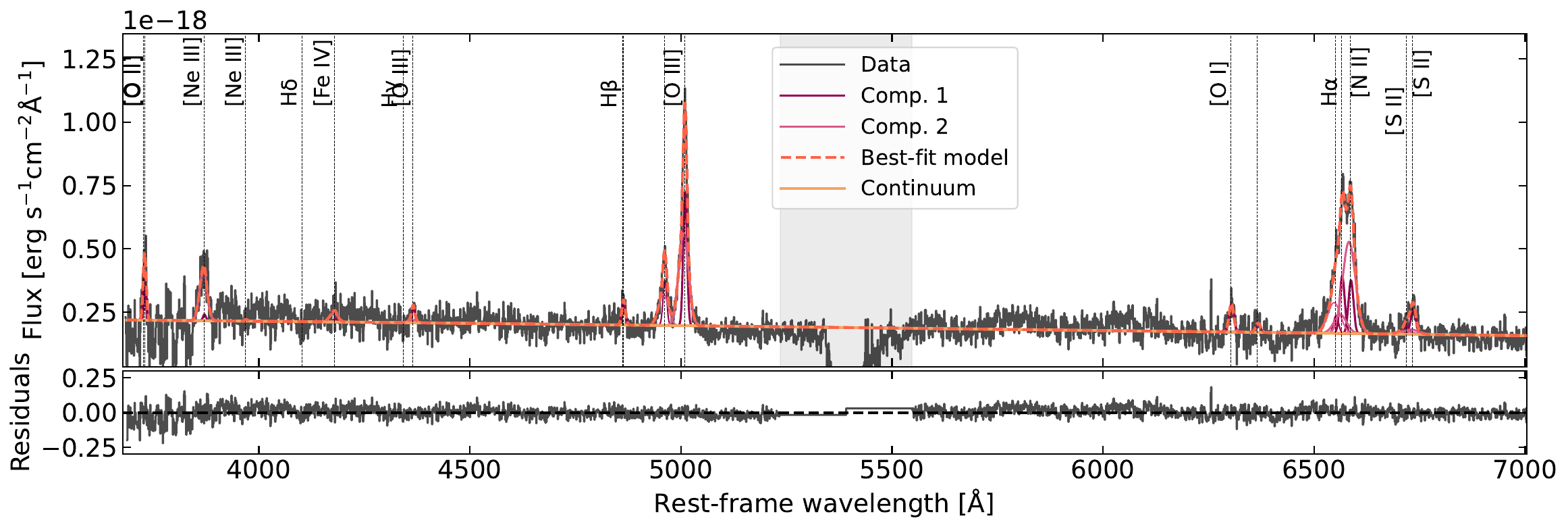}\\
      \includegraphics[width=\textwidth]{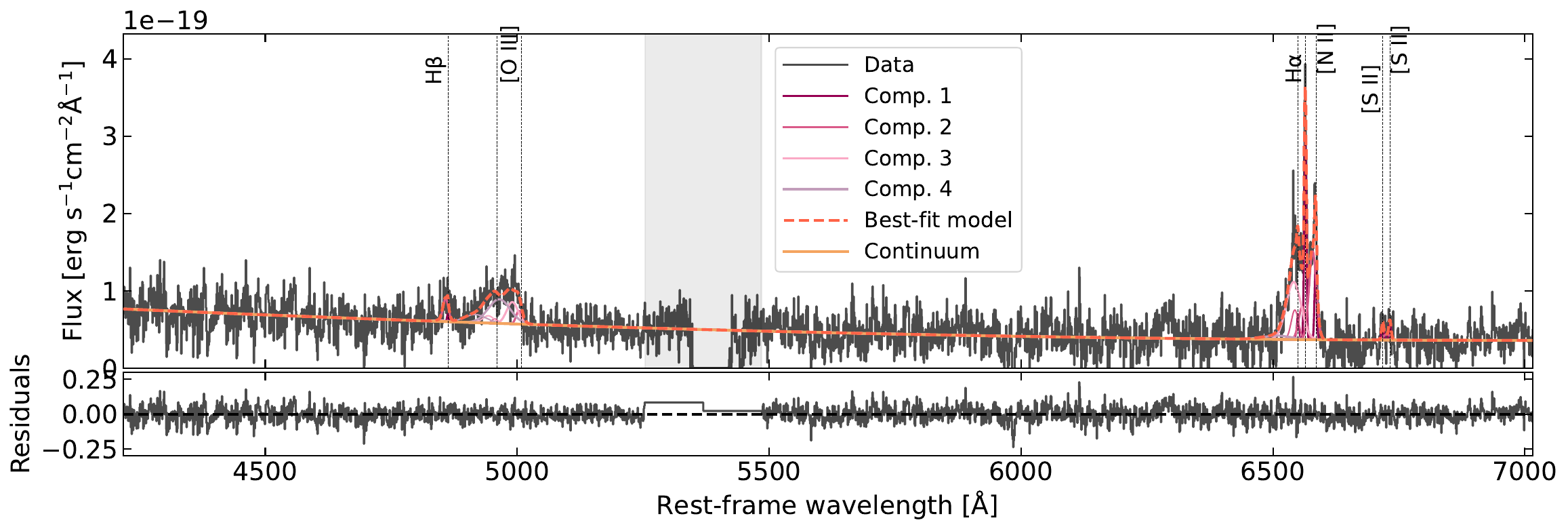}\\
    \caption{Spectra and best-fit models of COS2949 (Top), COS1656-A (Middle) and COS1638-B (Bottom) integrated over the outflow region (S/N>3 mask of the broad emission). Data and residuals are in black, total model is in dashed red, continuum is in orange, additional Gaussian emission line components are shown as dark to light purple. The gray shaded region marks the wavelength range that was excluded in the fitting process. }
    \label{fig:spectra_type_2}
\end{figure*}

\begin{figure*}[!h]
    \centering
    \includegraphics[width=0.9\textwidth]{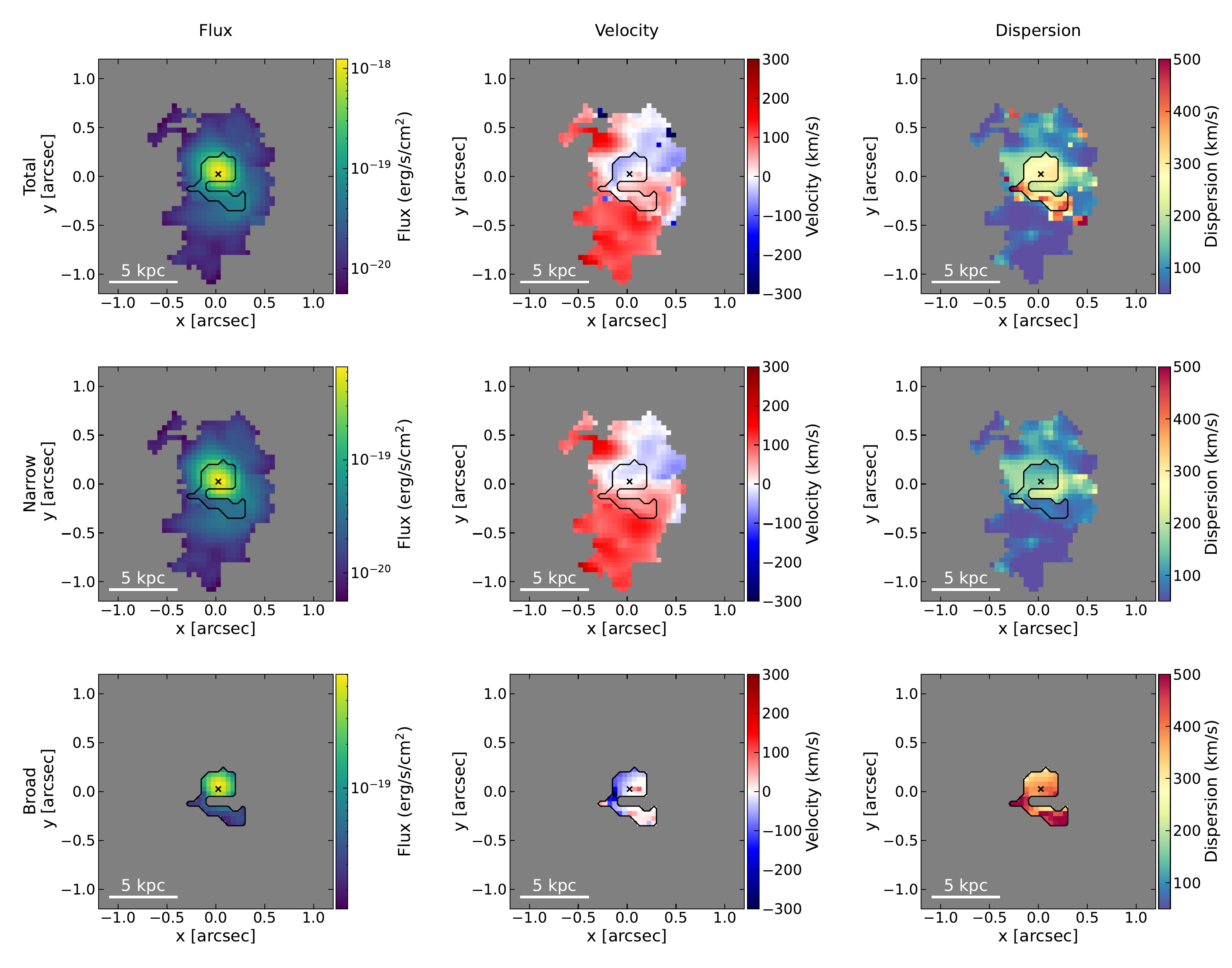}
    \\
    \includegraphics[width=0.45\textwidth]{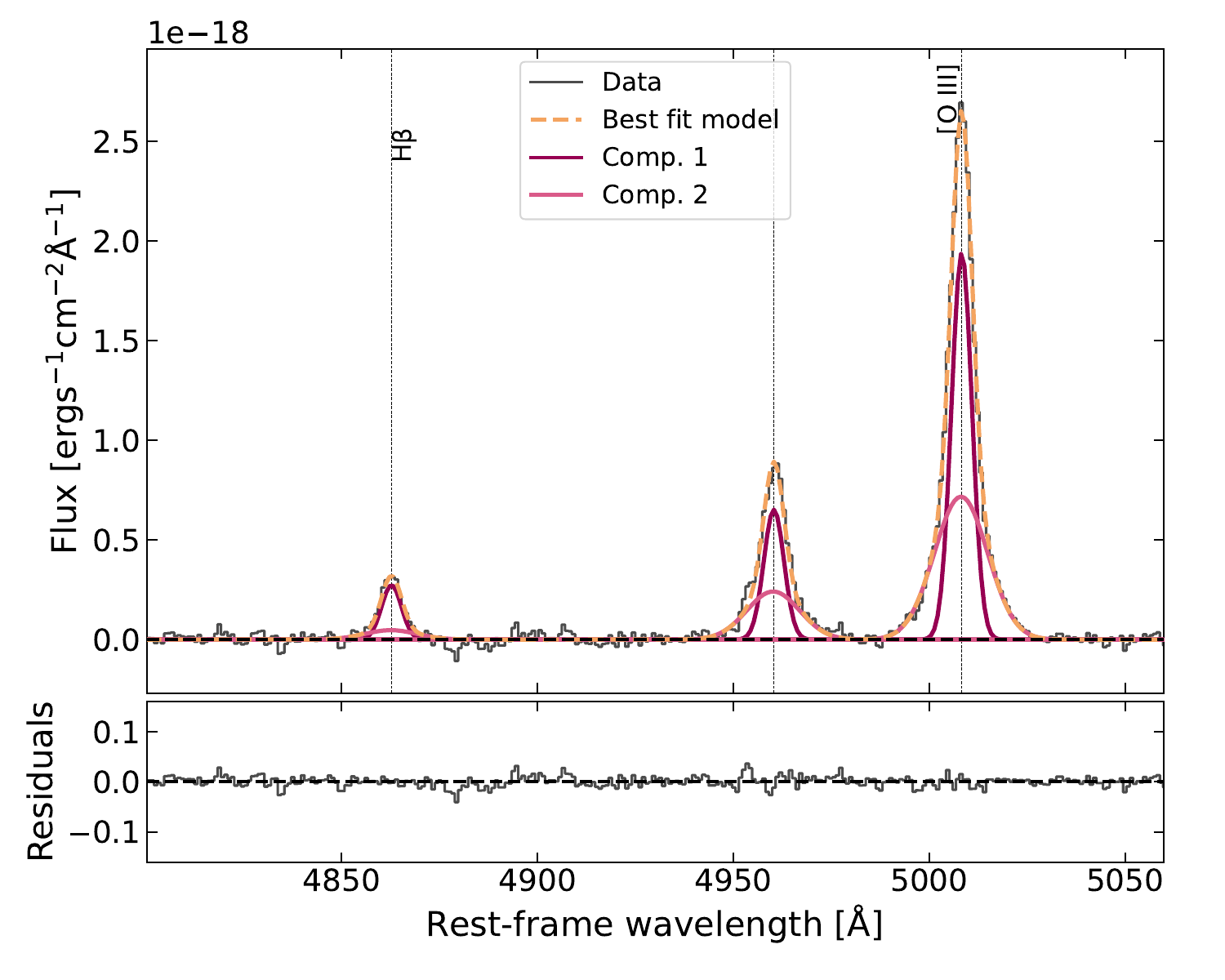}
    \hfil
    \includegraphics[width=0.45\textwidth]{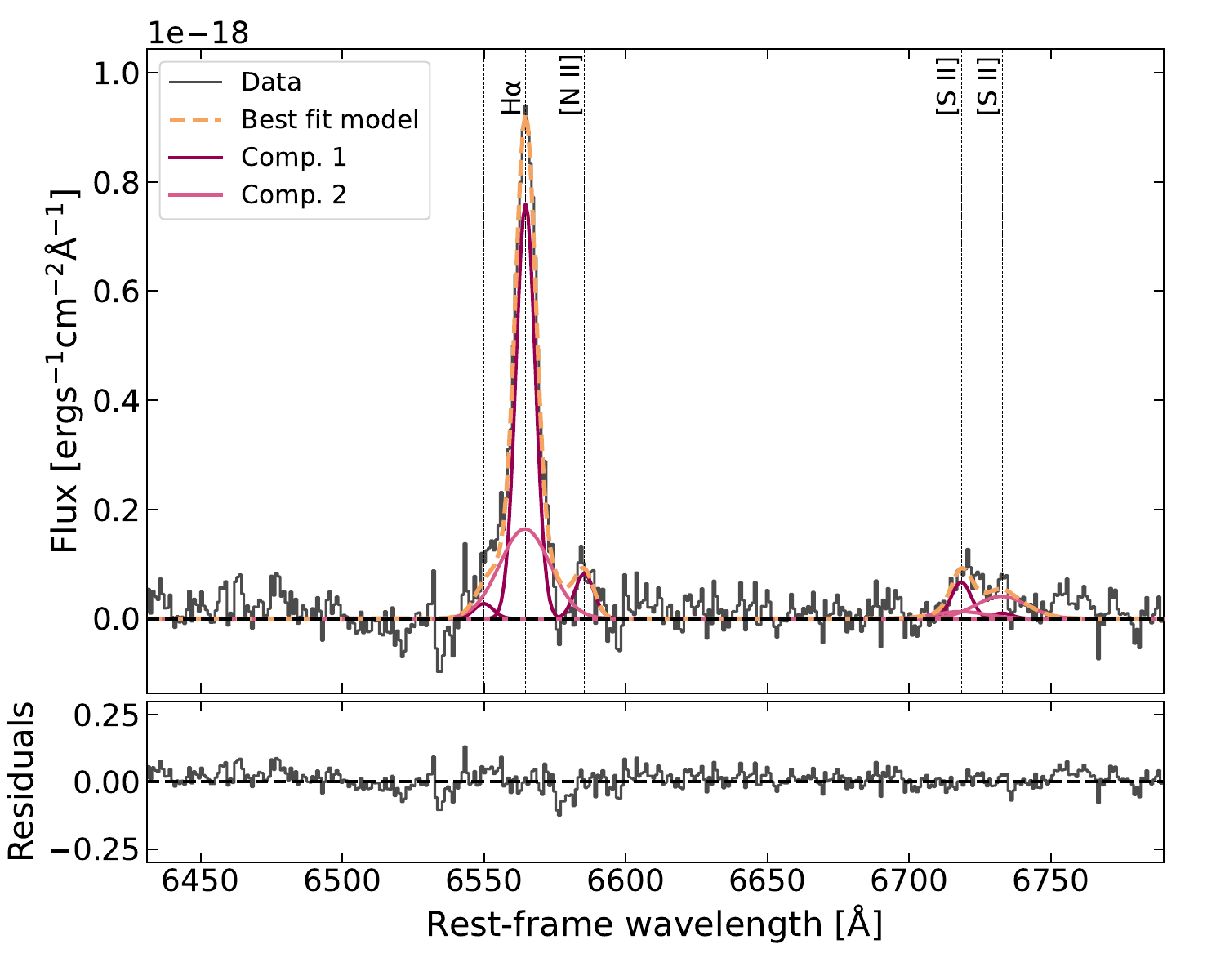}
    \caption{\textit{Top}: \oiii maps of COS1118. From top to bottom: total, narrow and broad components. From Left to Right: flux, velocity and velocity dispersion. Black contours mark the outflow region used to produce the outflow-integrated spectra. \textit{Bottom}: Close-up on the \hbe+\oiii complex (Left) and the \hal+\nii+\sii complex (Right) of the continuum- and BLR-subtracted spectrum of COS1118 integrated over the 3$\sigma$ mask of the broad \oiii emission. Data and residuals are in black, total model is in dashed orange, single line components are shown as dark to light purple. Component 1 corresponds to the narrow component, additional components add up to the broad component. }
    \label{fig:maps2_cos1118_oiii}
\end{figure*}

\begin{figure*}[!h]
    \centering
    \includegraphics[width=0.9\textwidth]{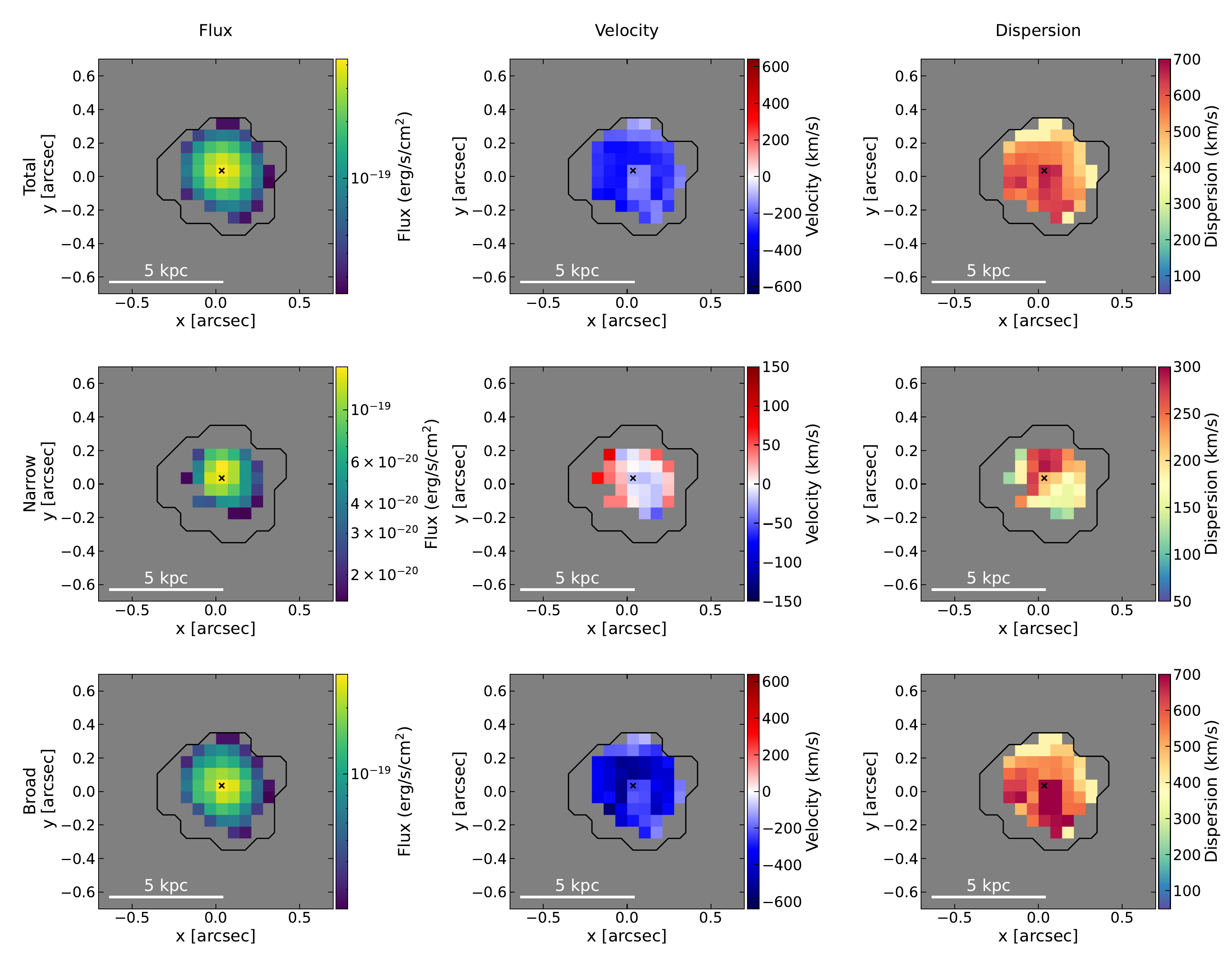}
    \\
    \includegraphics[width=0.45\textwidth]{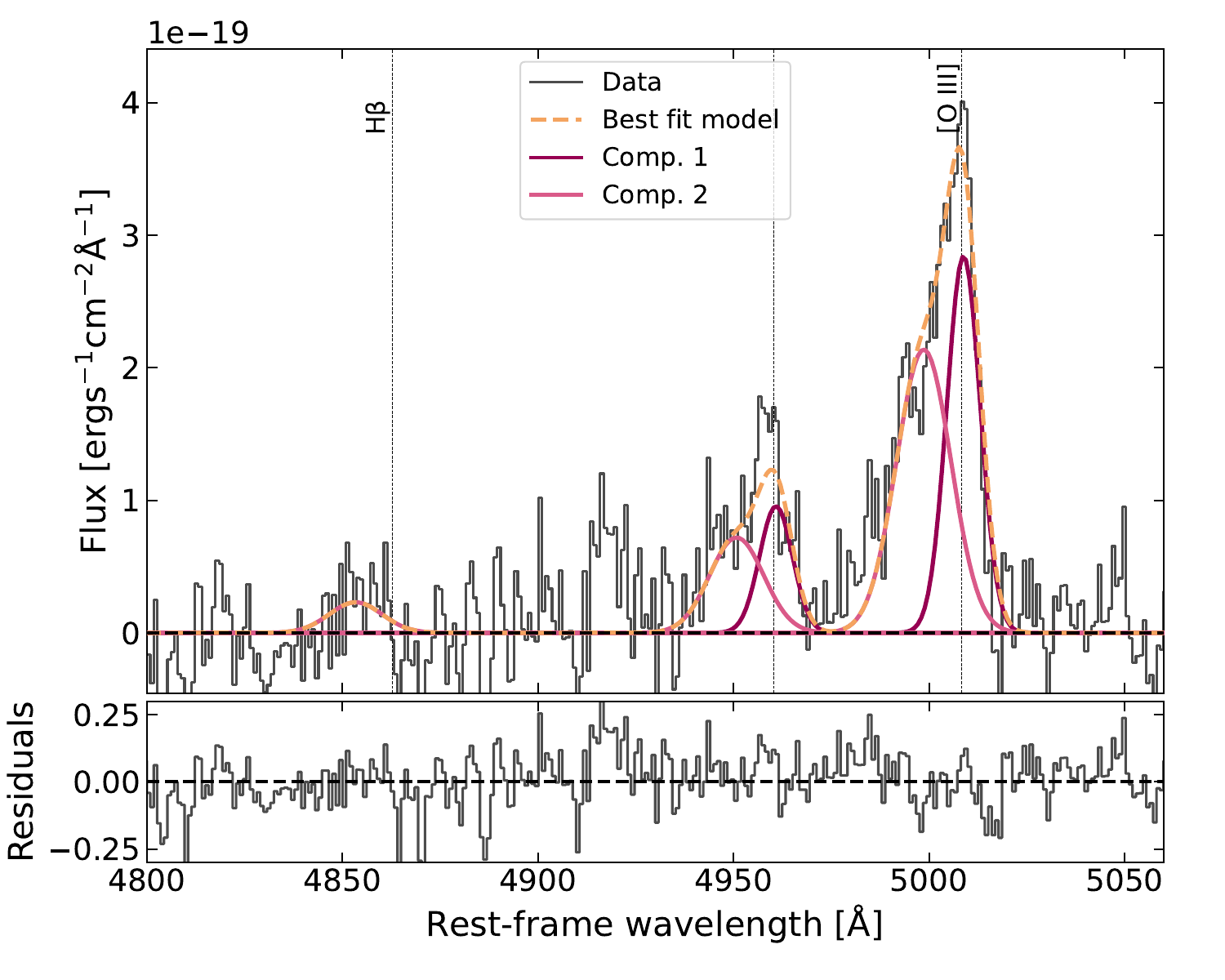}
    \hfil
    \includegraphics[width=0.45\textwidth]{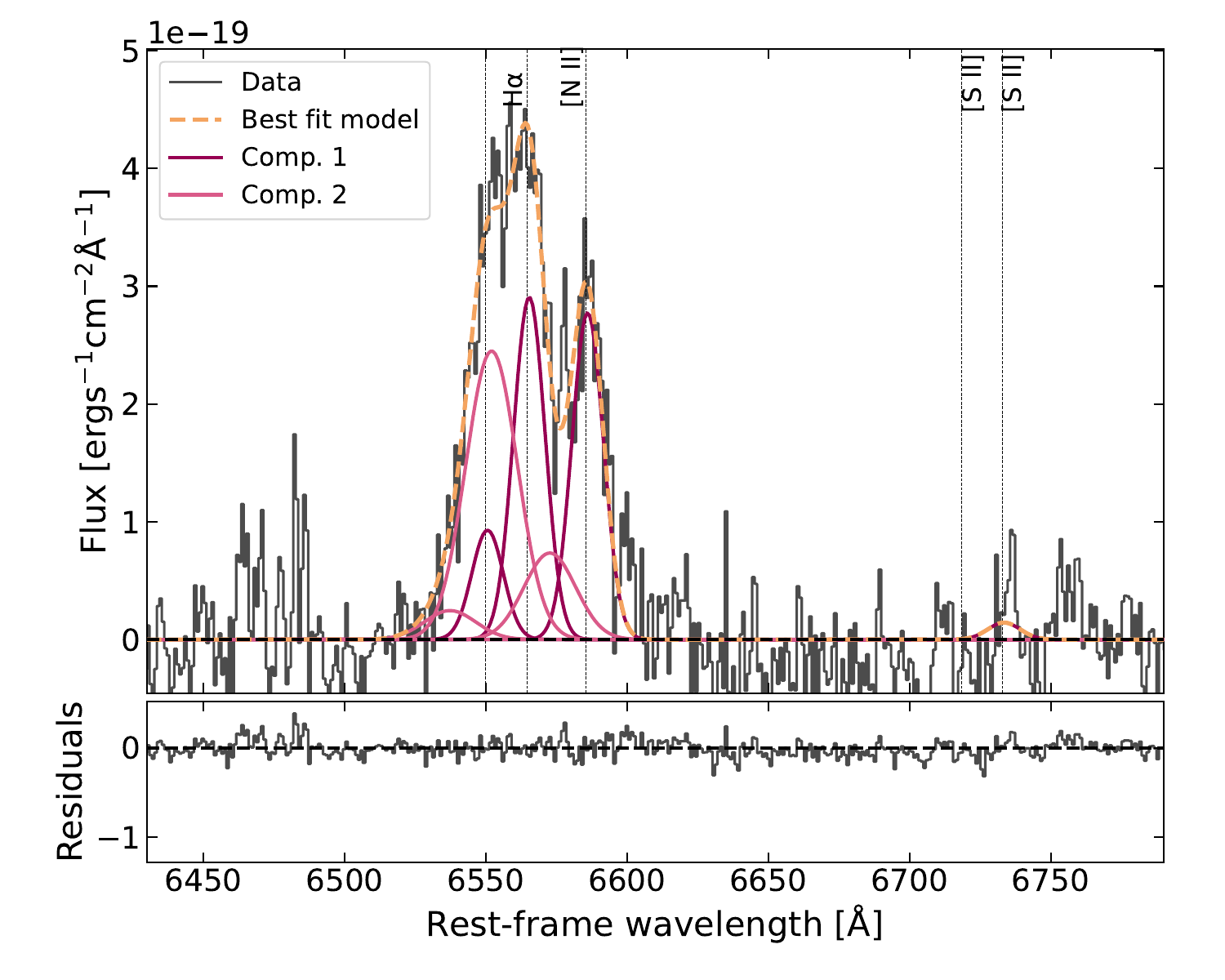}
    \caption{\textit{Top}: \oiii maps of COS349. From top to bottom: total, narrow and broad components. From Left to Right: flux, velocity and velocity dispersion. Black contours mark the outflow region used to produce the outflow-integrated spectra. \textit{Bottom}: Close-up on the \hbe+\oiii complex (Left) and the \hal+\nii+\sii complex (Right) of the continuum- and BLR-subtracted spectrum of COS349 integrated over the 3$\sigma$ mask of the broad \oiii emission. Data and residuals are in black, total model is in dashed orange, single line components are shown as dark to light purple. Component 1 corresponds to the narrow component, additional components add up to the broad component. }
    \label{fig:maps2_cos349_oiii}
\end{figure*}

\begin{figure*}[!h]
    \centering
    \includegraphics[width=0.9\textwidth]{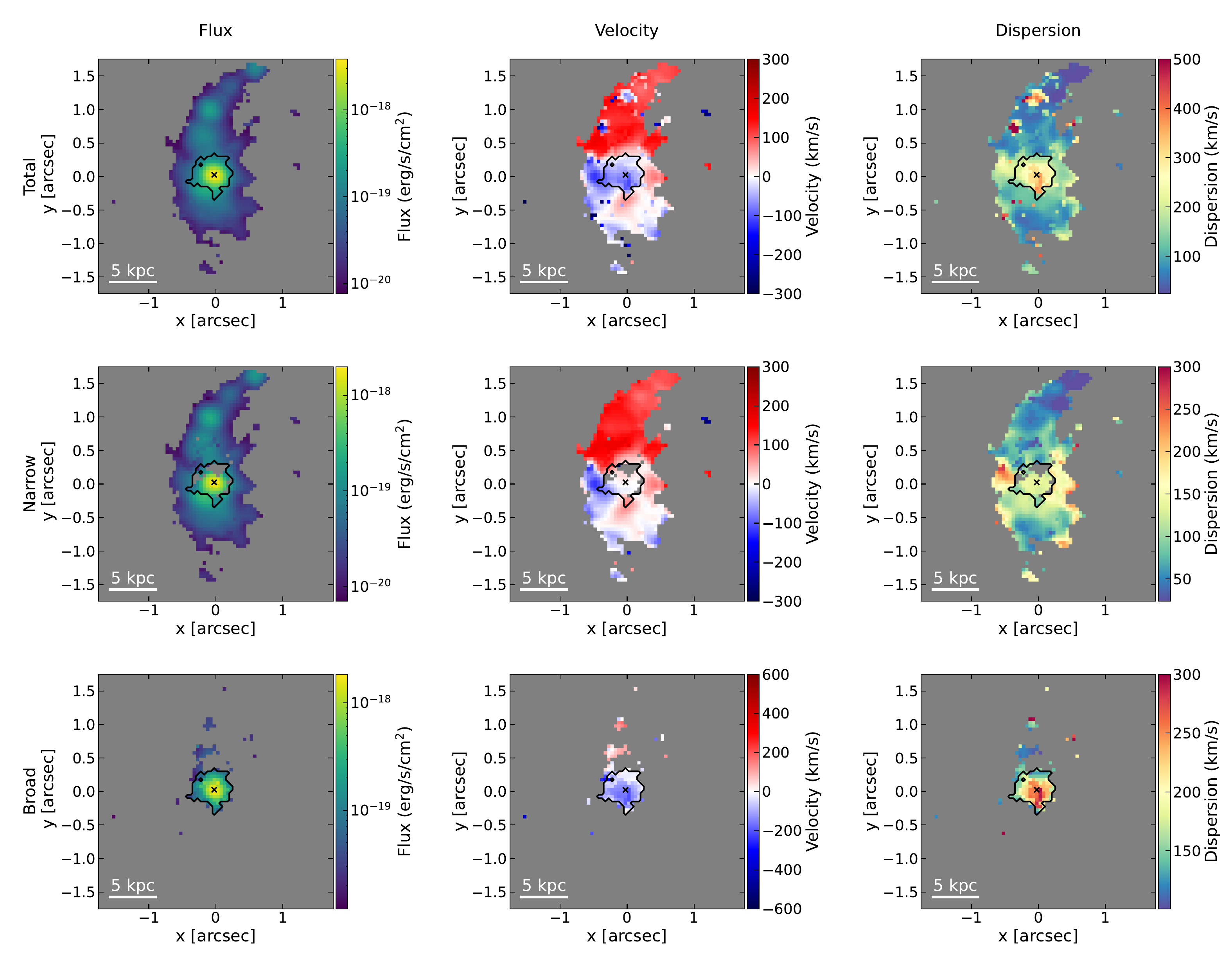}
    \\
    \includegraphics[width=0.45\textwidth]{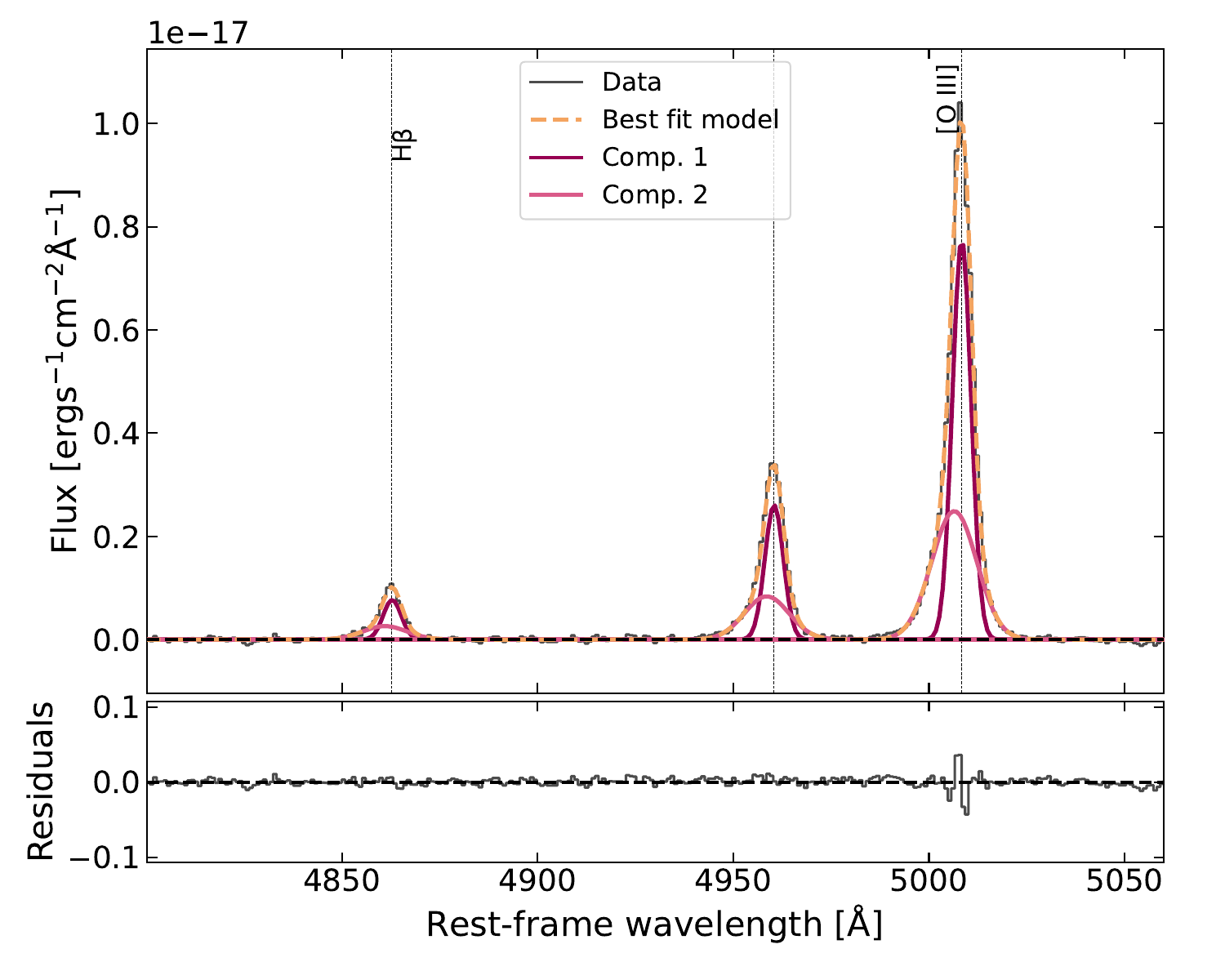}
    \hfil
    \includegraphics[width=0.45\textwidth]{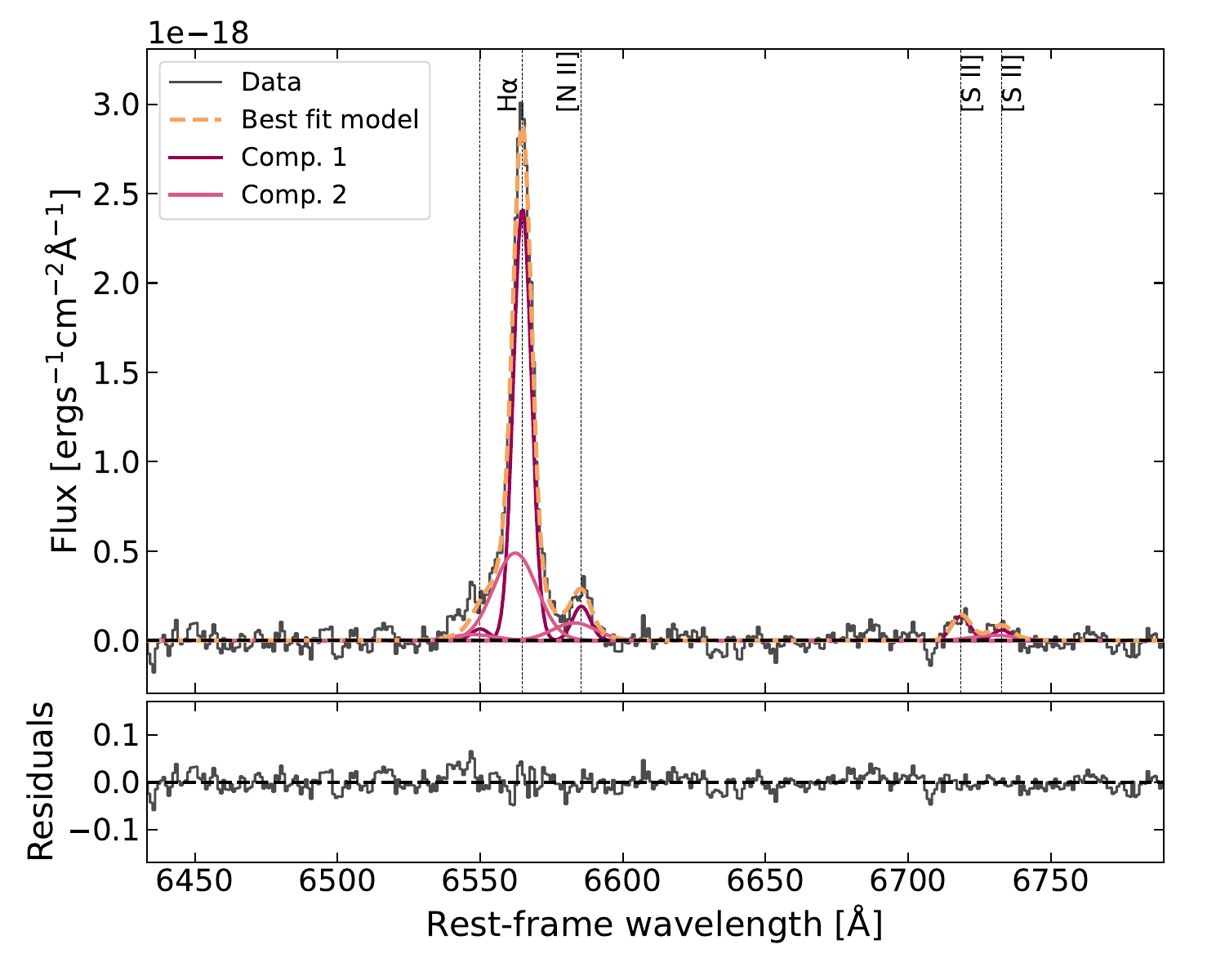}
    \caption{\textit{Top}: \oiii maps of COS590. From top to bottom: total, narrow and broad components. From Left to Right: flux, velocity and velocity dispersion. Black contours mark the outflow region used to produce the outflow-integrated spectra. \textit{Bottom}: Close-up on the \hbe+\oiii complex (Left) and the \hal+\nii+\sii complex (Right) of the continuum- and BLR-subtracted spectrum of COS590 integrated over the 3$\sigma$ mask of the broad \oiii emission, excluding the spaxels at the position of the other components northern of the AGN because possibily associated to tidal tails and not outflows. Data and residuals are in black, total model is in dashed orange, single line components are shown as dark to light purple. Component 1 corresponds to the narrow component, additional components add up to the broad component. }
    \label{fig:maps2_cos590_oiii}
\end{figure*}

\begin{figure*}[!h]
    \centering
    \includegraphics[width=\textwidth]{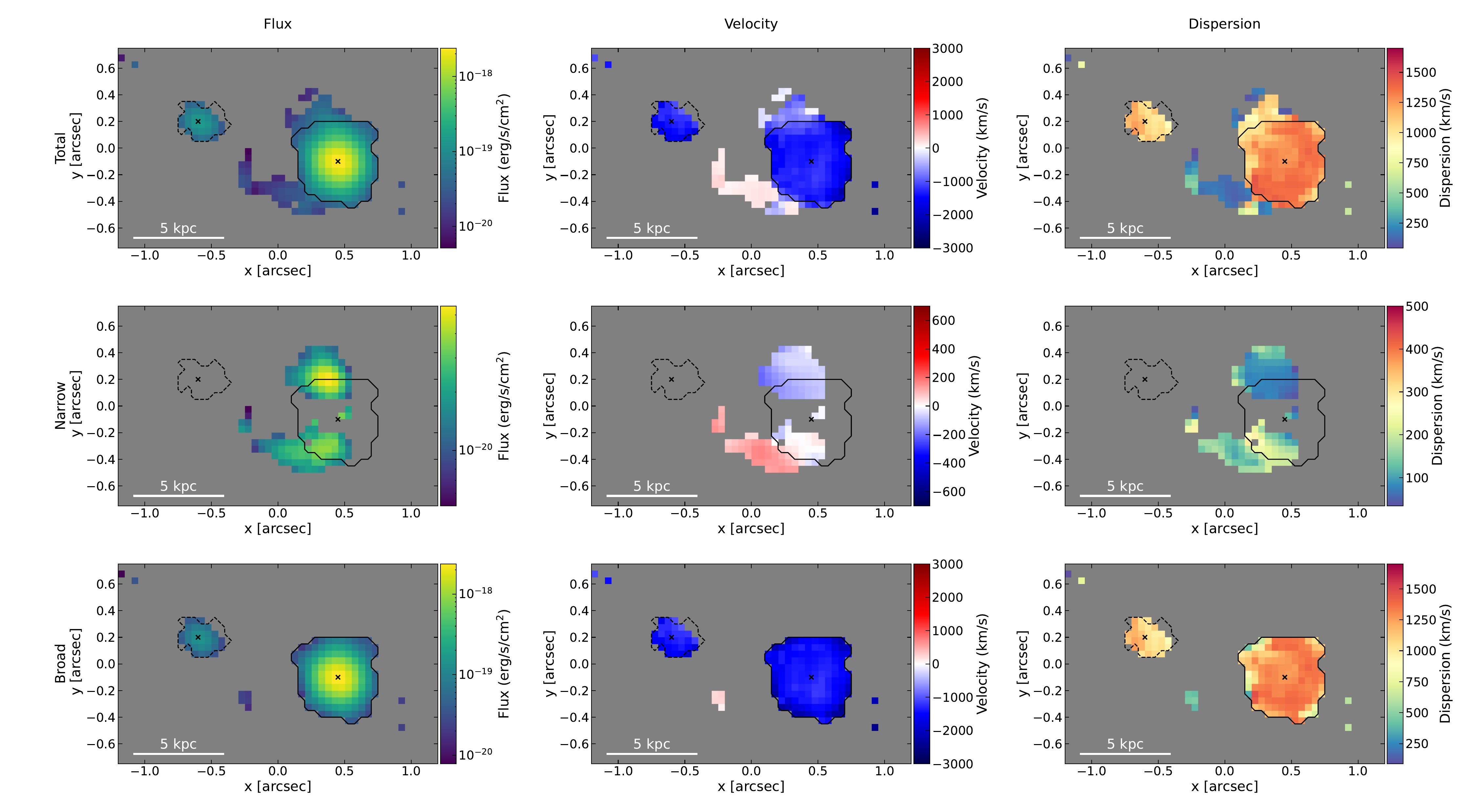}
    \\
    \includegraphics[width=0.45\textwidth]{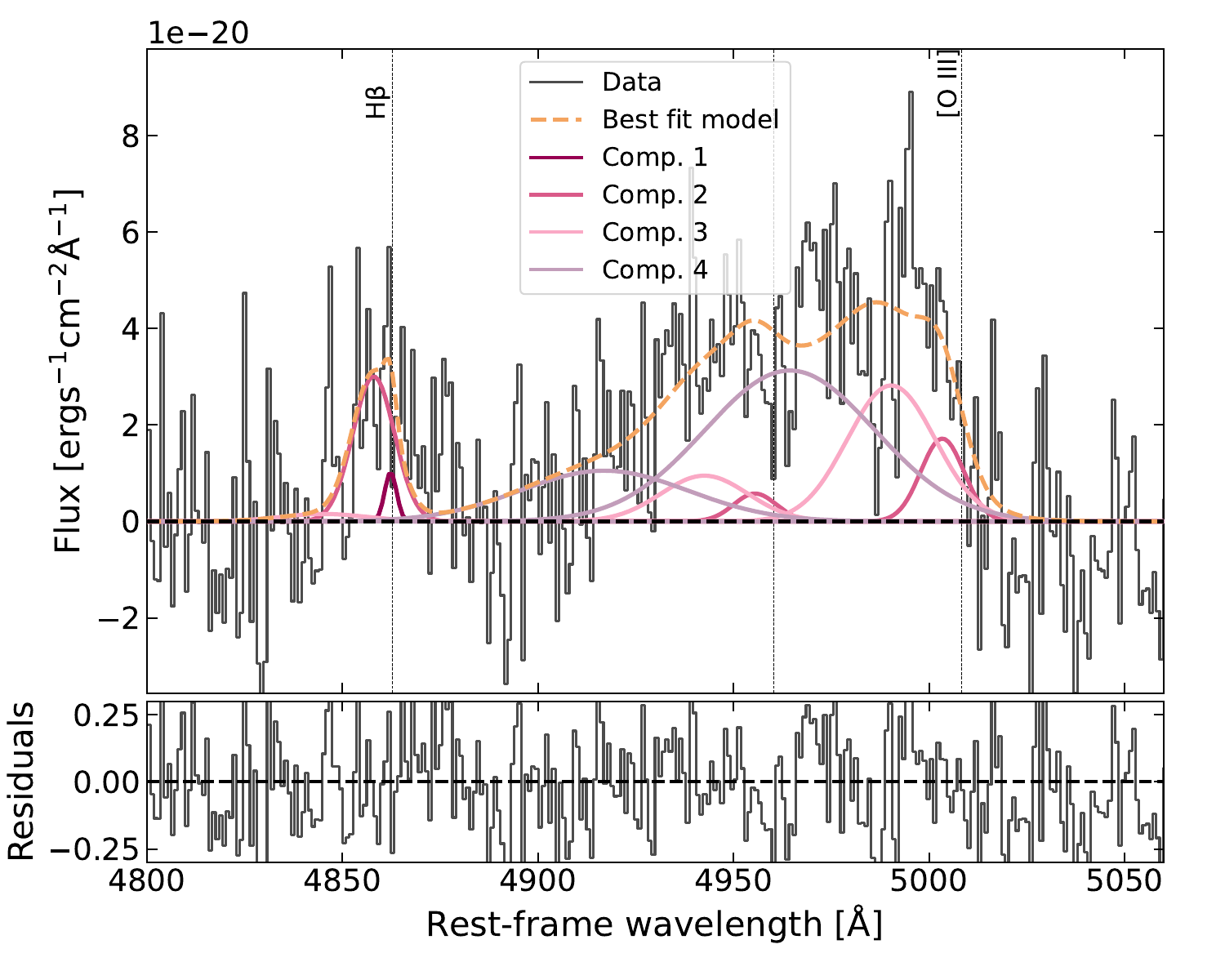}
    \hfil
    \includegraphics[width=0.45\textwidth]{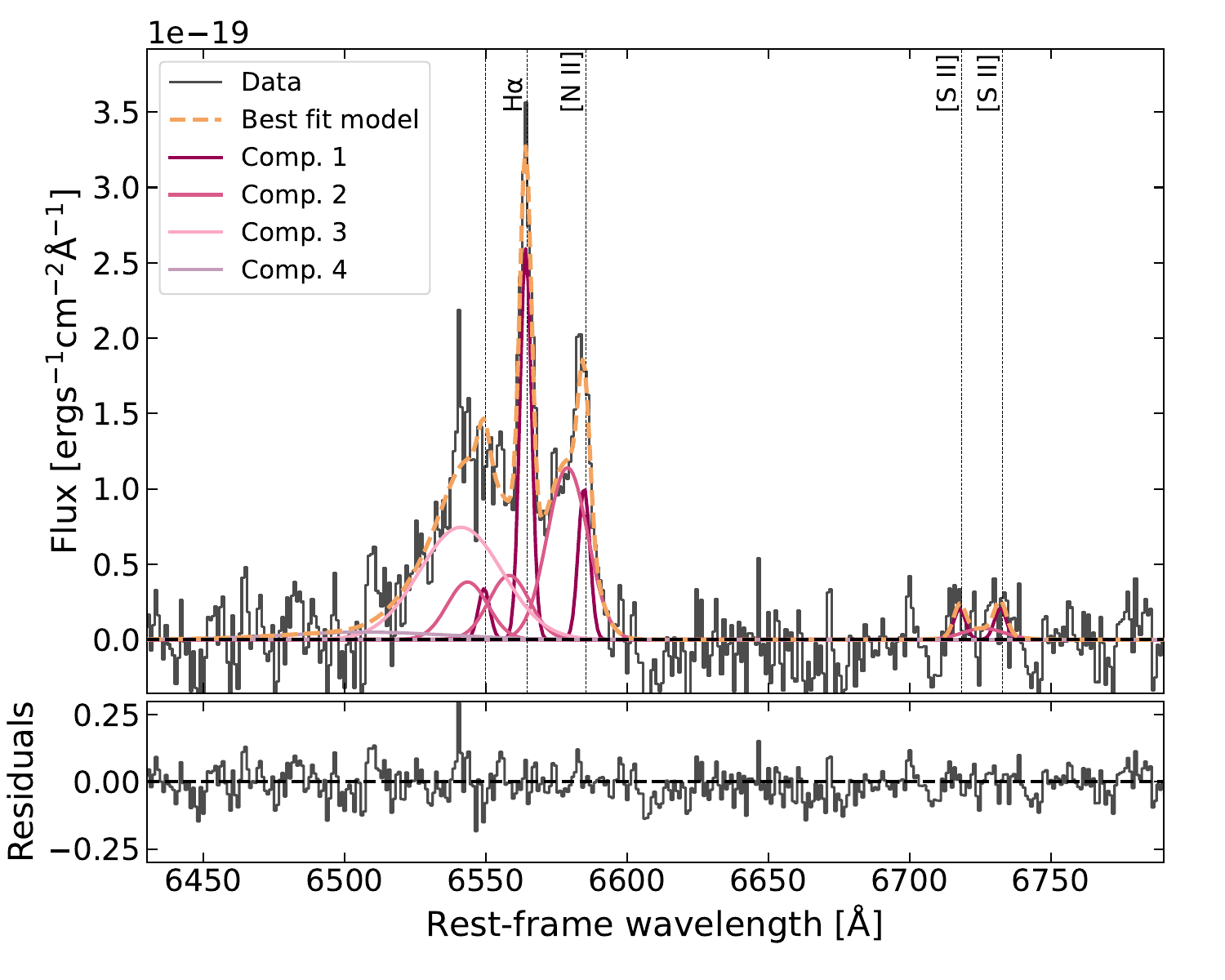}
    \caption{\textit{Top}: \oiii maps of COS1638 AGNi A and B. From top to bottom: total, narrow and broad components. From Left to Right: flux, velocity and velocity dispersion. Solid (dashed) black contours mark the outflow region used to produce the outflow-integrated spectra of COS1638-A (COS1638-B). \textit{Bottom}: Close-up on the \hbe+\oiii complex (Left) and the \hal+\nii+\sii complex (Right) of the continuum- and BLR-subtracted spectrum of COS1638 AGN B integrated over the 3$\sigma$ mask of the broad \oiii emission at the position of AGN A. Data and residuals are in black, total model is in dashed orange, single line components are shown as dark to light purple. Component 1 corresponds to the narrow component, additional components add up to the broad component. }
    \label{fig:maps2_cos1638_oiii}
\end{figure*}

\begin{figure*}[!h]
    \centering
    \includegraphics[width=0.9\textwidth]{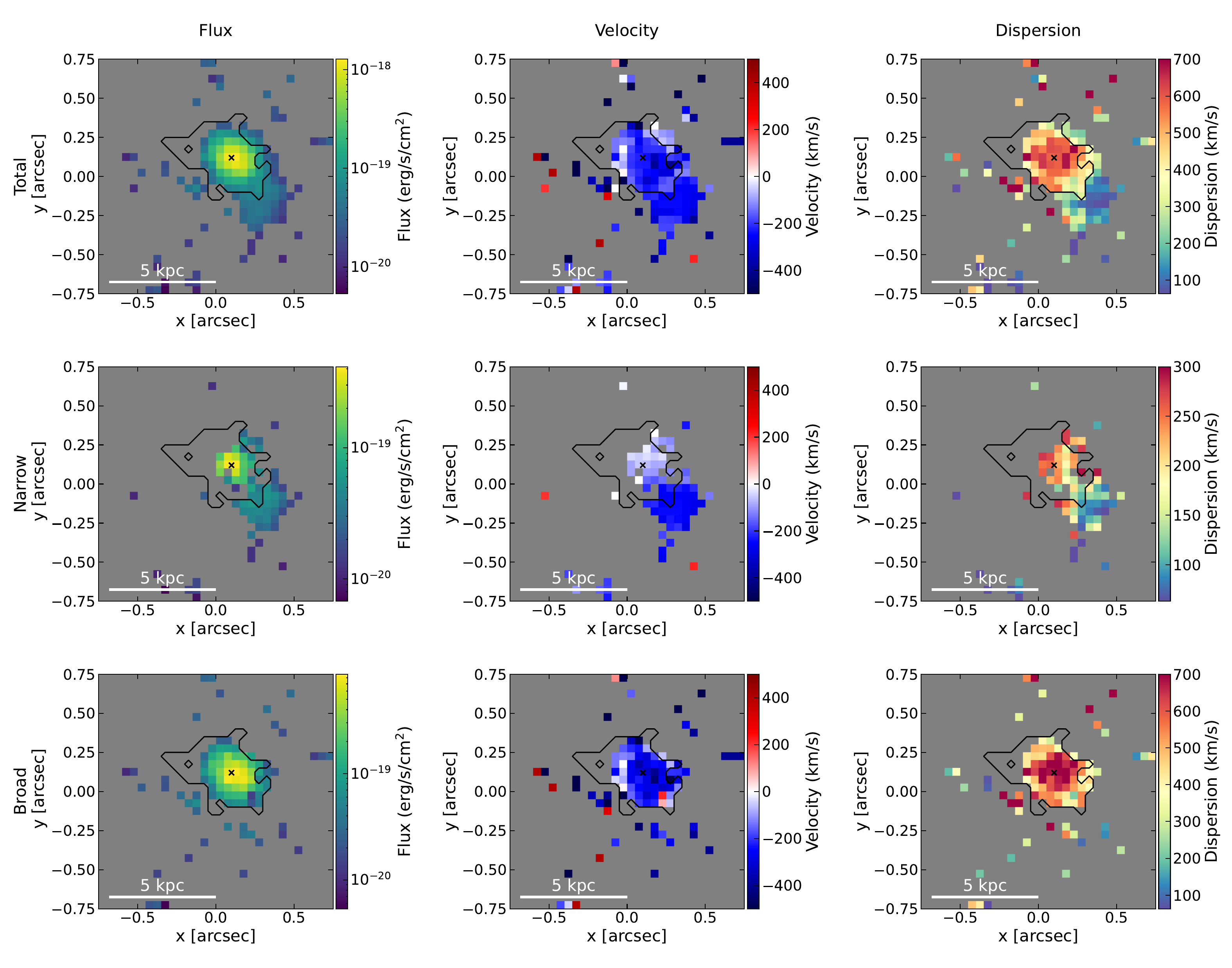}
    \\
    \includegraphics[width=0.45\textwidth]{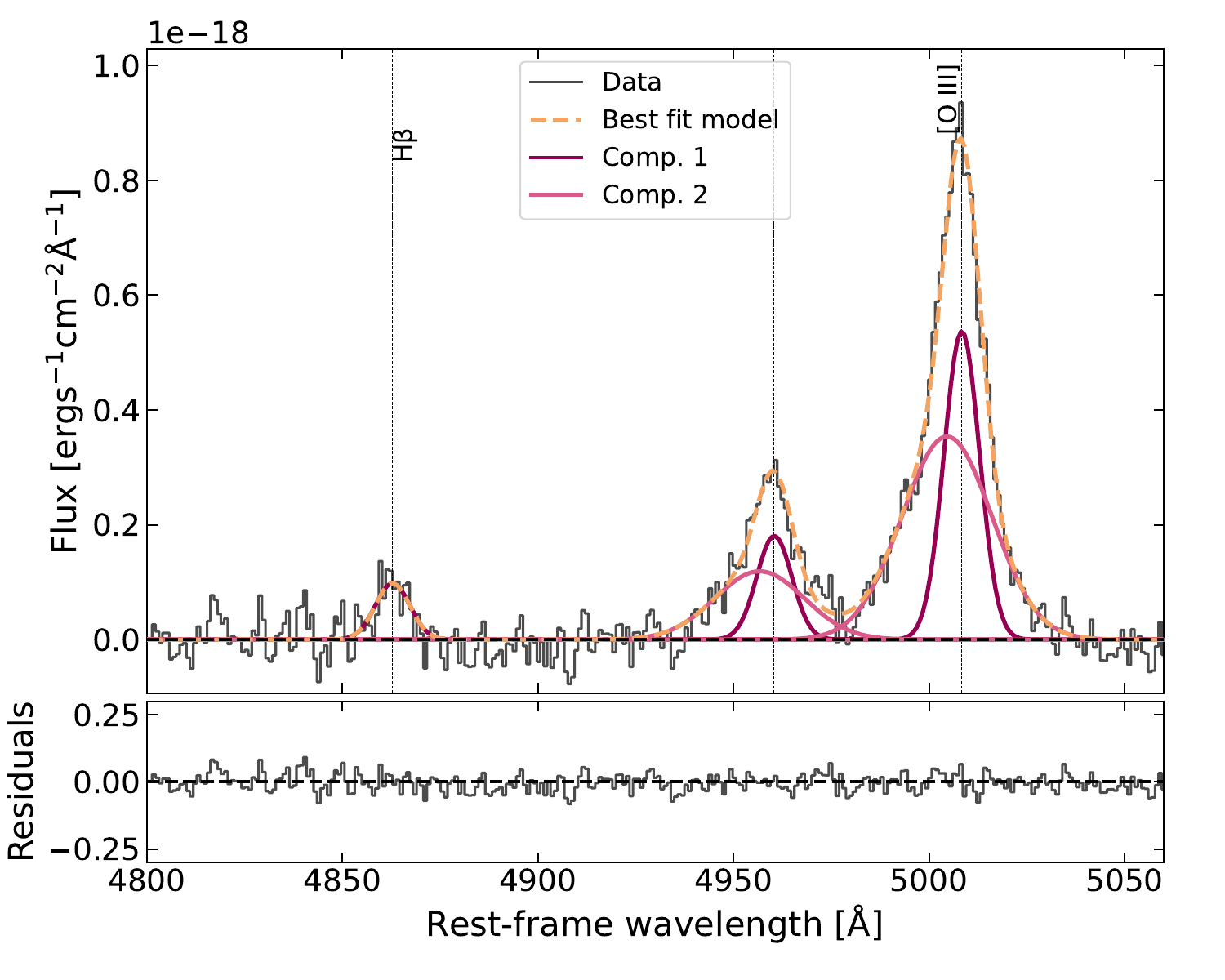}
    \hfil
    \includegraphics[width=0.45\textwidth]{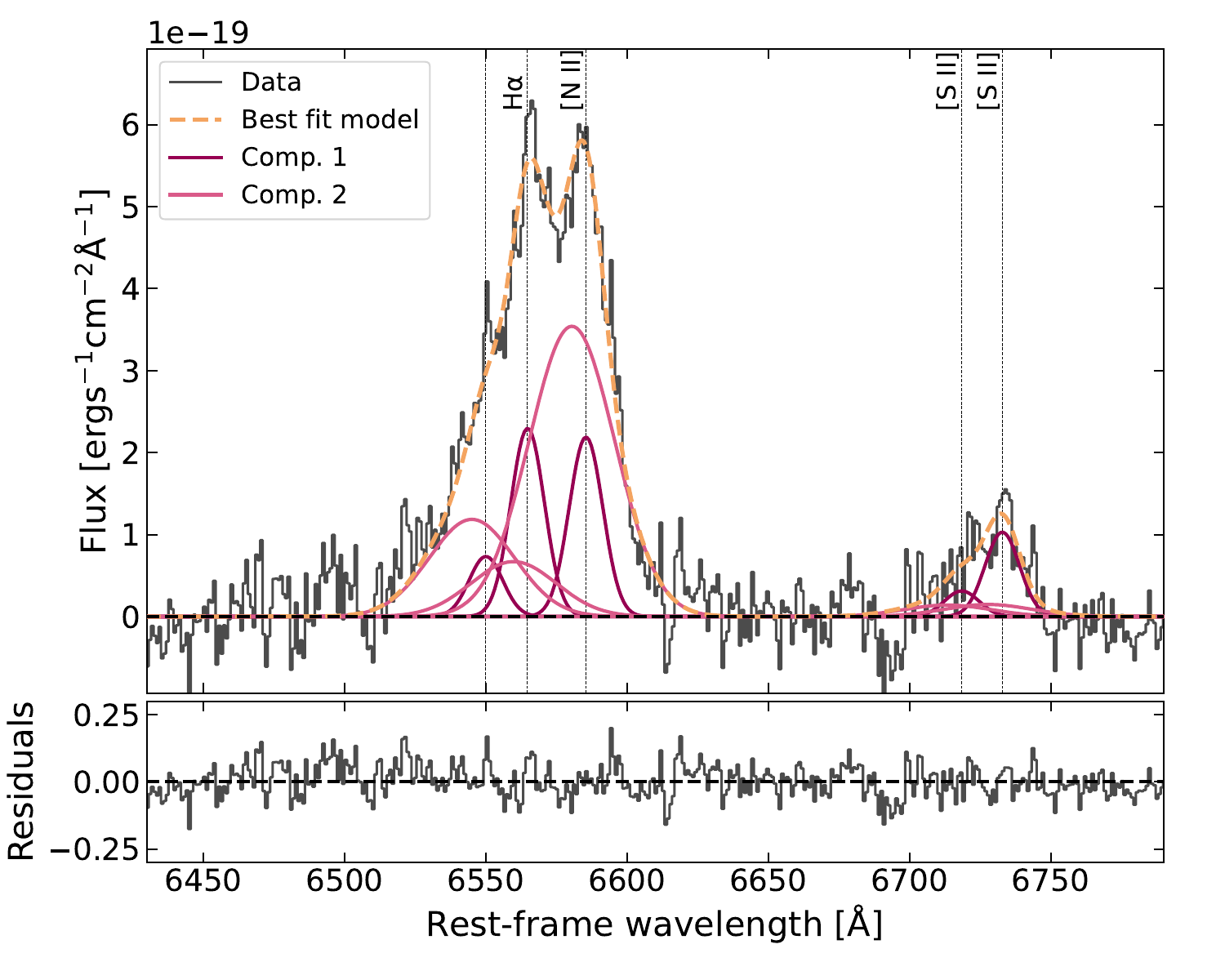}
    \caption{\textit{Top}: \oiii maps of COS1656 AGN A. From top to bottom: total, narrow and broad components. From Left to Right: flux, velocity and velocity dispersion. Black contours mark the outflow region used to produce the outflow-integrated spectra. \textit{Bottom}: Close-up on the \hbe+\oiii complex (Left) and the \hal+\nii+\sii complex (Right) of the continuum- and BLR-subtracted spectrum of COS1656 AGN A integrated over the 3$\sigma$ mask of the broad \oiii emission. Data and residuals are in black, total model is in dashed orange, single line components are shown as dark to light purple. Component 1 corresponds to the narrow component, additional components add up to the broad component. }
    \label{fig:maps2_cos1656a_oiii}
\end{figure*}

\begin{figure*}[!h]
    \centering
    \includegraphics[width=0.9\textwidth]{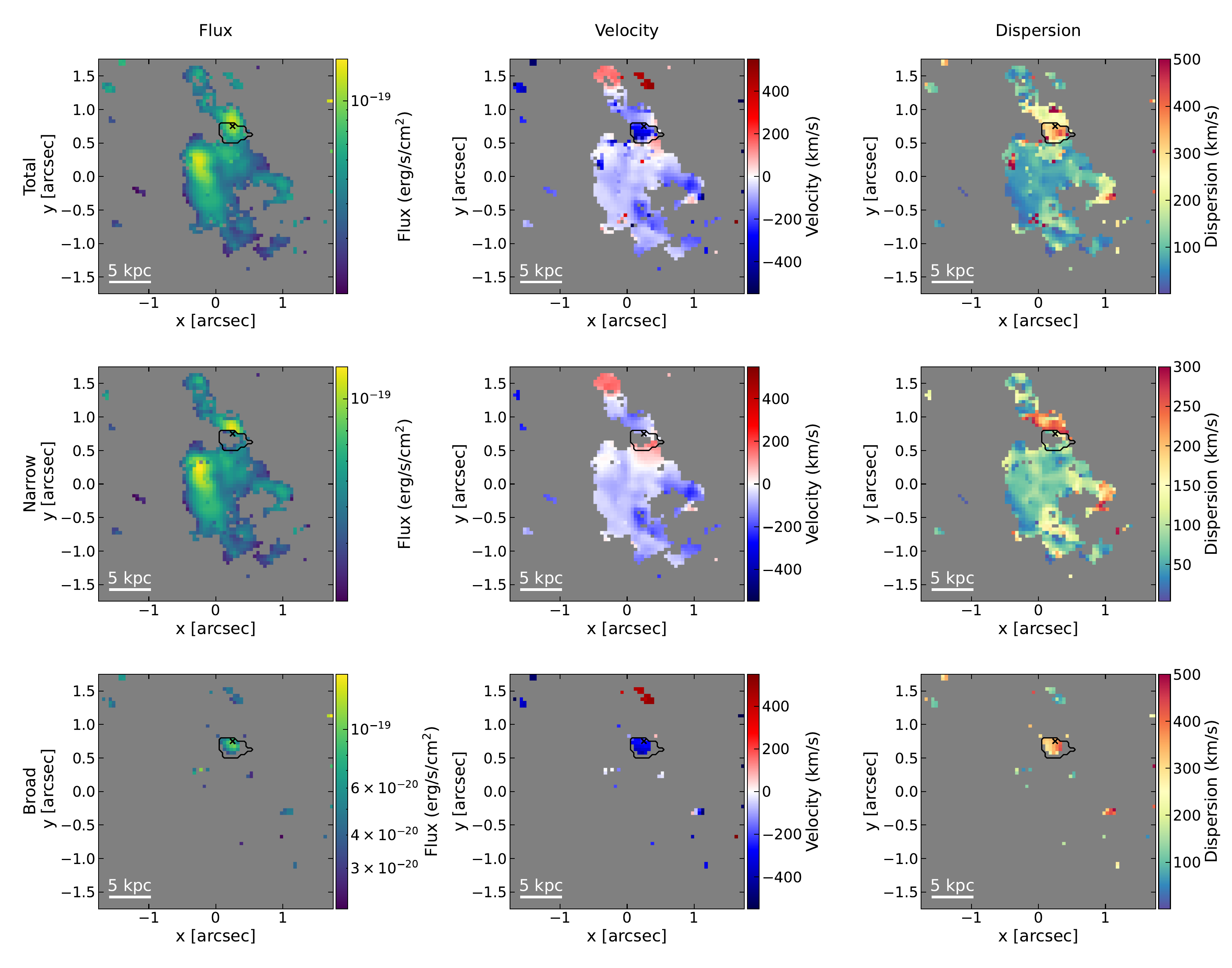}
    \\
    \includegraphics[width=0.45\textwidth]{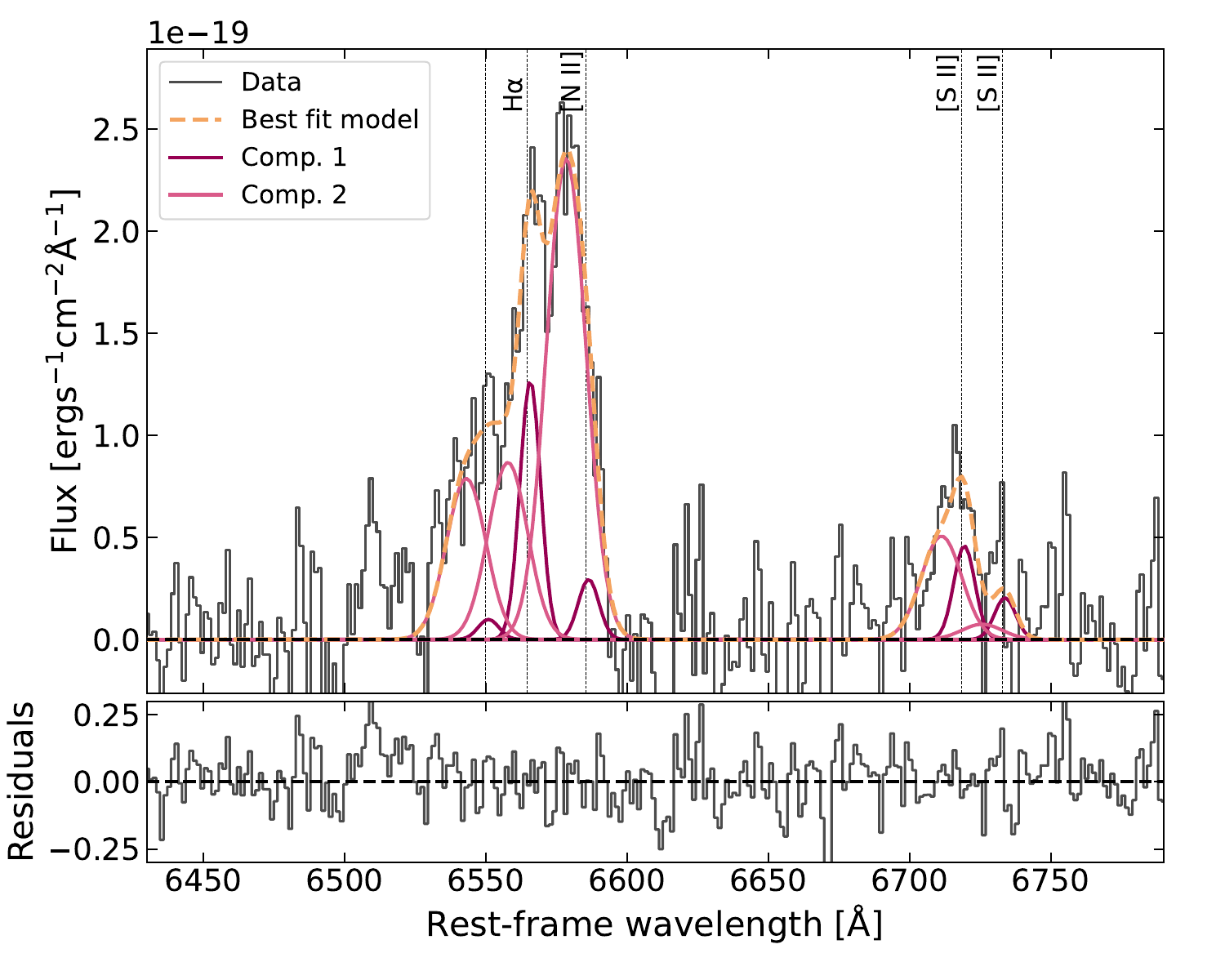}
    \caption{\textit{Top}: \hal maps of COS2949. From top to bottom: total, narrow and broad components. From Left to Right: flux, velocity and velocity dispersion. Black contours mark the flux levels of the corresponding kinematic component (total, broad, narrow) and they start at 3$\sigma$ and increase by 5$\times$. \textit{Bottom}: Close-up on the \hal+\nii+\sii complex (Right) of the continuum-subtracted spectrum of COS2949 integrated over the 3$\sigma$ mask of the broad \nii emission. Data and residuals are in black, total model is in dashed orange, single line components are shown as dark to light purple. Component 1 corresponds to the narrow component, additional components add up to the broad component. }
    \label{fig:maps1_cos2949_ha}
\end{figure*}

We present in this appendix the full spectra (Fig. \ref{fig:blr_of_1}-\ref{fig:spectra_type_2}), the maps of total, narrow and broad \oiii emission line components plus the close-up of ionized-gas integrated spectra of COS-AGNi (Fig. \ref{fig:maps2_cos1118_oiii}-\ref{fig:maps2_cos1656a_oiii}; see Sect. \ref{sec:data_analysis}). For COS2949, we show the \hal maps (Fig. \ref{fig:maps1_cos2949_ha}) since \oiii is outside of the wavelength range probed by JWST/\nirspec IFU. For COS1638, we present the \hal maps in the main text (see Fig. \ref{fig:maps2_cos1638_ha}) given that the filament connecting the two AGNi is not visible in \oiii. Figures  \ref{fig:maps2_cos1118_oiii}-\ref{fig:maps2_cos1656a_oiii} are complemented with the closeup of ionized-gas integrated spectra of the corresponding target between 4800-5060$\AA$ and 6460-6790$\AA$ (rest frame) showing the main rest-frame optical emission lines and their best fit models. Again, for COS2949 we only show the 6460-6790$\AA$ rest-frame range given that the JWST band starts at $\simeq5460\AA$ rest frame.

\end{appendix}

\end{document}